\documentclass[%
 reprint,
 amsmath,amssymb,
 aps,
floatfix,
]{revtex4-2}
\usepackage{afterpage}
\usepackage{graphicx}% Include figure files
\usepackage{dcolumn}% Align table columns on decimal point
\usepackage{bm}% bold math
\usepackage[tracking=true]{microtype}
\usepackage{braket}
\usepackage{siunitx}

\usepackage[dvipsnames]{xcolor}
\usepackage{hyperref}

\hypersetup{
	colorlinks=true,       % false: boxed links; true: colored links
	linkcolor=VioletRed,  % color of internal links (change box color with linkbordercolor)
    citecolor=JungleGreen,        % color of links to bibliography
    filecolor=Orchid,      % color of file links
    urlcolor=black           % color of external links
}

\newcommand{\RN}[1]{%
  \textup{\uppercase\expandafter{\romannumeral#1}}%
  }
\usepackage{tikz}

\newcommand{\dbarw}[1]{d\hspace*{-0.08em}\bar{}\hspace*{0.1em}\omega_{#1}}
\DeclareMathOperator{\sinc}{sinc}

\usepackage{comment}
\usepackage{fancyhdr}
\fancyhfoffset[L]{1cm} % left extra length
\fancyhfoffset[R]{1cm} % right extra length
\numberwithin{equation}{section}
\begin{document}
\thispagestyle{fancy}
\preprint{APS/123-QED}

\title{Aspects of two-photon absorption of squeezed light: the CW limit}% Force line breaks with \\

\author{C. Drago}
 \email{christian.drago@mail.utoronto.ca}

\author{J. E. Sipe}%

\affiliation{%
 Department of Physics, University of Toronto, 60 St. George Street, Toronto, Ontario, Canada, M5S 1A7
}%
\date{\today}

\begin{abstract}
We present a theoretical analysis of two-photon absorption of classical and squeezed light valid when one-photon absorption to an intermediate state is either resonant or far-detuned from resonance, and in both the low and high intensity regimes. In this paper we concentrate on continuous-wave excitation, although the approach we develop is more general. We calculate the energy removed from an incident field for typical experimental parameters and consider the limiting cases when the photon pairs are narrowband or broadband compared to the molecular linewidths. We find an enhancement of the two-photon absorption due to resonant contributions from the large squeezed light bandwidth and due to photon bunching in the low intensity regime.  However, in both cases, for the parameters we choose, the one-photon absorption is the dominant process in the region of parameter space where a large enhancement of the two-photon absorption is possible. 
\end{abstract}

\maketitle
\thispagestyle{fancy}
\section{introduction \label{sec:Introduction}}
With sources of non-classical light becoming widely available, the study of advantages such light might offer in spectroscopic and sensing methods is an active area of research \cite{schlawin2018entangled,schlawin2017entangled,dorfman2016nonlinear,szoke2020entangled,mukamel2020roadmap}. One current research topic is whether or not the use of broadband photon pairs can lead to enhanced rates for two-photon absorption, thereby increasing the 
effectiveness of applications that rely on classical two-photon absorption, such as two-photon microscopy, photodynamic cancer therapy, and 3D photopolymerization \cite{ashworth2007two,bhawalkar1997two,ostendorf2006two}.

Photon pairs can be generated such that their
correlation time is much less than molecular decay times, while also being anti-correlated in frequency such that pairs of photons have the correct total energy to lead to two-photon absorption. It was first argued in the 1980s that correlated pairs of photons would lead to a scaling of two-photon absorption linear with the incident intensity \cite{klyshko1982transverse,gea1989two,javanainen1990linear}. 
This was shortly confirmed with experiments on trapped atomic cesium \cite{georgiades1995nonclassical}, and followed by many theoretical studies of two-photon absorption \cite{schlawin2018entangled,schlawin2017entangled,dorfman2016nonlinear,dayan2007theory,schlawin2017entangled,fei1997entanglement,schlawin2013photon,oka2018two} and its potential advantage in applications.

Other experiments also followed on two-photon absorption \cite{dayan2004two,tabakaev2021energy,villabona2018two,li2020squeezed} and sum frequency generation \cite{dayan2005nonlinear,sensarn2009resonant}, and some researchers have advocated for the development of two-photon microscopy \cite{varnavski2020two}. In each of these studies it is claimed that two-photon absorption can be achieved with the use of photon pairs at photon fluxes many orders of magnitude lower than would be required were classical light used. However, these results have been called into question by recent work, where
no enhancement due to photon pairs is observed \cite{parzuchowski2020setting,landes2020experimental,corona2022experimental}, and an alternate explanation for the experimental data can be provided \cite{mikhaylov2021hot,hickam2022single}.

A new theoretical initiative has been led by Raymer et al. \cite{raymer2020two,raymer2021large,landes2021quantifying,raymer2021entangled}, where the low flux (``isolated pair'') regime was considered.  This was later generalized to the high flux regime \cite{raymer2022theory}. In both studies they considered the ``far detuned limit," i.e., for the incident light used it is assumed that there are no one-photon resonances involving the ground state and intermediate states of the molecule. Their conclusion is that in the 
low flux regime the predicted rate of two-photon absorption should be undetectable with current technologies, while in the high flux regime, where the squeezed light bandwidth is much broader than molecular linewidths, squeezed light does not provide any enhancement. These new theoretical results are in agreement with the results of recent experiments \cite{parzuchowski2020setting,landes2020experimental,corona2022experimental,mikhaylov2021hot}.

In the far detuned limit, the low flux regime can be treated by usual perturbation approaches. For the high flux regime, there are, broadly speaking, two strategies that can be employed. The first is to ``begin at the beginning," with a nonlinear interaction Hamiltonian governing the generation of broadband squeezed light generated via a narrowband pump \cite{dayan2007theory}. This was the approach of 
Dayan, who found that in the high flux regime there are two quadratic dependencies on intensity, labeled as ``coherent'' and ``incoherent''.   Alternatively, if the incident field is squeezed light generated by an ultrashort pump pulse, one can employ a Schmidt decomposition of the squeezed state, allowing for the calculation of the time-frequency correlation functions governing the absorption process \cite{schlawin2013photon}. Here one must sum over the Schmidt modes, but the structure of the calculation is close in form to the structure of the calculation following from the approach of Dayan. Indeed, in their recent generalization to the high flux regime Raymer et al. \cite{raymer2022theory} used a method closely related to Dayan's with similar results.

What has generally not been addressed is the ``resonance limit," by which we mean that the incident light is on or close to a one-photon resonance between the ground state and one of the excited states of the molecule. In this article we present a new description of two-photon absorption of squeezed light (and classical light) valid in both the far-detuned and resonance limits, in both the low and high intensity regimes, and for both CW and pulsed excitation.

Our formalism differs from past treatments in that we work with Heisenberg operators instead of employing a density operator approach \cite{gerry2005introductory,mukamel1999principles,raymer2021entangled}. Within our formalism we include the effects of both non-radiative decay and dephasing in our calculations, the first of which is treated dynamically in the Markov limit, but could be generalized to beyond that limit. Further, we calculate the total energy removed from the incident field instead of the final state populations. This has the advantage that it is directly related to what is measured by experiment, although a simple distinction between one- and two-photon absorption is not always apparent.

A central quantity in squeezed light two-photon absorption is the photon correlation functions. Our calculation of the squeezed light correlation functions differs from those of Dayan \cite{dayan2007theory} and Raymer et al. \cite{raymer2022theory} in that we begin with a joint spectral amplitude for the incident squeezed light, and derive a field operator transformation that is used to evaluate the time-frequency correlation functions leading to the absorption. We use a simple model of the joint spectral amplitude of the squeezed light to derive explicit functions and scalings with intensity, and consider them in situations where the photons are narrowband compared to the molecular linewidths, and in situations where they are broadband compared to molecular linewidths. However, more complicated models for the squeezed light could also be implemented using our approach.

In this first communication we focus on CW excitation, and compare the energy absorption of classical light with that of squeezed light. In agreement with the results of earlier studies, we find the two scalings with intensity previously labeled as ``coherent'' and ``incoherent''. We find that when the incident squeezed light is narrow band with respect to molecular broadening, the squeezed light cross-sections are equivalent to those of classical light and the ratio of absorption is given by the normalized second order correlation function $g^{(2)}(0,0)$, in agreement with recent work \cite{raymer2022theory,cutipa2020measurement,spasibko2017multiphoton,popov2002ionization,boitier2013two}.

We then turn to a consideration of how the classical and broadband squeezed light cross-sections vary with the detuning from one-photon resonances for typical experimental parameters, extending past treatments \cite{schlawin2018entangled,schlawin2017entangled,dorfman2016nonlinear,dayan2007theory,schlawin2017entangled,fei1997entanglement,schlawin2013photon,oka2018two}. We show that the near resonant behaviour is nontrivial, and there is a significant enhancement of the two-photon absorption cross-section of broadband squeezed light when the magnitude of detuning is less than half the squeezed light bandwidth. However, for the typical parameters we adopt and in the region of parameter space where this enhancement is visible, the one-photon absorption cross-section is also significantly enhanced, and overpowers the two-photon absorption. 

In the far detuned limit, which for squeezed light is when the magnitude of detuning is greater than half the squeezed light bandwidth, and when the squeezed light bandwidth is greater than molecular linewidths, the ratio of absorption is \emph{approximately} given by $g^{(2)}(0,0)$. Our results agree with the calculation made by Raymer et al. \cite{raymer2022theory} when the squeezed light bandwidth is \emph{much} greater than molecular linewidths. Further, for broadband squeezed light in the low flux regime, where $g^{(2)}(0,0)\gg 1$ we do find that the squeezed light could provide an advantage. However, here again we find that the one-photon absorption dominates.

The outline of the paper is as follows: Working in the rotating-wave-approximation (RWA), in Section \ref{sec:model Hamiltonian} we present a model of a multilevel molecule (see Fig. \ref{fig:molecule diagram}) that allows one to explore a wide range of the parameter space. The incident field can be either far-detuned from, or near a one-photon resonance with, the excited states.

To properly model a molecule in typical experimental settings we include two sources of broadening, the nonradiative decay of excited states and dephasing. For large molecules with many degrees of freedom there are many ways in which absorbed energy can decay. To model these processes we couple the molecule to a quantum reservoir for each nonradiative decay pathway.  For typical experiments, molecules are in solution and thus subjected to broadening due to molecular collisions. To include this broadening affect we include in our Hamiltonian a stochastic fluctuation of the energy levels of the molecule.

In many previous treatments, the calculations are referred to as addressing the ``probability of two-photon absorption.'' However, long after an exciting pulse has passed the molecule it returns to its ground electronic state, and some fraction of the energy extracted from the exciting pulse has been re-emitted as fluorescence. In perhaps the most proper meaning of the word ``absorption," that fraction of the incident energy has not been ``absorbed," since it has been returned to the electromagnetic field. In Section \ref{sec:total absorption} we derive a quantum mechanical equation for the amount of energy removed from the electromagnetic field, perhaps most properly the \emph{absorption}. However, in this first communication we treat the fluorescence phenomenologically, by extending the use of the reservoir to include the effects of radiation from fluorescent molecules with high quantum efficiencies that 
are used for current squeezed light two-photon absorption experiments. Thus the ``absorption" we then calculate has its more colloquial meaning as the energy removed from the incident field.

In Sections \ref{sec:Input-output theory}-\ref{sec:absorption expressions} we develop the interaction picture in which we will be working, derive the equations of motion, and calculate the absorption for any incident field (pulsed or CW) near or far from resonance. Then in Section \ref{sec:Three level molecule} we specialize to a three level model for which we can construct closed form results and set up the calculations to follow. 

Finally, with the formalism developed and general expressions for the absorption presented, we specify to different incident fields. In Sections \ref{sec:CW Coherent state} and \ref{sec:Entangled States of Light} we provide an analysis of CW coherent and CW squeezed light in various regimes. Technical details are presented in appendices. 

\section{model Hamiltonian \label{sec:model Hamiltonian}}
\begin{figure}
    \centering
    \includegraphics[width=0.4\textwidth]{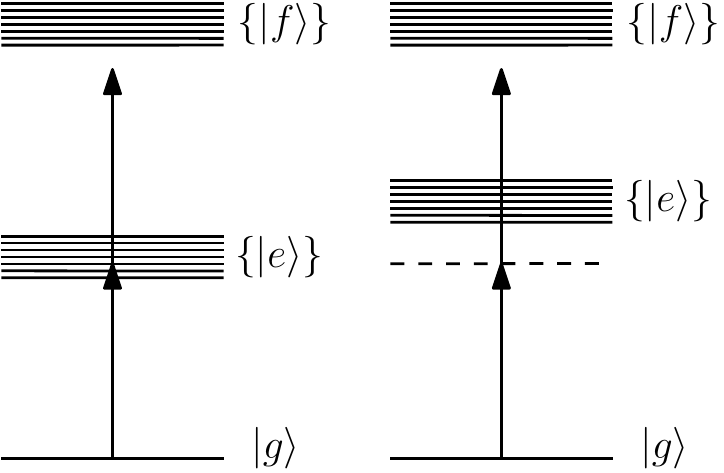}
    \caption{Level diagram for the molecular system we are considering. We make no assumptions to where the intermediate set of levels $\{\ket{e}\}$ and final levels $\{\ket{f}\}$ are, they can be near or far from resonance (within the RWA). We include the resonant and non-resonant diagram. We assume the only allowed transitions from the external field are between $\ket{g}\leftrightarrow\{\ket{e}\}$ and $\{\ket{e}\}\leftrightarrow\{\ket{f}\}$.}
    \label{fig:molecule diagram}
\end{figure}

We begin with a simple model of a molecule which has a ground state $\ket{g}$, a set of states $\{\ket{e}\}$ that can serve as intermediate states in two-photon absorption, and a set of states $\{\ket{f}\}$ that can serve as final states. For a review of quantum light-matter interactions, see, e.g. Gerry and Knight \cite{gerry2005introductory}, and Mukamel \cite{mukamel1999principles}. The level diagram for such a molecule is shown in Fig. \ref{fig:molecule diagram}. The free molecule Hamiltonian, $H_\text{M}$, is then
\begin{equation}
\label{eq:HM}
    H_\text{M} = \hbar\omega_g \sigma_{gg} + \sum_e\hbar\omega_{e} \sigma_{ee} + \sum_f\hbar\omega_{f} \sigma_{ff}
\end{equation}
where $\sigma_{ij} = \ket{i}\bra{j}$. We assume these states can be approximated as a complete basis,

\begin{equation}
    \label{eq:identity relation}
    \hat 1 =\sigma_{gg} + \sum_e\sigma_{ee} + \sum_f\sigma_{ff}. 
\end{equation}
From here on we do not explicitly indicate sums over intermediate $(e)$ or final $(f)$ states, and use the convention that whenever such state indices appear on the right-hand side of an equation, but not on the left-hand side, they are summed over. 

The free EM field Hamiltonian, $H_\text{EM}$, is given by
\begin{equation}
    H_\text{EM} =  \sum_{I}\int d\boldsymbol{k} \hbar\omega_{k} a_I^\dagger(\boldsymbol{k})a_I(\boldsymbol{k}),
\end{equation}
with $k = |\boldsymbol{k}|$, where 
\begin{equation}
    [a_I(\boldsymbol{k}),a_J^\dagger(\boldsymbol{k}^\prime)] = \delta(\boldsymbol{k} - \boldsymbol{k}^\prime)\delta_{IJ},
\end{equation}
and where the subscript labels the polarization of the field.  We treat the coupling between the molecule and the field within the electric dipole approximation, neglecting certain terms that lead to renormalization effects \cite{craig1998molecular}, by using an interaction Hamiltonian 
\begin{equation}
    H_{\text{M-EM}} = -\boldsymbol{\mu} \cdot \boldsymbol{E},
    \label{eq:H_M-EM v1}
\end{equation}
where $\boldsymbol{\mu}$ is the dipole moment operator and $\boldsymbol{E}$ is the electric field operator at the position of the molecule. These two operators commute, so we are free to order them in any way. 

We expand each operator in terms of its positive and negative frequency components $\boldsymbol{E} = \boldsymbol{E_+} + \boldsymbol{E_-}$ and $\boldsymbol{\mu} = \boldsymbol{\mu_+} + \boldsymbol{\mu_-}$, with $(\boldsymbol{E_+})^\dagger = \boldsymbol{E_-}$, $(\boldsymbol{\mu_+})^\dagger = \boldsymbol{\mu_-}$. We assume the relevant incident field components are close enough to resonances that the rotating-wave approximation (RWA) is valid; within this approximation the interaction is given by

\begin{equation}
    H_{\text{M-EM}} = -\boldsymbol{\mu_-}\cdot\boldsymbol{E_+} - \boldsymbol{E_-}\cdot \boldsymbol{\mu_+},
    \label{eq:H_M-EM v3}
\end{equation}
 where we have chosen the normally ordered form, which is the simpler to work with. In Appendix \ref{app:sec:Absorption, Extinction and Scattering} we confirm that the results are independent of the order chosen. 

We assume that the only dipole allowed transitions of interest are $\ket{g}\leftrightarrow\{\ket{e}\}$ and $\{\ket{e}\}\leftrightarrow\{\ket{f}\}$. The transitions are sketched in Fig. \ref{fig:molecule diagram}, where we indicate one scenario where there is resonance with an intermediate state and one where there is not.
For atomic systems, where parity is a good quantum number, if the transitions $\ket{g}\leftrightarrow\{\ket{e}\}$ and $\{\ket{e}\}\leftrightarrow\{\ket{f}\}$ are allowed then the transition $\ket{g}\leftrightarrow\{\ket{f}\}$  would be forbidden; for large florescent molecules the situation is more complicated. In any case, in this paper we treat any direct transition $\{\ket{f}\}\leftrightarrow \ket{g}$, or indeed any transition from $\ket{f}$ to other states that might exist with energies close to that of $\ket{g}$ but not shown in Fig. \ref{fig:molecule diagram}, in a phenomenological way, if they are relevant, as we discuss in Section III below. The expansion of the dipole moment $\boldsymbol{\mu_+}$ using the molecular basis states is then
\begin{equation}
\label{eq:muplus}
    \boldsymbol{\mu_+} = \boldsymbol{\mu}_{ge}\sigma_{ge} + \boldsymbol{\mu}_{ef}\sigma_{ef},
\end{equation}
where $\boldsymbol{\mu}_{ij} = \bra{i}\boldsymbol{\mu}\ket{j}$.  

In many two-photon absorption experiments \cite{tabakaev2021energy,villabona2018two,varnavski2020two,parzuchowski2020setting,landes2020experimental,corona2022experimental,mikhaylov2021hot}, large molecules are put in solution and then irradiated. Such molecules have many degrees of freedom that can be involved in the non-radiative decay of optically excited electronic states, leading to a loss of energy from the electromagnetic field. As well, the molecule suffers fluctuations in the local environment that can lead to fluctuations in its energy levels, and thus to dephasing of induced dipole moments. We include both effects in the following sections. 

\subsection{Non-radiative decay}
\label{sec:Non-radiative decay}
To model the first of these we couple the molecule to a quantum reservoir. Such a reservoir is often taken to be a set of harmonic oscillators, with the coupling to the system of interest treated in the Markov limit; this leads to a Lindblad equation for the reduced density operator of the system \cite{breuer2002theory}.  However, since here we want to keep the details of correlations between the molecule and the electromagnetic field in our equations, only the reservoir would be ``traced over," and the reduced density operator of the system of interest -- molecule and electromagnetic field -- would act on the combined Hilbert space of the molecule \emph{and} the electromagnetic field. To avoid such a complicated entity we work instead in the Heisenberg picture, and use the result of Fischer \cite{fischer2018derivation} that dynamics at the Markov level can be modeled by coupling each transition of interest to the excitations in a formal 1D waveguide, which is taken as a reservoir at zero temperature. In such a model, quantum fluctuations in the waveguide propagate towards the molecule before interacting with it, and then away from it after the interaction, assuring that the molecule is effectively always interacting with the vacuum state of the quantum reservoir. This idealizes the physical assumption that the coherence generated between the electronic transitions of the molecule and the other degrees of freedom, which is responsible for the loss of energy from the electromagnetic field, quickly decays. 

In our implementation of this approach, we assume there are independent non-radiative decay pathways of three types, involving transitions between $\ket{g}\leftrightarrow \{\ket{e}\}$, $\{\ket{e}\}\leftrightarrow \{\ket{f}\}$ and $\ket{g}\leftrightarrow\{\ket{f}\}$ respectively.  We use the form for a 1D waveguide Hamiltonian introduced earlier \cite{helt2010spontaneous}, where the propagation of light in integrated photonic structures was discussed.  For each decay process we introduce the waveguide “reservoir operators” $\psi_{ij}(z)$ and $\psi^\dagger_{ij}(z)$, which will lead to transitions between $\ket{i}$ and $\ket{j}$, with $\omega_i>\omega_j$.  For a single waveguide, the free reservoir Hamiltonian is given by
\begin{equation}
\begin{split}
    \label{eq:reservoir hamlitonian for ij}
        H_\text{R}^{ij} = \hbar \omega_{ij}&\int dz \psi^\dagger_{ij}(z)\psi_{ij}(z) \\
        & +\frac{i\hbar v_{ij}}{2}\int dz \left(\frac{d\psi^\dagger_{ij}(z)}{dz}\psi_{ij}(z) - \text{H.c.}\right),
\end{split}
\end{equation}
where $\omega_{ij}=\omega_{i}-\omega_{j}$ is taken as a center frequency and $v_{ij}$ is the group velocity of excitation propagation in the waveguide. We take the waveguide operators to satisfy the usual commutation relation
\begin{gather}
    [\psi_{ij}(z),\psi_{mn}^\dagger(z^\prime)] = \delta_{im}\delta_{jn}\delta (z - z^\prime),
\end{gather}
which ensures the dynamics of all the reservoir operators are independent. The full reservoir Hamiltonian is then written as the sum of three contributions, 
\begin{equation}
    \label{eq:reservoir hamlitonian}
    H_\text{R}  = H_\text{R}^{eg} + H_\text{R}^{fe} + H_\text{R}^{fg},
\end{equation}
where $H_\text{R}^{eg}$, $H_\text{R}^{fe}$ and $H_\text{R}^{fg}$ contain reservoir operators that facilitate transitions between $\ket{g}\leftrightarrow \{\ket{e}\}$, $\{\ket{e}\}\leftrightarrow \{\ket{f}\}$ and $\ket{g}\leftrightarrow\{\ket{f}\}$, respectively. Again, and below, there is an implicit sum over the intermediate and final states.

For the interaction between a transition $\sigma_{ij}$ and its associated reservoir, we assume a point coupling between the molecule and the waveguide modeling the reservoir. After applying the RWA, the coupling responsible for transitions between $\ket{i}\leftrightarrow\ket{j}$ due to the reservoir is given by \cite{helt2010spontaneous}
\begin{equation}
    \label{eq:reservoir interaction ij}
    H_\text{M-R}^{ij} = \hbar \eta_{ij}^*\sigma_{ij}\psi_{ij}(0) + \hbar \eta_{ij}\psi^\dagger_{ij}(0)\sigma_{ji},
\end{equation}
taking the ``position" of the molecule to be the origin of the fictitious waveguide, with the full coupling Hamiltonian then given by
\begin{equation}
    \label{eq:reservoir interaction}
    H_\text{M-R} = H_\text{M-R}^{eg} + H_\text{M-R}^{fe} + H_\text{M-R}^{fg}.
\end{equation}

In equation \eqref{eq:reservoir hamlitonian for ij} we introduced a nominal reference frequency and group velocity, and in Eq. \eqref{eq:reservoir interaction ij} we introduced a nominal coupling constant $\eta_{ij}$. In fact, the group velocity $v_{ij}$ and coupling $\eta_{ij}$ will lead to a new constant that will identify the non-radiative decay widths associated with each transition $\ket{i}\leftrightarrow\ket{j}$; it will be the physical parameter that can be identified by experiment, and there will be one of these physical parameters for each coupling of a transition to a waveguide reservoir. 

\subsection{Dephasing}
\label{sec:Dephasing}
We now turn to the effects of fluctuations in the energy levels of the molecule that arise due to fluctuations in its environment. The dephasing that results could be captured by a density operator formalism with a quantum reservoir model such as that described above, but one that would involve terms $H^{ij}_\text{M-R}$ with $i=j$ \cite{skinner1986pure}. In the low temperature regime such a treatment is necessary, but one can consider a simpler model of dephasing that reproduces the fully quantum treatment at higher temperatures \cite{skinner1986pure,kubo1969stochastic,mukamel1983nonimpact,mukamel1999principles}. In that simpler model the molecule is subject to classical stochastic fluctuations in its energy eigenvalues, described with a contribution to the Hamiltonian given by   
\begin{equation}
    \label{eq:H_fluc}
    H_{\text{fluc}}(t) = \hbar\tilde{\omega}_g(t) \sigma_{gg} + \hbar\tilde{\omega}_{e}(t) \sigma_{ee} + \hbar\tilde{\omega}_{f}(t) \sigma_{ff},
\end{equation}
where the $\tilde{\omega}_i(t)$ are classical stochastic functions of time. We make four assumptions concerning these random variables \cite{skinner1986pure,kubo1969stochastic}: i) They have zero mean; ii) The correlation between state $\ket{i}$ and $\ket{j}$ is characterized by $c_{ij}$; iii) the correlation between two variables at different times follows a fast exponential decay with a time scale $\tau_c$; and iv) the random variables follow a Gaussian distribution. We denote the classical expectation value as $[\![\cdot ]\!]$, and then the first three assumptions are  summarized by
\begin{subequations}
    \label{eq:properties of stochastic variables}
    \begin{gather}
        [\![ \tilde{\omega}_i(t) ]\!] = 0\\
        [\![ \tilde{\omega}_i(t)\tilde{\omega}_j(t^\prime) ]\!] = c_{ij}e^{-|t-t^\prime|/\tau_c} \label{eq:stochastic correlation function}.
    \end{gather}
\end{subequations}
In the full problem we calculate all quantities of interest, and then finally take the stochastic averages as given above to describe the effects of dephasing.

Then the full Hamiltonian for our model is
\begin{equation}
\label{eq:Hfull}
H(t) = H_0 + V(t),
\end {equation}
where $H_0$ is the free Hamiltonian given by
\begin{equation}
    \label{eq:H_0}
    H_0 = H_\text{M}  + H_\text{EM} + H_\text{R},
\end{equation}
and $V(t)$ the time dependent interaction term given by
\begin{equation}
    \label{eq:V(t)}
    V(t) = H_\text{M-EM} + H_\text{M-R} + H_{\text{fluc}}(t).
\end{equation}
We note that although $V(t)=V^\dagger(t)$ is time dependent it is still a Schr\"{o}dinger operator.

To write the interaction Hamiltonian in a compact form we define the operators
\begin{equation}
\label{eq:Fij+}
    F_+^{ij} = \frac{\boldsymbol{\mu}_{ij}}{\hbar}\cdot\boldsymbol{E_+} - \eta_{ij}^*\psi_{ij}(0),
\end{equation}
and their adjoints,
\begin{equation}
    (F_+^{ij})^\dagger = \frac{\boldsymbol{\mu}_{ji}}{\hbar}\cdot\boldsymbol{E_-} - \eta_{ij}\psi_{ij}^\dagger(0) \equiv F_-^{ji}.
\end{equation}
The interaction $V(t)$ can then be written as
\begin{equation}
\begin{split}
    \label{eq:interaction with F definition}
        V(t) =& -\hbar\sigma_{eg}F_+^{eg}-\hbar\sigma_{fe}F_+^{fe} -\hbar\sigma_{fg}F_+^{fg} + \text{H.c.}\\
        &+ H_{\text{fluc}}(t).
\end{split}
\end{equation}
We note that there are three $F_+^{ij}$ operators that contribute to the interaction: $F_+^{eg}$, $F_+^{fe}$, and $F_+^{fg}$. However, the last term $F_+^{fg}$ is \emph{sui} \emph{generis}. It only contains reservoir operators, as we have assumed there are no one-photon transition between $\ket{g}\leftrightarrow \{\ket{f}\}$, i.e., $\boldsymbol{\mu}_{fg} = 0$. The form of the first line in Eq. \eqref{eq:interaction with F definition} allows a simple interpretation: the energy of the molecule is increased (decreased) by annihilating (creating) energy in the external field or reservoir. 

\section{absorption(s) \label{sec:total absorption}}
The term \emph{absorption} is used in a number of different ways in optics. In one usage it refers to the removal of energy from the electromagnetic field, and in our discussion here we begin with that meaning. In Appendix A we show that, within the RWA, the absorption of light between $t_I$ and $t_F>t_I$ by a molecule is given by 
\begin{equation}
    \label{eq:quantum absorption}
    \mathcal{A} = \int\limits_{t_I}^{t_F}dt\bra{\Psi(t_0)} \boldsymbol{E_-}^H(t)\cdot\frac{d\boldsymbol{\mu_+}^H(t)}{dt}\ket{\Psi(t_0)} + \text{c.c.},
\end{equation}
where $\ket{\Psi(t_0)}$ is the initial ket ($t_0<t_I$) of the full molecule-field-reservoir system, and the superscript $H$ indicates that the operators evolve according to the full Hamiltonian. (For a discussion of the special case where the state of the electromagnetic field is a single photon, see Valente et al. \cite{valente2018work}). Although the result \eqref{eq:quantum absorption} is exact we have yet to solve for the Heisenberg operators  $\boldsymbol{\mu_\pm}^H(t)$ and $\boldsymbol{E_\pm}^H(t)$.

Besides renormalization contributions, which we neglect, the solution for these quantities also involve radiation reaction effects \cite{dalibard1982vacuum}. These we also neglect in this paper. For typical one-photon absorption processes in molecules this is well-justified, for the natural linewidth is much less than that due to non-radiative decay and dephasing. But for typical fluorescent molecules used for two-photon absorption experiments \cite{dayan2004two,tabakaev2021energy,villabona2018two}, there can be large fluorescence quantum efficiencies that would make this neglect suspect. To compensate for this we expand the use the reservoir introduced in Section \ref{sec:model Hamiltonian} to also provide a phenomenological model of the decay of the states $\{\ket{f}\}$ to the ground state $\ket{g}$ due to fluorescent processes; we leave a more complete description with radiation reaction to later work

Then the expression for the absorption is given by 
\begin{equation}
    \label{eq:quantum extinction approx}
     \mathcal{A} \rightarrow \int\limits_{t_I}^{t_F}dt\bra{\Psi(t_0)}\boldsymbol{E_-}^{H_0}(t)\cdot\frac{d\boldsymbol{{\mu}_+}^{H'}(t)}{dt}\ket{\Psi(t_0)} + c.c.,
\end{equation}
where by $\boldsymbol{\mu_\pm}^{H'}(t)$ we mean the evolution of those quantities according to the Heisenberg equations, but with $\boldsymbol{E_\pm}^H(t)$ in those equations replaced by $\boldsymbol{E_\pm}^{H_0}(t)$. 
 Since this expression for the ``absorption" only \emph{models} the fluorescent contributions to the extinction as loss from the full electromagnetic field, it includes effects that physically are not absorptive in the strict sense introduced at the start of this section.  Nonetheless, in the more colloquial sense of ``absorption" implicit in the phrase ``two-photon absorption," this can be understood as describing the removal of energy from the components of the electromagnetic field near the fundamental frequency.   
For the rest of this paper we use Eq. \eqref{eq:quantum extinction approx} to calculate \emph{this}
``absorption," but to avoid clutter in the notation we henceforth drop the primes on the $H'$ in $\boldsymbol{\mu_\pm}^{H'}(t)$,
and also drop the inverted commas around ``absorption."

We then proceed by using the expression for $\boldsymbol{\mu_\pm}^{H}(t)$. The calculated absorption then has two contributions, the first due to the coherence operator $\sigma_{ge}^H(t)$ and the second is due to $\sigma_{ef}^H(t)$. Since within our approximations Eq. \eqref{eq:quantum extinction approx} 
identifies the \emph {full} absorption, we will see that a simple division of the result into ``one-photon" and ``two-photon" terms is not always possible, especially when the fundamental frequency is on resonance with a transition to an intermediate level. 

We proceed by expanding each coherence operator $\sigma_{ij}^H(t)$ in a perturbative series, using the interaction picture we present below.
%--------------------------------------------------------%
\section{Interaction Picture \label{sec:Input-output theory}}
%--------------------------------------------------------%
In our full Hamiltonian \eqref{eq:Hfull}, $V(t)$ can be written
as a function of a set of Schrödinger operators $\left\{ O_{\alpha}\right\}$ and time, $V\left(\left\{ O_{\alpha}\right\} ;t\right)$. The time evolution of a ket from $t_{a}$ to $t_{b}$ in the absence of $V\left(\left\{ O_{\alpha}\right\} ;t\right)$ is formally described by the unitary evolution operator $\mathcal{U}_{0}(t_{b},t_{a})$, which satisfies the Schrödinger equation
\begin{equation}
  i\hbar\frac{d}{dt_{b}}\mathcal{U}_{0}(t_{b},t_{a})=H_{0}\mathcal{U}_{0}(t_{b},t_{a}),
\end{equation}
and has a solution 
\begin{equation}
  \mathcal{U}_{0}(t_{b},t_{a})=e^{-iH_{0}(t_{b}-t_{a})/\hbar}\,.
\end{equation}
Including $V\left(\left\{ O_{\alpha}\right\} ;t\right)$, the full
evolution operator $\mathcal{U}(t_{b},t_{a})$ satisfies the equation
\begin{equation}
  i\hbar\frac{d}{dt_{b}}\mathcal{U}(t_{b},t_{a})=H(t_{b})\mathcal{U}(t_{b},t_{a}),
\end{equation}
with the initial condition $\mathcal{U}(t_{a},t_{a})=\hat{1}$ for
all $t_{a}$.

Now consider times $t_\text{min}$ and $t_\text{max}$ such that for $t \leq t_\text{min}$ or $t \geq t_\text{max}$ the interaction between the electromagnetic field and the molecule vanishes. In general, of course, such times do not exist; even when there is no pulse of light, if initially a molecule were in its ground state or any other eigenstate of $H_\text{M}$ (Eq. \eqref{eq:HM}), it would interact with the quantized electromagnetic field. However, for the RWA form of the Hamiltonian we adopt, such ``vacuum state'' interactions vanish; see Appendix \ref{app:sec:Absorption, Extinction and Scattering}. So for times $t_{a},t_{b}\leq t_\text{min}$ or $t_{a},t_{b}\geq t_\text{max}$ we have
\begin{equation}
  \mathcal{U}(t_{a},t_{b})=\mathcal{U}_{0}(t_{a},t_{b}).\label{eq:UUnought}
\end{equation}

It is then useful to introduce the operator 
\begin{equation}
  U(t_{b},t_{a})=\mathcal{U}_{0}(0,t_{b})\mathcal{U}(t_{b},t_{a})\mathcal{U}_{0}(t_{a},0),\label{eq:Udef}
\end{equation}
which satisfies all the properties of a unitary time evolution operator, including the condition that $U(t_{a},t_{a})=\hat{1}$ for all $t_{a}$. We can then write 
\begin{equation}
  \left|\Psi(t)\right\rangle =\mathcal{U}_{0}(t,0)U(t,t_0)\left|\Psi_\text{in}\right\rangle ,\label{eq:psiF}
\end{equation}
where the ``asymptotic-in'' ket 
\begin{equation}
  \left|\Psi_\text{in}\right\rangle \equiv\mathcal{U}_{0}(0,t_0)\left|\Psi(t_0)\right\rangle 
\end{equation}
is what the ket would be at $t=0$ were there no interaction $V(t)$ and $t_0<t_I,t_\text{min}$ is the initial start time. Note that since $t_0$ is long before any interaction occurs, $\left|\Psi_\text{in}\right\rangle$ is independent of $t_0$.
From the definition (\ref{eq:Udef}) we can show, using
(\ref{eq:UUnought}), that for $t_{a},t_{a}'\leq t_\text{min}$ we have $U(t,t_{a})=U(t,t_{a}'$), and thus both must be equal to $U(t,-\infty)$; if we put 
\begin{equation}
  U(t)\equiv U(t,-\infty)\label{eq:Utdef}
\end{equation}
we can write (\ref{eq:psiF}) as 
\begin{equation}
  \left|\Psi(t)\right\rangle =\mathcal{U}_{0}(t,0)U(t)\left|\Psi_\text{in}\right\rangle.
\end{equation}
We now move to the Heisenberg picture, and develop an interaction representation related to this.

To begin we introduce an interaction representation of each Schrödinger operator $O_{\alpha}$, 
\begin{equation}
  \hat{O}_{\alpha}(t)\equiv\mathcal{U}_{0}(0,t)O_{\alpha}\mathcal{U}_{0}(t,0).\label{eq:OHdef}
\end{equation}
 For our operators of interest we will have, schematically, 
\begin{equation}
  \hat{O}_{\alpha}(t)=\sum_{\beta}g_{\alpha\beta}(t)O_{\beta},\label{eq:OHexpression}
\end{equation}
where the $g_{\alpha \beta}(t)$ are (classical) functions of time. For
example, if $O_{\alpha}$ is $\sigma_{ij}$ or $a(\boldsymbol{k})$
then there will be only one such $g_{\alpha\beta}(t)$ and we have
\begin{equation}
\begin{gathered}
  \hat{\sigma}_{ij}(t)=\sigma_{ij}e^{-i\omega_{ji}t},\\
  \hat{a}_{I}(\boldsymbol{k},t)=a_{I}(\boldsymbol{k})e^{-i\omega_{k}t},
\end{gathered}
\end{equation}
but if $O_{\alpha}$ is $\psi_{ij}(z),$ we have 
\begin{equation}
\label{eq:propnew}
\begin{split}
  \hat{\psi}_{ij}(z,t)&=\psi_{ij}(z-v_{ij}t)e^{-i\omega_{ij}t}\\
  &=\int g_{zz'}(t)\psi_{ij}(z')dz',
\end{split}
\end{equation}
where $g_{zz'}(t)=\delta(z'-z+v_{ij}t)\exp(-i\omega_{ij}t)$.

With the expressions (\ref{eq:OHdef},\ref{eq:OHexpression}), and
the definitions of $\mathcal{U}_{0}(t_{a},t_{b})$, $\mathcal{U}(t_{a},t_{b})$,
and $U(t)$, it is easy to confirm that the expectation value of any operator $O_\alpha$ is
\begin{equation}
    \bra{\Psi(t)}O_\alpha\ket{\Psi(t)} = \bra{\Psi_\text{in}}\bar O_\alpha(t)\ket{\Psi_\text{in}},
\end{equation}
where
\begin{equation}
\begin{split}
  \bar{O}_{\alpha}(t)&=U^{\dagger}(t)\mathcal{U}_{0}(0,t)O_{\alpha}\mathcal{U}_{0}(t,0)U(t)\label{eq:bar_from_caret}\\
  &=\sum_{\beta}g_{\alpha\beta}(t)\check{O}_{\beta}(t),
\end{split}
\end{equation}
and
\begin{equation}
  \check{O}_{\beta}(t)\equiv U^{\dagger}(t)O_{\beta}U(t).\label{eq:caret_def}
\end{equation}

Since absorption only occurs for times $t$ between the ``interaction region'' $t_\text{min}\le t \le t_\text{max}$, as long as $t_0<t_I<t_\text{min}$ and $t_F>t_\text{max}$ we can extend the limits of integration to infinity because the absorption is zero for those times and write Eq. \eqref{eq:quantum extinction approx} as 
\begin{equation}
  \mathcal{A}=\int\limits_{-\infty}^{\infty}dt\langle \Psi_\text{in}|\boldsymbol{\hat{E}}_{-}(t)\cdot\frac{d\boldsymbol{\bar{\mu}}_{+}(t)}{dt}|\Psi_\text{in}\rangle +c.c.,
\end{equation}
where $\boldsymbol{\hat{E}}_{-}(t)\equiv\boldsymbol{E}_{-}(\{ \hat{a}_{I}^{\dagger}(\boldsymbol{k},t)\})=\boldsymbol{E}_{-}(\{ a_{I}^{\dagger}(\boldsymbol{k})e^{i\omega_{k}t}\})$,
and since in calculating $\boldsymbol{\mu}_{+}^{H}(t)$ we are to
replace $\boldsymbol{E}_{+}^{H}(t)$ by $\boldsymbol{E}_{+}^{H_{0}}(t)$,
in calculating $\bar{\boldsymbol{\mu}}_{+}(t)$ we are to replace
$\boldsymbol{\bar{E}}_{+}(t)\equiv\boldsymbol{E}_{+}(\{ \bar{a}_{I}(\boldsymbol{k},t)\})=\boldsymbol{E}_{+}(\{ \check{a}_{I}(\boldsymbol{k},t)e^{-i\omega_{k}t}\})$
by $\boldsymbol{\hat{E}}_{+}(t)\equiv\boldsymbol{E}_{+}(\{ \hat{a}_{I}(\boldsymbol{k},t)\})=\boldsymbol{E}_{+}(\{a_{I}(\boldsymbol{k})e^{-i\omega_{k}t}\})$. 
This we will do in the equations of motion we derive below.

\section{equations of motion \label{sec:effective Heisenberg equations}}
The operator $\boldsymbol{\bar{\mu}}_{+}(t)$ involves the operators $\bar{\sigma}_{ge}(t)$ and $\bar{\sigma}_{ef}(t)$, and
we will see below that the dynamics for those operators involves other operators $\bar{\sigma}_{ij}$(t), and so the $\bar{\sigma}_{ij}(t)$ will be our operators $\bar{O}_{\gamma}(t)$ of interest. Now the complicated part of the dynamics of an operator $\bar{O}_{\gamma}(t)$ is due to its dependence on the $\check{O}_{\gamma}(t)$, and the
evolution of an operator $\check{O}_{\gamma}(t)$ follows from the
evolution of $U(t).$ From its definition (\ref{eq:Utdef}) the dynamical equation for $U(t)$ can be found,
\begin{equation}
  i\hbar\frac{d}{dt}U(t)=V\left(\left\{ \hat{O}_{\alpha}(t)\right\} ;t\right)U(t),
\end{equation}
from which follows the dynamical equation for $\check{O}_{\gamma}(t),$
\begin{equation}
  i\hbar\frac{d}{dt}\check{O}_{\gamma}(t)=\left[\check{O}_{\gamma}(t),V\left(\left\{ \bar{O}_{\alpha}(t)\right\} ;t\right)\right].\label{eq:dynamics}
\end{equation}

In these equations the $\bar{O}_{\alpha}(t)$ are functions of
the $\check{O}_{\beta}(t)$, as given by (\ref{eq:bar_from_caret}).
Noting this relation is the same as that between the $\hat{O}_{\alpha}(t)$
operators and the Schrödinger operators $O_{\beta}$ (\ref{eq:OHexpression}),
we have 
\begin{equation}
\label{eq:bar_results}
\begin{gathered}
  \bar{\sigma}_{ij}(t)=\check{\sigma}_{ij}(t)e^{-i\omega_{ji}t},\\
  \bar{a}_{I}(\boldsymbol{k},t)=\check{a}_{I}(\boldsymbol{k},t)e^{-i\omega_{k}t}, \\
  \bar{\psi}_{ij}(z,t)=\check{\psi}_{ij}(z-v_{ij}t,t)e^{-i\omega_{ij}t}. 
\end{gathered}
\end{equation}
Since the relation between $\bar{\sigma}_{ij}(t)$ and $\check{\sigma}_{ij}(t)$
is so simple, and it is the operators $\bar{O}_{\alpha}(t)$
appearing in $V\left(\left\{ \bar{O}_{\alpha}(t)\right\} ;t\right)$
in (\ref{eq:dynamics}), we begin by constructing dynamical equations
for the $\bar{\sigma}_{ij}(t)$ using (\ref{eq:dynamics},\ref{eq:bar_results}),
\begin{equation}
  i\hbar\left(\frac{d}{dt}+i\omega_{ji}\right)\bar{\sigma}_{ij}(t)=\left[\bar{\sigma}_{ij}(t),V\left(\left\{ \bar{O}_{\alpha}(t)\right\} ;t\right)\right].\label{eq:sigma_bar_dynamics}
\end{equation}
They involve the quantities 
\begin{equation}
\label{eq:Fbar}
\begin{gathered}
  \bar{F}_{+}^{ij}(t)\equiv\frac{\boldsymbol{\mu}_{ij}}{\hbar}\cdot\boldsymbol{E}_{+}(\{\bar{a}_{I}(\boldsymbol{k},t)\})-\eta_{ij}^{*}\bar{\psi}_{ij}(0,t),
\end{gathered}
\end{equation}
for $\omega_{i}>\omega_{j}$, and their adjoints (see Eq. \eqref{eq:Fij+}).  
Note that the dynamical equations (\ref{eq:sigma_bar_dynamics}) for
the $\bar{\sigma}_{ij}(t)$ take the same structure as the Heisenberg
equations $\sigma_{ij}^{H}(t)$. However, while the ket relevant for
expectation values of the $\sigma_{ij}^{H}(t)$ is $\left|\Psi(t_0)\right\rangle $,
that relevant for expectation values of the $\bar{\sigma}_{ij}(t)$
is $\left|\Psi_\text{in}\right\rangle $. And while the $\sigma_{ij}^{H}(t)$
equal the corresponding Schrödinger operators $\sigma_{ij}$ at $t_0$,
it is the operators $\check{\sigma}_{ij}(t)$ that equal the corresponding
Schrödinger operators $\sigma_{ij}$ at a special time given by
$t=-\infty$. We return to this point below.

Since the relation between $\bar{\psi}_{ij}(z,t)$ and $\check{\psi}_{ij}(z,t)$
is not as simple as that between $\bar{\sigma}_{ij}(t)$ and
$\check{\sigma}_{ij}(t)$ (see (\ref{eq:bar_results})), we simply
construct the dynamical equation for $\check{\psi}_{ij}(z,t)$ directly
from (\ref{eq:dynamics}); again, since an $\check{O}_{\alpha}(t)$
is related to the corresponding Schrödinger operator $O_{\alpha}$
by a unitary transformation (\ref{eq:caret_def}), the commutation
relations are preserved, and we find 
\begin{equation}
  \frac{\partial}{\partial t}\check{\psi}_{ij}(z,t)=-i\eta_{ij}\check{\sigma}_{ji}(t)\delta(z+v_{ij}t).\label{eq:psis_dynamics}
\end{equation}
The equations (\ref{eq:sigma_bar_dynamics},\ref{eq:psis_dynamics})
are the set of dynamical equations we wish to solve; note that (\ref{eq:sigma_bar_dynamics})
could easily be rewritten as equations for the $\check{\sigma}_{ij}(t)$,
and we would be solving for the set $\left\{ \check{O}_{\alpha}(t)\right\} $
of operators.

\section{Formal solutions \label{sec:solving the equations of motion}}
The equation (\ref{eq:psis_dynamics}) that describes how the $\check{\sigma}_{ji}(t)$
affect the $\check{\psi}_{ij}(z,t)$ can be solved immediately. Integrating
from $t=-\infty$ to $t$, we recall that $\check{\psi}_{ij}(z,-\infty)=\psi_{ij}(z),$
the Schrödinger operator, and we find 
\begin{equation}
  \check{\psi}_{ij}(-v_{ij}t,t)=\psi_{ij}(-v_{ij}t)-\frac{i\eta_{ij}}{2v_{ij}}\check{\sigma}_{ji}(t).
\end{equation}
Multiplying by $\exp(-i\omega_{ij}t)$, using (\ref{eq:bar_results})
we can also write this as 
\begin{equation}
\begin{split}
 & \bar{\psi}_{ij}(0,t)=\hat{\psi}_{ij}(0,t)-\frac{i\eta_{ij}}{2v_{ij}}\bar{\sigma}_{ji}(t),\label{eq:psi_bar_result}
\end{split}
\end{equation}
where we have also used (\ref{eq:propnew}). Note that $\hat{\psi}_{ij}(0,t)$
can be written in terms of Schrödinger operators, and as such can
be thought of as the ``input'' of the reservoir field at the site
of the molecule; $\bar{\psi}_{ij}(0,t)$ can then be thought
of as the ``output,'' which contains a contribution from the molecule.

With this in hand we can simplify the expression (\ref{eq:Fbar})
for the $\bar{F}^{ij}_+(t)$. For $\omega_{i}>\omega_{j}$ we have
\begin{equation}
\begin{split}
 \bar{F}_{+}^{ij}(t)=&\frac{\boldsymbol{\mu}_{ij}}{\hbar}\cdot\boldsymbol{E}_{+}[\{\check{a}_{I}(\boldsymbol{k},t)e^{-i\omega_{k}t}\}]-\eta_{ij}^{*}\hat{\psi}_{ij}(0,t)\\
 &+i\frac{\left|\eta_{ij}\right|^{2}}{2v_{ij}}\bar{\sigma}_{ji}(t)
\end{split}
\end{equation}
Neglecting the radiation reaction (and renormalization) effects on
the electric field operator, as discussed in section III, we put $\check{a}_{I}(\boldsymbol{k},t)\rightarrow a_{I}(\boldsymbol{k})$
(see discussion after (Eq. \eqref{eq:quantum absorption}), and so 
\begin{equation}
\boldsymbol{E}_{+}[\{\check{a}_{I}(\boldsymbol{k},t)e^{-i\omega_{k}t}\}]\to\boldsymbol{E}_{+}[\{a_{I}(\boldsymbol{k})e^{-i\omega_{k}t}\}]=\boldsymbol{\hat{E}}_{+}(t),
\end{equation}
and we can write 
\begin{equation}
\label{eq:Fbarsolved}
 \bar{F}_{+}^{ij}(t)=\hat{F}_+^{ij}(t)+i\Gamma_{ij}\bar{\sigma}_{ji}(t),
\end{equation}
where we have put 
\begin{equation}
\hat{F}_+^{ij}(t)\equiv\frac{\boldsymbol{\mu}_{ij}}{\hbar}\cdot\boldsymbol{\hat{E}}_{+}\left(t\right)-\eta_{ij}^{*}\hat{\psi}_{ij}(0,t),
\end{equation}
and 
\begin{equation}
  \Gamma_{ij}=\frac{\left|\eta_{ij}\right|^{2}}{2v_{ij}}.
\end{equation}
The reservoir parameters that contribute to calculation are now combined into the constant $\Gamma_{ij}$, and we have written $\bar{F}_+^{ij}(t)$ as the sum of two contributions. The first is from the operator $\hat F_+^{ij}(t)$, which is a Schr\"{o}dinger operator, and therefore a  ``known'' quantity which drives transitions. The second contribution depends on the molecule operator $\bar\sigma_{ji}(t)$ which is yet to be determined. Finally, the constant $\Gamma_{ij}$ which has units of inverse time, will turn out to be the decay rate associated with transitions between states $\ket{i}\leftrightarrow\ket{j}$. 

Unlike the subscripts on  $\sigma_{ij}$ and $\boldsymbol{\mu}_{ij}$, the subscript of $\Gamma_{ij}$ follows from the label we gave to each reservoir operator $\psi_{ij}(z)$, which was only defined when $\omega_i>\omega_j$; thus $\Gamma_{ij}$ is also only defined then. However, in the following it will be convenient to define $\Gamma_{ji}$ when $\omega_i>\omega_j$ to be $\Gamma_{ji}\equiv \Gamma_{ij}$, i.e. the decay constant $\Gamma_{ij}$ is defined to be symmetric. Although $\Gamma_{ij}$ is defined to be symmetric, it always represents the decay rate of energy of the molecule transitioning from a state of high energy to one of lower energy. The definition is convenient in introducing the Green function below.

Using the solution \eqref{eq:Fbarsolved} for each  $\bar{F}_+^{ij}(t)$ in the equation \eqref{eq:sigma_bar_dynamics} for the corresponding $\bar\sigma_{ij}(t)$,  damping terms $\Gamma_{ij}$ are introduced which we group on the left-hand side of each equation and define the total relaxation constants due to the reservoir for each molecule operator by 
\begin{subequations}
\label{eq:Gamma bar definition}
\begin{gather}
    \bar\Gamma_{eg} \equiv \Gamma_{eg},\\
    \bar\Gamma_{gg} \equiv 2\Gamma_{eg},\\
    \bar\Gamma_{e^\prime e} \equiv \Gamma_{e^\prime g} + \Gamma_{eg},\\
    \bar\Gamma_{fe} \equiv \Gamma_{eg} + \sum_{e^\prime}\Gamma_{fe^\prime} +  \Gamma_{fg},\\
    \bar\Gamma_{fg} \equiv \sum_{e^\prime}\Gamma_{fe^\prime} +  \Gamma_{fg},
\end{gather}
\end{subequations}
where each barred decay constant $\bar{\Gamma}_{ij}$ ($=\bar{\Gamma}_{ji}$) is associated with $\bar\sigma_{ji}(t)$. We find 
\begin{widetext}
\begin{subequations}
\label{eq:equations of motion v2}
\begin{gather}
        \label{eq:e.o.m sigma_ge 2}
        \left(\frac{d}{dt} {+} \bar\Gamma_{eg} {+} i\omega_{eg} {+} i\tilde{\omega}_{eg}(t)\right)\bar\sigma_{ge}(t) = i[\bar\sigma_{gg}(t)\hat F_+^{eg}(t) 
        {-} \bar\sigma_{e^\prime e}(t)\hat F_+^{e^\prime g}(t) {+} \hat F_-^{ef}(t)\bar\sigma_{gf}(t)],\\
        \label{eq:e.o.m sigma_ef 2}
        \left(\frac{d}{dt} {+} \bar\Gamma_{fe} {+} i\omega_{fe} {+} i\tilde{\omega}_{fe}(t)\right)\bar\sigma_{ef}(t) = 
        i[\bar\sigma_{ee^\prime}(t)\hat F_+^{fe^\prime}(t) {-} \hat F_-^{ge}(t)\bar\sigma_{gf}(t){-} \bar\sigma_{f^\prime f}(t)\hat F_+^{f^\prime e}(t)],\\
        \label{eq:e.o.m sigma_gf 2}
        \left(\frac{d}{dt} {+}\bar\Gamma_{fg} {+}  i\omega_{fg} {+}i\tilde{\omega}_{fg}(t)\right)\bar\sigma_{gf}(t) = 
        i[\bar\sigma_{ge}(t)\hat F_+^{fe}(t) {-} \bar\sigma_{ef}(t)\hat F_+^{eg}(t)],\\
        \label{eq:e.o.m sigma_e eprime hat 2}
        \left(\frac{d}{dt} {+} \bar\Gamma_{e^\prime e}{+}i\omega_{e^\prime e} {+} i\tilde{\omega}_{e^\prime e}(t)\right)\bar\sigma_{ee^\prime}(t) =i[
        \bar\sigma_{eg}(t)\hat F_+^{e^\prime g}(t) 
        {-}\hat F_-^{ge}(t)\bar\sigma_{ge^\prime}(t)
        {-}\bar\sigma_{fe^\prime}(t)\hat F_+^{fe}(t) {+}\hat F_-^{e^\prime f}(t)\bar\sigma_{e f}(t){-}2i\Gamma_{fe}\bar\sigma_{ff}(t)\delta_{ee^\prime}],
\end{gather}
\end{subequations}
\end{widetext}
where $\tilde{\omega}_{ij}(t)=\tilde{\omega}_{i}(t)-\tilde{\omega}_{j}(t)$. Further, we have dropped terms involving $\hat F_{+}^{fg}(t)$, which can be done with impunity: Since the equations here are normally ordered, when we solve for each molecule operator perturbatively they will remain normally ordered; $\hat F_{+}^{fg}(t)$ is a known quantity that only involves reservoir operators, so it will always annihilate the ket $\ket{\Psi_\text{in}}$, since we take the reservoir to be initially the vacuum state. Thus the dropped terms will lead to no contribution to the absorption. For the same reason, at this point we could replace each $\hat F_{+}^{ij}(t) \to \boldsymbol{\mu}_{ij}/\hbar\cdot \boldsymbol{E}_+(t)$ for $ij \neq fg$, but for now we choose to stay in the $\hat F_{+}^{ij}(t)$ notation to keep the equations more condensed. We note that Eq. \eqref{eq:equations of motion v2}, with the replacement of $\hat F_{+}^{ij}(t) \to \boldsymbol{\mu}_{ij}/\hbar\cdot \boldsymbol{E}_+(t)$, is the usual Heisenberg time evolution for the coherence operators; see, for example, Gerry and Knight \cite{gerry2005introductory}.

The equations in \eqref{eq:equations of motion v2} and their adjoints, together with the corresponding equations for $\bar\sigma_{gg}(t)$ and $\bar\sigma_{f^\prime f}(t)$ -- not written down because they are not necessary to calculate the absorption in a perturbative calculation -- form a closed set of coupled differential equations for the molecule operators $\bar\sigma_{ij}(t)$, since each $\hat F_\pm^{ij}(t)$ is a Schr\"{o}dinger operator which is known. 

We now consider the general form of each equation in \eqref{eq:equations of motion v2}, which
is given by 
\begin{equation}
  \left(\frac{d}{dt}+\bar{\Gamma}_{ij}+i\omega_{ij}+i\tilde{\omega}_{ij}(t)\right)\bar{\sigma}_{ji}(t)=K_{ij}(t),\label{eq:first}
\end{equation}
or 
\begin{equation}
\label{eq: general diff eq}
  \left(\frac{d}{dt}+\bar{\Gamma}_{ij}+i\tilde{\omega}_{ij}(t)\right)\check{\sigma}_{ji}(t)=e^{i\omega_{ij}t}K_{ij}(t).
\end{equation}
The second form has a formal solution that includes a homogeneous
solution satisfying the initial condition at $t=-\infty$, where each
$\check{\sigma}_{ji}(t)$ is equal to the corresponding Schrödinger
operator $\sigma_{ji}$. Because of the damping term $\bar{\Gamma}_{ij}$,
that homogeneous solution will vanish for finite times. Correspondingly,
in a formal solution of (\ref{eq:first}) the homogeneous solution
will vanish as well, and introducing a Green function
\begin{equation}
    \label{eq:Green's function}
    G_{ij}(t,t_1) \equiv i e^{-(\bar\Gamma_{ij} + i\omega_{ij})(t-t_1)}e^{-i\int\limits_{t_1}^{t} dt^\prime\tilde{\omega}_{ij}(t^\prime)},
\end{equation}
the exact solution to Eq. \eqref{eq:first} and each operator in Eq. \eqref{eq:equations of motion v2} is
\begin{equation}
    \label{eq:exact solution to sigma_ij}
    \bar\sigma_{ji}(t) = \int\limits_{-\infty}^{t}dt_1G_{ij}(t,t_1)K_{ij}(t_1),
\end{equation}
for the appropriate $K_{ij}(t)$ on the right-hand side. We note that the Green function for each equation satisfies the property that
\begin{equation}\label{eq:Gij^* = -Gji}
    G_{ij}^*(t,t_1) = -G_{ji}(t,t_1),
\end{equation}
where we used the property that $\bar\Gamma_{ij}$ is symmetric. 

To process the terms further we need to evaluate the time integral in the definition of $G_{ij}(t,t_1)$. The time integral can be evaluated by taking the classical average of each coherence operator
\begin{equation}
\label{eq:sigma_ge expectation value}
    [\![\bar\sigma_{ji}(t) ]\!] = \int\limits_{-\infty}^{t}dt_1 [\![G_{ij}(t,t_1)K_{ij}(t_1) ]\!],
\end{equation}
where the classical average must include $K_{ij}(t_1)$ since it is a function of molecule operators, which also vary stochastically. For this reason the expectation value in Eq. \eqref{eq:sigma_ge expectation value} cannot be solved. Further, the right-hand side involves $K_{ij}(t)$, which includes unknown molecule operators. Therefore, to solve for the molecule operators where the right-hand side involves ``known" quantities we  need to resort to perturbation theory. It is at this point where each term will only depend on $\tilde{\omega}_{ij}(t)$ explicitly, allowing us to evaluate the expectation value.

Before resorting to perturbation theory we consider the exact formal expressions for $\bar\sigma_{ge}(t)$ and $\bar\sigma_{ef}(t)$ that will be used to calculate the absorption. Using the definition of $G_{ij}(t,t_1)$, exact expressions for  $\bar\sigma_{ge}(t)$ and $\bar\sigma_{ef}(t)$ are
\begin{subequations}
\label{eq:exact soltuions body}
\begin{gather}
    \label{eq:exact solution sigma_ge}
    \hspace{-15mm}\bar\sigma_{ge}(t) =\hspace{-2mm} \int\limits_{-\infty}^t \hspace{-2mm}dt_1[G_{eg}(t,t_1)\bar\sigma_{gg}(t_1)\hat F_+^{eg}(t_1) \\
    \hspace{2mm}+G_{ge}^*(t,t_1) \bar\sigma_{e^\prime e}(t_1)\hat F_+^{e^\prime g}(t_1) + G_{eg}(t,t_1)\hat F_-^{ef}(t_1)\bar\sigma_{gf}(t_1)]\nonumber,\\
    \label{eq:exact solution sigma_ef}
    \hspace{-15mm}\bar\sigma_{ef}(t) = \hspace{-2mm}
    \int\limits_{-\infty}^t \hspace{-2mm}dt_1[G_{f e}(t,t_1)\bar\sigma_{ee^\prime}(t_1)\hat F_+^{fe^\prime}(t_1) \\
    \hspace{2mm}+G^*_{ef}(t,t_1) \hat F_-^{ge}(t_1)\bar\sigma_{gf}(t_1)+G^*_{ef}(t,t_1) \bar\sigma_{f^\prime f}(t_1)\hat F_+^{f^\prime e}(t_1)]\nonumber,
\end{gather}
\end{subequations}
where we used Eq. \eqref{eq:Gij^* = -Gji} and which both have three contributions on the right-hand side, each corresponding to a distinct physical process that we will discuss. 
\begin{figure*}
    \centering
    \includegraphics[width = \linewidth]{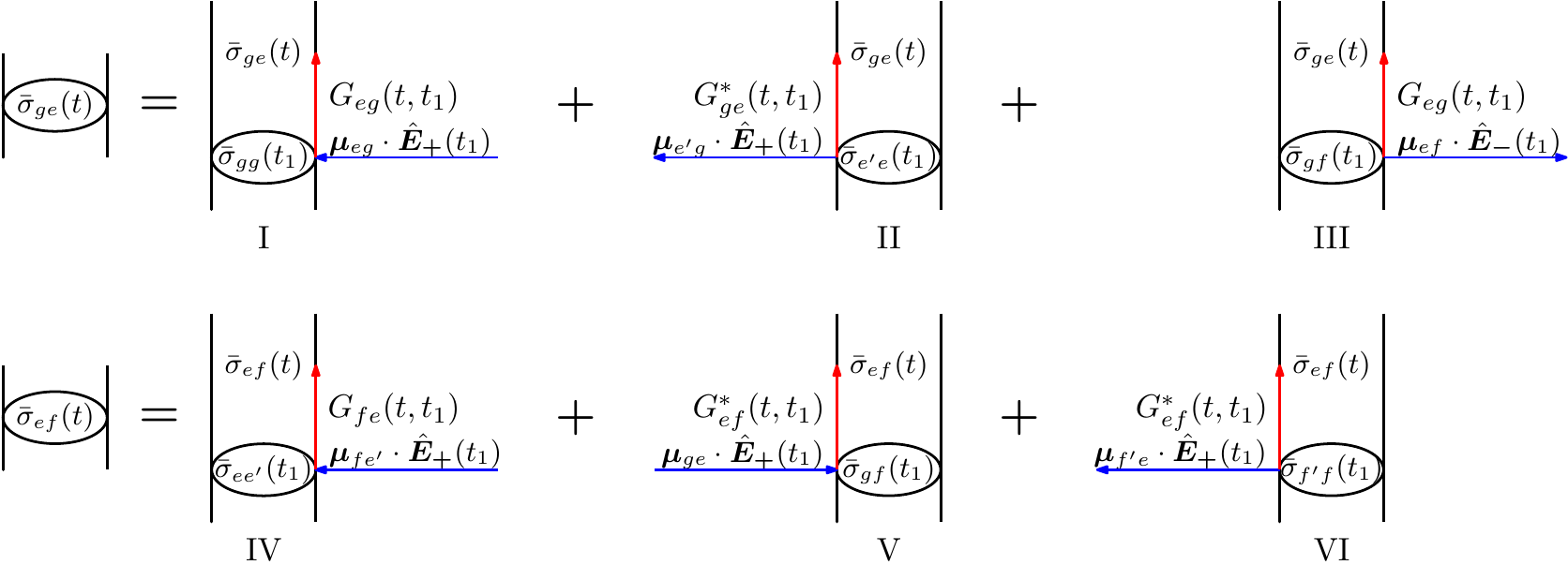}
    \caption{Exact diagrams for $\bar\sigma_{ge}(t)$ and $\bar\sigma_{ef}(t)$. Time flows upwards such that $t\ge t_1$.  Each diagram starts with the exact form of an operator in a bubble at $t_1$ and interacts with the field. The time evolution from $t_1\to t$ evolves according to the Green function $G_{ij}(t,t_1)$. Diagram $\RN{1}$ and $\RN{4}$ contribute to one- and two-photon absorption respectively. Diagrams $\RN{2}$ and $\RN{5}$ are saturation terms of the one- and two-photon absorption respectively. The physics of diagrams $\RN{3}$ and $\RN{6}$ depend on the detuning, but far from resonance the diagrams contribute to two-photon absorption.}
    \label{fig:exact diagrams split}
\end{figure*}

It will be useful to introduce diagrammatic representations of the exact expressions in Eqs. \eqref{eq:exact solution sigma_ge} and \eqref{eq:exact solution sigma_ef}; see Fig. \ref{fig:exact diagrams split}. These diagrams are similar to the double-sided Feynman diagrams considered in many studies \cite{schlawin2017entangled,dorfman2016nonlinear,boyd2020nonlinear} and they follow similar rules. However, in our approach we work in the Heisenberg picture as apposed to a density operator formalism, and include decay processes from the beginning; this sets each of the molecule operators in Eq. \eqref{eq:equations of motion v2} to zero until the interaction with the electromagnetic field is introduced (see discussion under Eq. \eqref{eq: general diff eq}.) When we begin perturbation theory, it follows that the zeroth order solution in our perturbation expansion is \emph{not} given by $\bar\sigma^{(0)}_{gg} \neq \ket{g}\bra{g}$, as we discuss in detail below. Thus, the starting point of the perturbative expansion of each diagram does not have a ``$\ket{g}\bra{g}$'' term. Our diagrams involve the exact form of each coherence operator at time $t$ given by $\bar\sigma_{ij}(t)$, and later when we begin perturbation theory we will include a superscript to denote the order of perturbation theory, with $\bar\sigma_{ij}^{(k)}(t)$, for example, denoting the $k$'th order in perturbation theory. 

To represent the fact that the coherence operators on the left- and right-hand side of Eq. \eqref{eq:exact solution sigma_ge} and \eqref{eq:exact solution sigma_ef} are formal solutions we put them in a ``bubble". When we begin to solve for them perturbatively we will provide a perturbative expansion of each ``bubble" diagram. 

The rules of each diagram are as follows. Time increases upward. Inward (outward) arrows on the right- and left-hand side represent absorption (emission). Inward arrows on the bra (ket) side raise the bra (ket) side of $\bar\sigma_{ij}(t_1)$ and outward arrows on the bra (ket) side lower the bra (ket) side of $\bar\sigma_{mn}(t_1)$. Each inward/outward arrow is connected to a vertical arrow and a Green function $G_{pq}(t,t_1)$. If the vertical arrow is on the right-hand (left-hand) side we include a factor of $G_{qp}(t,t_1)$ ($G^*_{pq}(t,t_1)$) and the subscript is opposite (same) of the $\bar\sigma_{qp}(t_1)$ that it connects to. We note that because the equations are normally ordered and the initial ket of the reservoir is the vacuum state we always take $\hat{F}^{ij}_{\pm}(t_1)\to\boldsymbol{\mu}_{ij}\cdot\hat{\boldsymbol{E}}_{\boldsymbol{\pm}}(t_1)$ in each diagram and read the dipole matrix elements on the bra (ket) side from right to left (left to right).

With the aid of each diagram in Fig. \ref{fig:exact diagrams split} we can describe each contribution on the right-hand side of $\bar\sigma_{ge}(t)$ and $\bar\sigma_{ef}(t)$ in Eq. \eqref{eq:exact soltuions body}; in later sections we will expand these simple descriptions with more details. 

First we consider the coherence operator $\bar\sigma_{ge}(t)$ (Eq. \eqref{eq:exact solution sigma_ge}) which has three contributions on the right-hand side. The first contribution to the dynamics (I in Fig. \ref{fig:exact diagrams split}) is given by the term $\bar\sigma_{gg}(t)\hat F_+^{eg}(t)$; it depends on the population in the ground state and an annihilation operator $\hat F_+^{eg}(t) $, which facilitates transitions between $g$ and $e$. Taking the expectation value, we understand the annihilation operator $\hat F_+^{eg}(t)$ as acting on the right to annihilate a photon from the field, which raises the bra side of $\bar\sigma_{gg}(t)\to\bar\sigma_{ge}(t)$. At lowest order in perturbation theory this term leads to one-photon absorption and its higher order contribution will be saturation.  

The second contribution to the dynamics of $\bar\sigma_{ge}(t)$ (II in Fig. \ref{fig:exact diagrams split}) is given by $\bar\sigma_{e^\prime e}(t)\hat F_+^{e^\prime g}(t)$; it depends on coherences between intermediate states and the annihilation operator $\hat F_+^{e^\prime g}(t)$, which facilitates transitions between $g$ and $e^\prime$. The annihilation operator $\hat F_+^{e^\prime g}(t)$ acts on the left creating an excitation in the field-reservoir system, which lowers the ket side of $\bar\sigma_{e^\prime e}(t)\to \bar\sigma_{ge}(t)$. At lowest order this term describes a saturation of the one-photon absorption.

The third and final contribution to the dynamics of $\bar\sigma_{ge}(t)$ (III in Fig. \ref{fig:exact diagrams split}) is given by $\hat F_-^{ef}(t)\bar \sigma_{gf}(t)$; it depends on the field-reservoir creation operator $\hat F_-^{ef}(t)$, which facilitates transitions between $e$ and $f$, and on the coherence between $g$ and $f$. Coherence between $g$ and $e$ is generated from coherence between $g$ and $f$ and the creation operator $\hat F_-^{ef}(t)$ acting on the right, which creates an excitation in the field-reservoir system and lowers the bra side of $\bar \sigma_{gf}(t)\to \bar \sigma_{ge}(t)$. In later sections we will find that this term's contribution will depend largely on the detunings from resonances, and a simple characterization of ``one-photon absorption" or ``two-photon absorption" seems not possible.  However, when the incident field is far detuned from resonance with an intermediate state, this term can be identified as a contribution to two-photon absorption.

Next we consider the coherence operator $\bar\sigma_{ef}(t)$ (Eq. \eqref{eq:exact solution sigma_ef}) which has three contributions on the right-hand side. The first contribution to the dynamics (IV in Fig. \ref{fig:exact diagrams split}) is given by the term $\bar\sigma_{ee^\prime}(t)\hat F_+^{fe^\prime}(t)$; it depends on coherence between excited states and the annihilation operator $\hat F_+^{fe^\prime}(t)$, which facilitates transitions between $e^\prime$ and $f$. Taking the expectation value, we understand the annihilation operator $\hat F_+^{fe^\prime}(t)$ as acting on the right to annihilate a photon from the field, which raises the bra side of $\bar\sigma_{ee^\prime}(t)\to\bar\sigma_{ef}(t)$. At lowest order in perturbation theory this term contributes to two-photon absorption. 

The second contribution to the dynamics of $\bar\sigma_{ef}(t)$ (V in Fig. \ref{fig:exact diagrams split}) is given by $\hat F_-^{ge}(t)\bar\sigma_{gf}(t)$; it depends on the field-reservoir creation operator $\hat F_-^{ge}(t)$, which facilitates transitions between $g$ and $e$, and on the coherence between $g$ and $f$. Coherence is generated between $e$ and $f$ from coherence between $g$ and $f$ and the creation operator $\hat F_-^{ge}(t)$ acting on the left, which annihilates a photon from the field and raises the ket side of $\bar \sigma_{gf}(t)\to \bar \sigma_{ef}(t)$.  Similar to the term shown in III in Fig. \ref{fig:exact diagrams split}, its contribution depends strongly on the detunings from resonances, and does not allow a simple characterization. But like III, when the  incident field is far detuned from a resonance with an intermediate state this term can be identified as a contribution  two-photon absorption.

The third and final contribution to the dynamics of $\bar\sigma_{ef}(t)$ (VI in Fig. \ref{fig:exact diagrams split}) is given by $\bar\sigma_{f^\prime f}(t)\hat F_+^{f^\prime e}(t)$; it depends on coherences between final states and the annihilation operator $\hat F_+^{f^\prime e}(t)$, which facilitates transitions between $e$ and $f^\prime$. We understand this term as modifying the coherence $\bar\sigma_{ef}(t)$ by the annihilation operator $\hat F_+^{f^\prime e}(t)$ acting on the left to create an excitation in the field-reservoir system, which lowers the ket side of $\bar\sigma_{f^\prime f}(t)\to\bar\sigma_{ef}(t)$. At lowest order this term describes a saturation of the two-photon absorption; however, at the order of perturbation we will be considering, it will not contribute.

\section{perturbative solution \label{sec:perturbative solution}}
To solve the system of equations \eqref{eq:equations of motion v2} we expand each $\bar{\sigma}_{ij}(t)$ in a perturbative expansion given by 
\begin{equation}
    \label{eq:perturbative expansion}
    \bar{\sigma}_{ij}(t) = \bar{\sigma}_{ij}^{(0)}(t) + \lambda \bar{\sigma}_{ij}^{(1)}(t) + \lambda^2 \bar{\sigma}_{ij}^{(2)}(t) +...
\end{equation}
and take $\hat{\boldsymbol{E}}_{\boldsymbol{\pm}}(t)\to\lambda\hat{\boldsymbol{E}}_{\boldsymbol{\pm}}(t)$ where $\lambda$  characterizes the order of the perturbation. We use the perturbative expansion \eqref{eq:perturbative expansion} with the exact formal solution \eqref{eq:exact solution to sigma_ij}, identify the zeroth order solution, and iterate order-by-order.

For the zeroth order solution, since the initial ket of the reservoir is the vacuum state and the equations are normal ordered, no coherence is generated and no population is moved out of the ground state. Then using the identity relation \eqref{eq:identity relation}, which is also satisfied in the interaction picture we work in, we solve for the last unknown molecule operator, and for finite times we have
\begin{equation}
\label{eq:ggnought}
    \bar\sigma_{gg}^{(0)}(t) = \hat 1.
\end{equation}
This occurs because the interaction between the molecule and reservoir is still operative, and since we take the initial ket of the reservoir to be vacuum, the molecule can only lose energy to the reservoir. So for $\ket{\Psi_\text{in}}$ describing the molecule in any state, for finite times  $t>-\infty$ the molecule will move to the ground state, and Eq. \eqref{eq:ggnought} necessarily results. 

To generate each higher order term we use the zeroth order term in the right-hand side of each exact result and then continue order by order. In Appendix \ref{appB:Perturbation theory} we begin with the exact solutions for each molecule operator and apply the steps outlined above to third order in $\bar\sigma_{ge}(t)$ and $ \bar\sigma_{ef}(t)$. The nonzero terms that contribute to the absorption are $\bar\sigma_{ge}^{(1)}(t),\bar\sigma_{ge}^{(3)}(t)$ and $\bar\sigma_{ef}^{(3)}(t)$ and are given in Eq. \eqref{eq:sigma_ge first order v1}, \eqref{eq:sigma_ge third order v1} and \eqref{eq:sigma_ef third order v1} respectively. 

Each coherence operator $\bar\sigma_{ge}^{(1)}(t),\bar\sigma_{ge}^{(3)}(t)$ and $\bar\sigma_{ef}^{(3)}(t)$ still involves the stochastic functions $\tilde\omega_{ij}(t)$, which are included in the definition of $G_{ij}(t,t_1)$. To push forward we need to take the average over the classical distributions. In Appendix \ref{appB:Stochastic average} we begin with the simplest case and work out the expectation value of $G_{ij}(t,t_1)$ using the assumptions after Eq. \eqref{eq:H_fluc}. Taking the ``impact limit'' in which correlations decay on fast time scales compared to the coupling strength \cite{mukamel1999principles, mukamel1983nonimpact}, $\tau_c^2c_{ij}\ll 1$, we find $[\![G_{ij}(t,t_1)]\!] = G_{ij}(t-t_1)$ where
\begin{equation}
    \label{eq:modified Green's function}
    G_{ij}(t-t_1) \equiv ie^{-(\gamma_{ij}+i\omega_{ij
    })(t-t_1)},
\end{equation}
where we defined the total decay width
\begin{equation}
    \gamma_{ij}\equiv \bar\Gamma_{ij} + \Lambda_{ij},
\end{equation}
where $\Lambda_{ij}$ is the dephasing decay rate between state $i$ and $j$ defined in Appendix \ref{appB:Stochastic average} and satisfies $\Lambda_{ij} = \Lambda_{ji}$ and $\Lambda_{ii}=0$. Since each contribution to $\gamma_{ij}$ is symmetric so is $\gamma_{ij}$. Then the new Green function defined in Eq. \eqref{eq:modified Green's function} also satisfies
\begin{equation}
    G^*_{ij}(t-t_1) = -G_{ji}(t-t_1). 
\end{equation}

The result for $[\![G_{ij}(t,t_1)]\!]$ can be directly applied to the stochastic average for $\bar\sigma_{ge}^{(1)}(t)$ because there is only one $G_{eg}(t,t_1)$ present. However, for the higher order molecule operators there are three Green functions present and we must take the expectation value over the product of them which can become quite complicated.
If we assume that the correlation between random variables which decays on time scales given by $\tau_c$ is much shorter than the dephasing rates $\Lambda_{ij}$, for all $i$ and $j$, i.e. $\Lambda_{ij}\tau_c \ll 1$, then we can apply the ``factorization approximation'' \cite{mukamel1983nonimpact,mukamel1999principles} where the expectation value of a product of green functions is equal to the product of expectation values, that is
\begin{equation}
    [\![G_{ij}(t,t_n) \dots G_{pq}(t_2,t_1)]\!] = G_{ij}(t-t_n) \dots G_{pq}(t_2-t_1).
\end{equation}
Thus to take the average value of any coherence operator we replace each Green function with Eq. \eqref{eq:modified Green's function}, which includes the total rate $\gamma_{ij}$.

After applying the above prescription and taking the stochastic average we have solved for the perturbative expansion of $\bar\sigma_{ge}(t)$ and $\bar\sigma_{ef}(t)$ up to third order and in doing so we also solved for $\bar\sigma_{gg}^{(2)}(t)$, $\bar\sigma_{e^\prime e}^{(2)}(t)$ and $\bar\sigma_{gf}^{(2)}(t)$.

\begin{figure*}
    \centering
    \includegraphics[width = 0.85\linewidth]{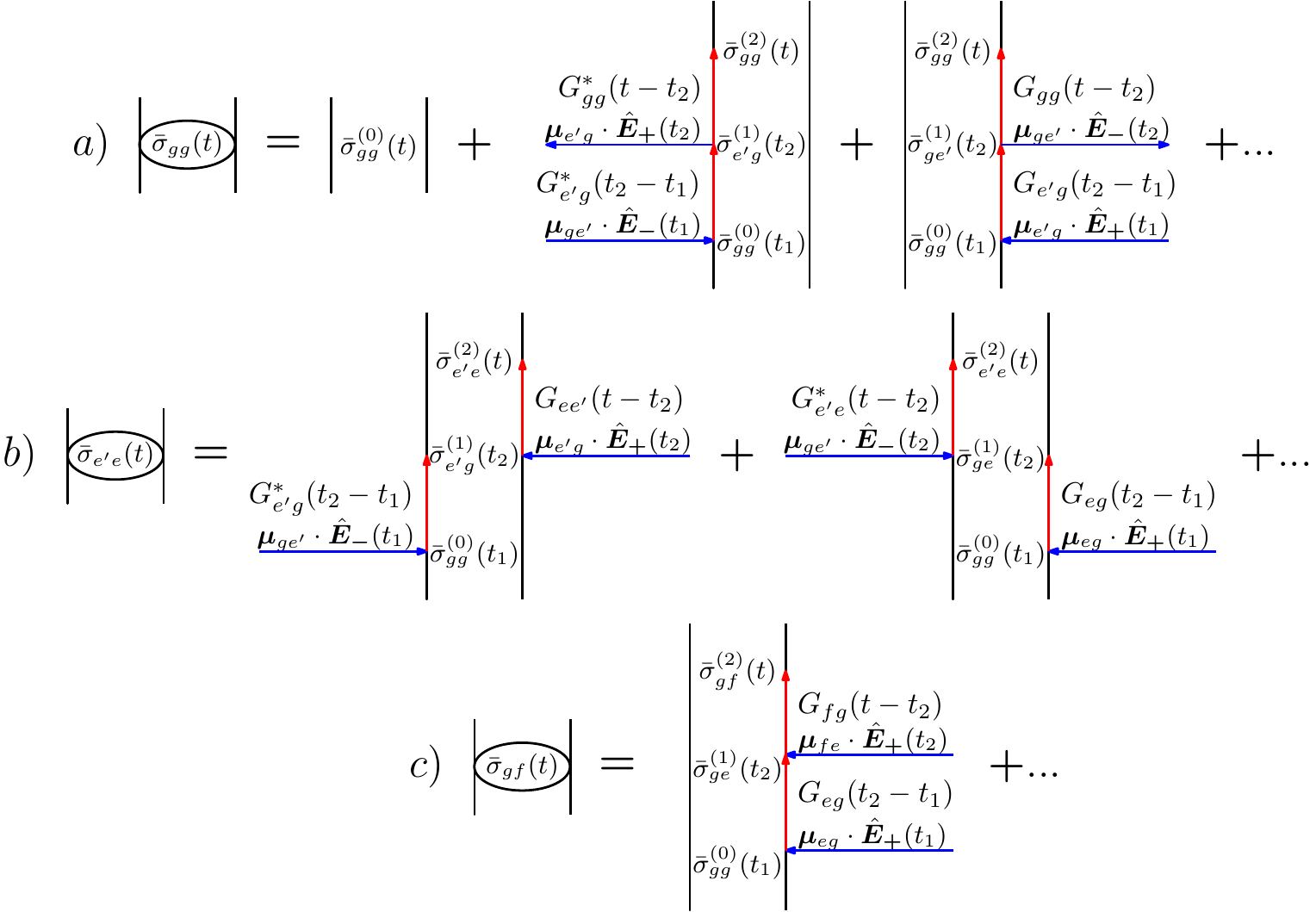}
    \caption{Perturbative expansion of each bubble diagram on the left.}
    \label{fig:Fig7_ipe.pdf}
\end{figure*}

In Fig. \ref{fig:exact diagrams split} we included bubbles to denote the formal solution of each coherence operator. Using the perturbative solution of $\bar\sigma_{gg}(t)$, $\bar\sigma_{e^\prime e}(t)$ and $\bar\sigma_{gf}(t)$ we can fill in each bubble with a perturbative diagram shown in Fig. \ref{fig:Fig7_ipe.pdf}, which we now describe. 

The first diagram is shown in Fig. \ref{fig:Fig7_ipe.pdf}a and is the perturbative expansion of the operator $\bar\sigma_{gg}(t)$. The zeroth order term is Eq. \eqref{eq:ggnought}, while the second order solution involves the sum of two contributions, absorption and emission on the bra and ket side. Since the result of the second order term involves an emission of a photon, the second order term $\bar\sigma_{gg}^{(2)}(t)$ is a correction to the zeroth order term, lowering the amplitude that the molecule is in the ground state. Thus when considering their contribution to the dynamics of $\bar\sigma_{ge}(t)$, the zeroth order term leads to one-photon absorption, and the second order term contributes to its saturation. 

The second diagram shown in in Fig. \ref{fig:Fig7_ipe.pdf}b is the perturbative expansion for $\bar\sigma_{e^\prime e}(t)$, which is only nonzero at second order and has two contributions. In the terminology of nonlinear spectroscopy the first term leads to the ``rephasing'' contribution and the second to the ``nonrephasing'' contribution \cite{Shen1984-fv,hamm2005principles}. Both of these terms combine to give coherence between intermediate states and when $e = e^\prime$ amplitude in the intermediate state. Thus when considering the contribution to the dynamics of $\bar\sigma_{ge}(t)$ we expect it to lead to saturation, and to that of $\bar\sigma_{ef}(t)$ we expect it to lead to two-photon absorption.

The third and final diagram is for the perturbative expansion of $\bar\sigma_{gf}(t)$ given by Fig. \ref{fig:Fig7_ipe.pdf}.c) and has one contribution which is given by the absorption of two photons on the bra side. This terms leads to a virtual two-photon transition to the excited state $f$, which has been referred to as the ``double
quantum coherence'' \cite{raymer2021entangled}, and the effect that this term has on $\bar\sigma_{ge}(t)$ and $\bar\sigma_{ef}(t)$ is complicated in general, but far from resonance will lead to two-photon absorption.

\section{absorption expressions \label{sec:absorption expressions}}
We are now at the point where we are able to calculate the absorption. The general expressions that result are given in Appendix \ref{sec:Absorption in time and frequency}. Here we define the terms involved and the steps necessary; in the next section we deal with simplified limits of the general expressions in an example calculation.

We begin by taking the classical expectation value which is done by replacing $G_{ij}(t,t_1)\to G_{ij}(t-t_1)$ and drop the average value symbol $[\![\cdot]\!]$ to avoid extra notation. Then we input the coherence operators $\bar\sigma_{ge}^{(1)}(t)$,$\bar\sigma_{ge}^{(3)}(t)$ and $\bar\sigma_{ef}^{(3)}(t)$ given in Eq. \eqref{eq:sigma_ge first order v1}, \eqref{eq:sigma_ge third order v1}, \eqref{eq:sigma_ef third order v1} respectively into Eq. \eqref{eq:quantum extinction approx} for the absorption. The result involves iterated time integrals (Eq. \eqref{eq:absorption in time}). 

To consider different incident states of light it is useful to move to frequency space to evaluate this expression. For any Hermitian operator $O(t)$, such as the electric field operator at the site of the molecule, we write 
\begin{equation}
    O(t) = \int\limits_{-\infty}^{\infty} \dbarw{} O(\omega)e^{-i\omega t},
\end{equation}
where we define $\dbarw{} = d\omega/\sqrt{2\pi}$. Introducing positive and negative frequency components in the usual way we then write all frequency integrals to range from $0\to \infty$, and leave these limits implicit in the expressions below. 

Evaluating each time integral introduces resonant frequency denominator contributions
\begin{equation}
    \label{eq:define Q}
    Q_{ij}(\omega) \equiv \omega_{ij} - \omega -i\gamma_{ij},
\end{equation}
which follow from the exponent of each Green function $G_{ij}(t-t_1)$. And the ``nonrephasing" and ``rephasing" contributions to $\bar\sigma_{e^\prime e}(t)$ mentioned above can be combined into a single term, involving a new function 

\begin{equation}
    \label{eq:definition R}
    R_{ee^\prime}(\omega) \equiv \frac{\omega_{ee^\prime} - \omega -i\gamma_{eg} - i\gamma_{e^\prime g}}{\omega_{ee^\prime} - \omega -i\gamma_{ee^\prime}}.
\end{equation}
The term $R_{ee^\prime}(\omega)$ deviates from unity only because we consider two broadening mechanisms, population decay and dephasing. In the limit when dephasing is negligible ($\gamma_{eg}\to \bar\Gamma_{eg}$), $R_{ee^\prime}(\omega) \to 1$. Physically, $R_{ee^\prime}(\omega)$ corresponds to contributions to the absorption that involve either a population term or a coherence between excited states. 

With the use of Eqs \eqref{eq:define Q} and \eqref{eq:definition R}, our expression involving iterated time integrals (Eq. \eqref{eq:absorption in time}) reduces to one involving frequency integrals (Eq. \eqref{eq:absorption in frequency}), which we write in terms of $\hat F_\pm^{ij}(\omega)$ to keep the notation simpler. But since the initial ket of the reservoir is the vacuum state,
\begin{equation}
    \hat F_{+}^{ij}(\omega)\ket{\Psi_\text{in}} = \frac{\boldsymbol{\mu}_{ij}}{\hbar}\cdot \hat{\boldsymbol{E}}_{+}(\omega)\ket{\Psi_\text{in}},
\end{equation} 
since each absorption term is normally ordered. Transitions are thus only made by the external field, as expected. 

This expression (Eq. \eqref{eq:absorption in frequency}) for the absorption of a molecule with many intermediate and final states; with possible resonant excitation to intermediate state and with two sources of broadening, is a main result of this paper. It is valid near resonance and for any incident state of light, in general propagating in three dimensions with any polarization, and either pulsed or CW. However, for our sample calculation in the following section we adopt a usual approximation and consider a field essentially uniform over a cross-section area $A$ in the $xy$  plane, and propagating in vacuum in the $z$ direction. The positive frequency components of the electric field operator within the cross-section area $A$, which in our case is sometimes called the ``entanglement area" \cite{lee2006entangled,parzuchowski2020setting,raymer2020two,landes2021quantifying,schlawin2018entangled}, are given by 
\begin{equation}
    \label{eq:E operator in frequency v1}
    \hat{\boldsymbol{E}}_{\boldsymbol{+}}(z,\omega) =\boldsymbol{e}\sqrt{\frac{\hbar \omega}{2\epsilon_0cA}}a(\omega)e^{ik(\omega)z},
\end{equation}
where we assumed the incident field has a single polarization given by $\boldsymbol{e}$, the wave-vector is given by $k(\omega) = \omega/c$, and the creation and annihilation operators satisfy the commutation relations 
\begin{equation}
    \left[a(\omega),a^\dagger(\omega^\prime)\right] = \delta(\omega - \omega^\prime),
\end{equation}
see, e.g., Blow et al. \cite{blow1990continuum}. In our sample calculations below, we consider a single molecule at the origin, and assume that the incident fields we consider have a centre frequency $\bar{\omega}$ and a bandwidth much smaller than their centre frequency. Then the positive frequency component of the electric field operator that appears in the dipole interaction can be taken as 
\begin{equation}
    \label{eq:E operator in frequency v2}
    \boldsymbol{\hat{E}_+}(\omega) = \boldsymbol{e}Ea(\omega),
\end{equation}
where 
\begin{equation}
    \label{eq:script E}
    E \equiv \sqrt{\frac{\hbar\bar{\omega}}{2\epsilon_0 c A}}.
\end{equation} 

\section{Three level molecule \label{sec:Three level molecule}}
As a sample calculation where the results can be given in closed form, we consider a three level molecule with a ground state $g$, single intermediate state $e$ and single final state $f$. 
As in Section \ref{sec:absorption expressions} we consider the incident field is given by a single polarization $\boldsymbol{e}$, and with that assumption we define the two constants
\begin{subequations}
\label{eq:dipole moment d_ij definition}
\begin{gather}
    d_{eg} \equiv \frac{\boldsymbol{\mu}_{eg}\cdot\boldsymbol{e}}{\hbar},\\
    d_{fe} \equiv \frac{\boldsymbol{\mu}_{fe}\cdot\boldsymbol{e}}{\hbar},
\end{gather}
\end{subequations}
which are the dipole transition elements scaled by $\hbar$.

Using these constants and our expression \eqref{eq:E operator in frequency v2} for the field operator, the absorption (Eq. \eqref{eq:absorption in frequency}) can be written as the sum of five contributions,  $\mathcal{A} = \mathcal{A}^1+\mathcal{A}^2+\mathcal{A}^3+\mathcal{A}^4+\mathcal{A}^5$, where
\begin{widetext}
\begin{subequations}
\label{eq:simplified absorption terms}
\begin{gather}
    \label{eq:simplified absorption terms one-photon}
        \mathcal{A}^1 = \int d\omega\text{Im}\left[\mathcal{R}^1(\omega)\right]\frac{\hbar\bar\omega\langle a^\dagger(\omega)a(\omega)\rangle }{A},\\
        \label{eq:simplified absorption terms two-photon}
        \mathcal{A}^i =2\pi\int \dbarw{1}\dbarw{2}\dbarw{3}\dbarw{4}\text{Im}\left[\mathcal{R}^i(\omega_1,\omega_2,\omega_3,\omega_4)\frac{(\hbar\bar\omega)^2\langle a^\dagger(\omega_4)a^\dagger(\omega_3)a (\omega_2)a(\omega_1)\rangle}{A^2}\right]\delta(\omega_1 + \omega_2 - \omega_3 - \omega_4),
\end{gather}
\end{subequations}
\end{widetext}
for $i=2, 3, 4, 5$, with the ``broadening functions'' given by
\begin{subequations}
\label{eq:resonant denominator functions}
\begin{gather}
    \mathcal{R}^1(\omega_1) = \frac{\hbar\bar\omega}{\epsilon_0c} \frac{|d_{eg}|^2}{Q_{eg}(\omega_1)},\\
    \mathcal{R}^2(\omega_1,\omega_2,\omega_3,\omega_4) = \frac{-\hbar\bar\omega|d_{eg}|^4 R_{ee}(\omega_2 - \omega_3)}{(\epsilon_0c)^2Q_{eg}(\omega_4)Q_{eg}^*(\omega_3)Q_{eg}(\omega_2)},\\
    \mathcal{R}^3(\omega_1,\omega_2,\omega_3,\omega_4) = \frac{\hbar\bar\omega|d_{fe}d_{eg}|^2}{2(\epsilon_0c)^2Q_{eg}(\omega_4)Q_{fg}(\omega_1 + \omega_2)Q_{eg}(\omega_2)},\\
    \mathcal{R}^4(\omega_1,\omega_2,\omega_3,\omega_4) = \frac{\hbar\bar\omega|d_{fe}d_{eg}|^2R_{ee}(\omega_2 - \omega_3) }{2(\epsilon_0c)^2Q_{fe}(\omega_4)Q_{eg}^*(\omega_3)Q_{eg}(\omega_2)},\\
    \mathcal{R}^5(\omega_1,\omega_2,\omega_3,\omega_4) = \frac{-\hbar\bar\omega|d_{fe}d_{eg}|^2}{2(\epsilon_0c)^2Q_{fe}(\omega_4)Q_{fg}(\omega_1 + \omega_2)Q_{eg}(\omega_2)}.
\end{gather}
\end{subequations}
Each absorption term is then given as a product of two functions. The lowest order term $\mathcal{A}^1$ consists of the imaginary part of the broadening function $\mathcal{R}^1(\omega)$ multiplied by the energy density/area of the incident field given by the one-photon correlation function $\langle a^\dagger(\omega)a(\omega)\rangle$, containing all the one-photon statistics. Each higher order term $\mathcal{A}^i$ ($i=2,3,4,5$) is given by the imaginary part of the higher order broadening function $\mathcal{R}^i(\omega_1,\omega_2,\omega_3,\omega_4)$ multiplied by the $\text{(energy density/area)}^2$ given by the four point correlation function $\langle a^\dagger(\omega_4)a^\dagger(\omega_3)a (\omega_2)a(\omega_1)\rangle$, containing all the two-photon statistics of the incident light.

The absorption terms $\mathcal{A}^1$ and $\mathcal{A}^2$ are generated from the perturbative expansion of the diagram in Fig. \ref{fig:exact diagrams split}.$\RN{1}$ and \ref{fig:exact diagrams split}.$\RN{2}$ and contribute to the one-photon absorption, the second term describing the onset of saturation. The fourth absorption term $\mathcal{A}^4$ is generated from the perturbative expansion of the diagrams in Fig. \ref{fig:exact diagrams split}.$\RN{4}$ and contributes to two-photon absorption. The third and fifth absorption terms, $\mathcal{A}^3$ and $\mathcal{A}^5$, are generated from the perturbative expansion of the diagrams in Fig. \ref{fig:exact diagrams split}.$\RN{3}$ and in Fig. \ref{fig:exact diagrams split}.$\RN{5}$, respectively, and are more complicated. But in the far detuned limit they will also contribute to two-photon absorption. 

We now consider different states of light and calculate each term. Our interest in this paper is in the CW limit.  

\section{CW coherent light \label{sec:CW Coherent state}}
We begin with coherent light, described by a state
\begin{equation}
    \label{eq:coherent state v1}
    \ket{\alpha} =
    D(\alpha) \ket{\text{vac}},
\end{equation}
where $\alpha$ is a complex coefficient and the displacement operator
\begin{equation}
    D(\alpha) \equiv e^{\alpha\int d\omega \varphi(\omega)a
    ^\dagger(\omega) - \text{H.c.}}.
\end{equation}
The spectral profile of the state is given by $\varphi(\omega)$ and is normalized to satisfy
\begin{equation}
    \int dt |\varphi(t)|^2 = \int d\omega |\varphi(\omega)|^2 = 1.
\end{equation}
To model coherent light in the CW limit, we consider a field oscillating at the frequency $\bar{\omega}$ for a length of time $T_p$. The normalized distribution in time and frequency is then
\begin{subequations}
\begin{gather}
    \varphi(t) = \frac{e^{-i\bar{\omega}t}}{\sqrt{T_p}}, \hspace{5mm} -\frac{T_p}{2}\le t\le \frac{ T_p}{2},\\
    \varphi(\omega) = \frac{1}{\sqrt{\Omega_p}}\sinc\left( \frac{(\omega - \bar{\omega})\pi}{\Omega_p}\right),
\end{gather}
\end{subequations}
where we take $\Omega_p = 2\pi/ T_p$ to identify an effective bandwidth. The CW limit is then $T_p\to \infty$, $\Omega_p\to0$, yielding a spectral profile $\varphi(\omega)$ strongly peaked at $\omega = \bar\omega$ with the squared spectral profile given by \cite{kusse2010mathematical}
\begin{equation}
        \lim_{\Omega_p \to 0} |\varphi(\omega)|^2 = \delta (\omega - \bar\omega).
\end{equation}
Taking the CW limit will lead to infinite absorption, since the light is always `on,' and in the calculation we divide each absorption term by $T_p$ to get a finite result as $T_p\to\infty$ that can be identified as an absorption rate. 

Using the transformation property
\begin{equation}
\label{eq:D opertor properties}
        D^\dagger(\alpha)a(\omega)D(\alpha) = a(\omega) + \alpha \varphi(\omega),
\end{equation}
 and the unitarity of the displacement operator, we can calculate the two correlation functions we need. The one-photon correlation function is given by
\begin{equation}
    \label{eq:coherent state expoectation values ada}
    \bra{\alpha} a^\dagger(\omega)a(\omega)\ket{\alpha} = |\alpha|^2\varphi^*(\omega)\varphi(\omega),
\end{equation}
and the two-photon correlation function by \begin{equation}
    \label{eq:coherent state expoectation values adadaa}
    \begin{split}
        &\bra{\alpha} a^\dagger(\omega_4)a^\dagger(\omega_3)a(\omega_2)a(\omega_1)\ket{\alpha}\\ &\hspace{35mm}=|\alpha|^4\varphi^*(\omega_4)\varphi^*(\omega_3)\varphi(\omega_2)\varphi(\omega_1).
    \end{split}
\end{equation}
In the CW limit both these quantities will peak and be significantly different from zero only when their argument(s) are close to $\bar\omega$.
 
From the one-photon correlation function we can immediately identify the energy of the light, which is given by
\begin{equation}
    \label{eq:energy in a coherent state}
    E_{\text{coh}} = \hbar\bar{\omega}|\alpha|^2,
\end{equation}
where as usual $|\alpha|^2$ is the average number of photons in the coherent state. We will let this diverge as $T_p\to\infty$ so the intensity in the pulse does not change in that limit, introducing a fixed parameter $\alpha_0$ by setting $\alpha = \alpha_0\sqrt{T_p}$. Then $F_\text{coh} \equiv|\alpha_0|^2/A$ is identified as the photon flux; the intensity of the beam is then
\begin{equation}
    I_\text{coh} = \hbar\bar\omega F_\text{coh}.
\end{equation}

Using the result of the correlation functions in Eq. \eqref{eq:coherent state expoectation values ada} and \eqref{eq:coherent state expoectation values adadaa}, and that in the CW limit $\varphi(\omega)$ is strongly peaked at $\omega = \bar\omega$, the rate of each absorption term for coherent light is given by
\begin{subequations}
\label{eq:final general coherent state absorption terms}
\begin{gather}
    \frac{\mathcal{A}^1_\text{coh}}{T_p} = I_{\text{coh}}  \sigma^1_\text{coh}\\
    \frac{\mathcal{A}^i_\text{coh}}{T_p}= I_\text{coh}^2 \sigma^i_\text{coh},
\end{gather}
\end{subequations}
for $i=2,3,4,5$, where each cross-section is
\begin{subequations}
    \label{eq:coherent state cross-section def.}
    \begin{gather}
        \sigma^1_\text{coh} =\text{Im}\left[ \mathcal{R}^1(\bar\omega)\right]\\
        \sigma^i_\text{coh} =\text{Im}\left[\mathcal{R}^i(\bar\omega,\bar\omega,\bar\omega,\bar\omega)\right].
    \end{gather}
\end{subequations}
The one-photon term scales with intensity and each higher order two-photon term scales with the intensity squared, as expected.
We note that each $\sigma^i_\text{coh}$ for $i=1,2,3,4,5$ is independent of the beam parameters aside from frequency. The quantity $\sigma^1_\text{coh}$ is the lowest order cross-section due to a one-photon transition, and has units of area. Each higher order cross-section $\sigma^i_\text{coh}$ for $i=2,3,4,5$ has units of $\text{area}/\text{intensity}$. 

Finally, we calculate the rate of absorption in full by evaluating and expanding each cross-section term in Eq. \eqref{eq:final general coherent state absorption terms} with the appropriate broadening functions in Eq. \eqref{eq:resonant denominator functions}, then each absorption term is given by
\begin{widetext}
\begin{subequations}
    \label{eq:coherent absorption terms in full}
    \begin{gather}
        \frac{\mathcal{A}^1_\text{coh}}{T_p} = I_\text{coh} \frac{|d_{eg}|^2\hbar\bar\omega}{\epsilon_0c} \frac{\gamma_{eg}}{(\omega_{eg} - \bar\omega)^2 + \gamma_{eg}^2},\\
        \frac{\mathcal{A}^2_\text{coh}}{T_p} =  -I_\text{coh}^2\frac{|d_{eg}|^4}{(\epsilon_0c)^2}\frac{\hbar\bar{\omega}}{\bar\Gamma_{eg}} \frac{\gamma_{eg}^2}{[(\omega_{eg} - \bar\omega)
        ^2 + \gamma_{eg}^2]^2},\\
        \label{eq:coherent absorption terms in full A3}
        \frac{\mathcal{A}^3_\text{coh}}{T_p} = I_\text{coh}^2\frac{|d_{eg}d_{fe}|^2\hbar\bar\omega}{2(\epsilon_0c)^2}\frac{(\omega_{eg} - \bar\omega)^2\gamma_{fg}+2(\omega_{eg} - \bar\omega)(\omega_{fg}-2\bar\omega)\gamma_{eg}-\gamma_{fg}\gamma_{eg}^2}{[(\omega_{eg} - \bar\omega)
        ^2 + \gamma_{eg}^2]^2[(\omega_{fg} - 2\bar\omega)
        ^2 + \gamma_{fg}^2]},\\
        \frac{\mathcal{A}^4_\text{coh}}{T_p}=I_\text{coh}^2 \frac{|d_{eg}d_{fe}|^2}{2(\epsilon_0c)^2}\frac{\hbar\bar{\omega}}{\bar\Gamma_{eg}}\frac{\gamma_{eg}}{(\omega_{eg} - \bar\omega)^2 + \gamma_{eg}^2}\frac{\gamma_{fe}}{(\omega_{fe} - \bar\omega)^2 + \gamma_{fe}^2},\\
        \label{eq:coherent absorption terms in full A5}
        \frac{\mathcal{A}^5_\text{coh}}{T_p} =I_\text{coh}^2 \frac{|d_{eg}d_{fe}|^2\hbar\bar{\omega}}{2(\epsilon_0c)^2}\frac{\gamma_{eg}\gamma_{fe}\gamma_{fg}-(\omega_{eg} - \bar\omega)(\omega_{fe} - \bar\omega)\gamma_{fg} - (\omega_{fg} - 2\bar\omega)[(\omega_{eg} - \bar\omega)\gamma_{fe} + (\omega_{fe} - \bar\omega)\gamma_{eg}]}{[(\omega_{eg} - \bar\omega)
        ^2 + \gamma_{eg}^2][(\omega_{fe} - \bar\omega)
        ^2 + \gamma_{fe}^2][(\omega_{fg} - 2\bar\omega)
        ^2 + \gamma_{fg}^2]}.
    \end{gather}
\end{subequations}
\end{widetext}
From equation \eqref{eq:coherent absorption terms in full} it is clear that the contribution of the terms $\mathcal{A}^3_\text{coh}$ and $\mathcal{A}^5_\text{coh}$ depends strongly on the relative magnitudes of the detunings from resonance and decay parameters. In certain regimes $\mathcal{A}^3_\text{coh}$ and $\mathcal{A}^5_\text{coh}$ can be either positive or negative, and thus can contribute to either absorption or saturation.

\subsection{Far detuned from intermediate state resonances} 
\begin{figure}
    \centering
    \includegraphics[width = 0.8\linewidth]{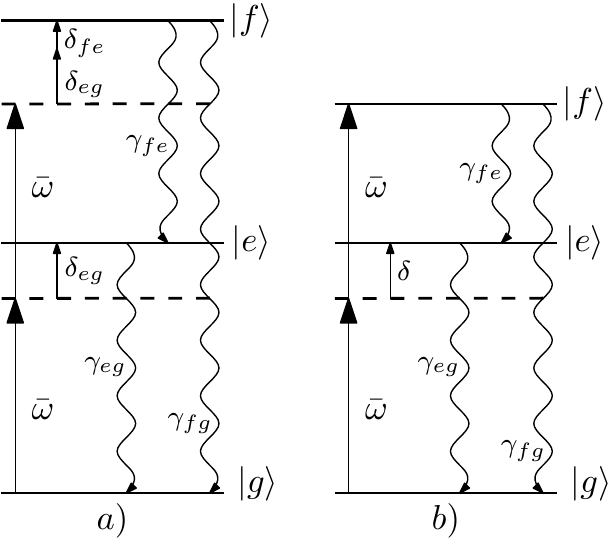}
    \caption{Three level molecule diagram for a) the general case, and b) when resonant with the two-photon transition.}
    \label{fig:figures/Fig4_ipe.pdf}
\end{figure}

To get a better understanding of $\mathcal{A}^3_\text{coh}$ and $\mathcal{A}^5_\text{coh}$, and to connect with the literature, we consider the limit when the field is far detuned from any resonances involving the intermediate level. That is, we assume $|\omega_{eg}-\bar\omega| \gg \gamma_{eg}$ and $|\omega_{fe}-\bar\omega| \gg \gamma_{fe}$; here one-photon absorption will be small, and two-photon absorption will dominate for large intensities.  Defining $\delta_{eg} \equiv \omega_{eg}-\bar\omega$ and $\delta_{fe} \equiv \omega_{fe}-\bar\omega$ to be detunings from the resonances (see Fig. \ref{fig:figures/Fig4_ipe.pdf}.a for molecule diagram), 
in the limit when $|\delta_{eg}| \gg \gamma_{eg}$ and $|\delta_{fe}| \gg \gamma_{fe}$, to good approximation the absorption terms are
\begin{subequations}
    \begin{gather}
        \frac{\mathcal{A}^1_\text{coh}}{T_p} = I_\text{coh} \frac{|d_{eg}|^2\hbar\bar\omega}{\epsilon_0c} \frac{\gamma_{eg}}{\delta_{eg}^2},\\
        \frac{\mathcal{A}^2_\text{coh}}{T_p} =  -I_\text{coh}^2\frac{|d_{eg}|^4}{(\epsilon_0c)^2}\frac{\hbar\bar{\omega}}{\bar\Gamma_{eg}} \frac{\gamma_{eg}^2}{\delta_{eg}^4},\\
        \frac{\mathcal{A}^3_\text{coh}}{T_p} = I_\text{coh}^2\frac{|d_{eg}d_{fe}|^2}{2(\epsilon_0c)^2}\frac{\hbar\bar\omega}{\delta_{eg}^2}\frac{\gamma_{fg}}{(\omega_{fg} - 2\bar\omega)
        ^2 + \gamma_{fg}^2},\\
        \frac{\mathcal{A}^4_\text{coh}}{T_p}=I_\text{coh}^2 \frac{|d_{eg}d_{fe}|^2}{2(\epsilon_0c)^2}\frac{\hbar\bar{\omega}}{\bar\Gamma_{eg}}\frac{\gamma_{eg}}{\delta_{eg}^2}\frac{\gamma_{fe}}{\delta_{fe}^2},\\
        \frac{\mathcal{A}^5_\text{coh}}{T_p} =I_\text{coh}^2 \frac{|d_{eg}d_{fe}|^2}{2(\epsilon_0c)^2}\frac{\hbar\bar{\omega}}{\delta_{eg}(-\delta_{fe})}\frac{\gamma_{fg}}{(\omega_{fg} - 2\bar\omega)
        ^2 + \gamma_{fg}^2}.
    \end{gather}
\end{subequations}
Here the first two terms describe one-photon absorption and its saturation, and the three remaining terms describe two-photon absorption. To maximize the two-photon absorption we take the incident field to be resonant with the two-photon transition, $2\bar\omega = \omega_{fg}$. In this limit the two detunings $\delta_{eg}$ and $\delta_{fe}$ simplify to a single detuning parameter $\delta$ which we define by $\delta \equiv \delta_{eg}$ and $\delta_{fe} = -\delta$ (see Fig. \ref{fig:figures/Fig4_ipe.pdf}.b for special case molecule diagram). Then each absorption term is further simplified to
\begin{subequations}
\label{eq:coherent state absorption far detuned}
    \begin{gather}
        \label{eq:coherent state absorption A^1}
        \frac{\mathcal{A}^1_\text{coh}}{T_p} = I_\text{coh} \frac{|d_{eg}|^2\hbar\bar\omega}{\epsilon_0c} \frac{\gamma_{eg}}{\delta^2},\\
        \label{eq:coherent state absorption A^2}
        \frac{\mathcal{A}^2_\text{coh}}{T_p} =  -I_\text{coh}^2\frac{|d_{eg}|^4}{(\epsilon_0c)^2}\frac{\hbar\bar{\omega}}{\bar\Gamma_{eg}} \frac{\gamma_{eg}^2}{\delta^4},\\
        \label{eq:coherent state absorption A^3}
        \frac{\mathcal{A}^3_\text{coh}}{T_p} = I_\text{coh}^2\frac{|d_{eg}d_{fe}|^2}{2(\epsilon_0c)^2}\frac{\hbar\bar\omega}{\delta^2}\frac{1}{\gamma_{fg}},\\
        \label{eq:coherent state absorption A^4}
        \frac{\mathcal{A}^4_\text{coh}}{T_p}=I_\text{coh}^2 \frac{|d_{eg}d_{fe}|^2}{2(\epsilon_0c)^2}\frac{\hbar\bar{\omega}}{\bar\Gamma_{eg}}\frac{\gamma_{eg}\gamma_{fe}}{\delta^4},\\
        \label{eq:coherent state absorption A^5}
        \frac{\mathcal{A}^5_\text{coh}}{T_p} =I_\text{coh}^2 \frac{|d_{eg}d_{fe}|^2}{2(\epsilon_0c)^2}\frac{\hbar\bar{\omega}}{\delta^2}\frac{1}{\gamma_{fg}}.
    \end{gather}
\end{subequations}
Each absorption term now has a simple scaling with the detuning $\delta$. Higher order population terms $\mathcal{A}^2_\text{coh}$ and $\mathcal{A}^4_\text{coh}$ scale as $\delta^{-4}$ and the virtual two-photon transitions terms $\mathcal{A}^3_\text{coh}$ and $\mathcal{A}^5_\text{coh}$ scale as $\delta^{-2}$.

In this limit we identify the rate of one-photon absorption
\begin{equation}
    \label{eq:coherent1PA}
    \frac{\mathcal{O}_\text{coh}}{T_p}= \frac{\mathcal{A}^1_\text{coh} + \mathcal{A}^2_\text{coh}}{T_p},
\end{equation}
where $\mathcal{A}^1_\text{coh}$ is the lowest order term and $\mathcal{A}^2_\text{coh}$ is the higher order correction due to saturation, and the rate of two-photon absorption by
\begin{equation}
    \frac{\mathcal{T}_\text{coh}}{T_p}=\frac{\mathcal{A}^3_\text{coh} + \mathcal{A}^4_\text{coh} + \mathcal{A}^5_\text{coh}}{T_p}.
\end{equation}
The second contribution to the two-photon absorption is $\mathcal{A}^4_\text{coh}$ which is generated from population in the intermediate state. The first and third contribution is $\mathcal{A}^3_\text{coh} + \mathcal{A}^5_\text{coh}$, which in this limit are identical, and are generated by coherence between $g$ and $f$ through a virtual two-photon transition.

For our perturbative expansion to be reasonable we must have $|\mathcal{A}^1_\text{coh}|\gg |\mathcal{A}^2_\text{coh}|$, so that the first saturation correction of the one-photon absorption is small and higher order corrections can be justifiably neglected. So we will restrict the possible detunings and intensities we consider so that the magnitude of $\mathcal{A}^2_\text{coh}$ satisfies
\begin{equation}
    \label{eq:valid PT}
    |\mathcal{A}^2_\text{coh}| \le 0.1 |\mathcal{A}^1_\text{coh}|,
\end{equation}
If this restriction holds, then the rate of one-photon absorption is to good approximation given by $\mathcal{O}_\text{coh}/T_p \simeq I_\text{coh}\sigma^{1}_\text{coh}$, where
\begin{equation}
    \sigma^1_\text{coh} =\hbar\bar\omega \frac{|d_{eg}|^2}{\epsilon_0c} \frac{\gamma_{eg}}{\delta^2},
\end{equation}
identifying the one-photon absorption cross-section $\sigma^\text{1PA}_\text{coh} \equiv \sigma^1_\text{coh}$, in agreement with past results \cite{boyd2020nonlinear,raymer2021entangled}. 

Moving to the two-photon absorption in the far detuned limit where $|\delta|\gg \gamma_{fe},\gamma_{ge}$, we find $\mathcal{A}^3_\text{coh} + \mathcal{A}^5_\text{coh}\gg \mathcal{A}^4_\text{coh}$ (assuming $\gamma_{fe}\simeq \gamma_{fg}$). So to good approximation $\mathcal{T}_\text{coh}/T_p \simeq I_\text{coh}^2 \sigma_\text{coh}^{\text{2PA}}$, where $\sigma_\text{coh}^{\text{2PA}}$ is the two-photon absorption cross-section given by
\begin{equation}
    \sigma_\text{coh}^{\text{2PA}} \simeq \sigma_\text{coh}^3 + \sigma_\text{coh}^5,
\end{equation}
which has units of area/intensity. Equivalently, we can define each cross-section by
\begin{equation}
    \acute{\sigma}_\text{coh}^i = \frac{\hbar\bar\omega}{2}\sigma_\text{coh}^i,
\end{equation}
and the two-photon absorption cross-section by $\acute{\sigma}_\text{coh}^{\text{2PA}}\simeq \acute{\sigma}_\text{coh}^3 + \acute{\sigma}_\text{coh}^5$, which has units of $\text{area}^4\text{s}/\text{photon}^2$, and is typically written using Goeppert Mayer units where $1\text{GM} = 10^{-58} \text{m}^4\text{s}/\text{photon}^{2}$. Then the rate of two-photon absorption is given by
\begin{equation}
    \begin{split}
        \frac{\mathcal{T}_\text{coh}}{T_p} &\simeq I_\text{coh}^2\sigma_\text{coh}^{\text{2PA}}\\
        &=2\hbar\bar\omega F_\text{coh}^2 \acute{\sigma}_\text{coh}^{\text{2PA}}\\
    \end{split}
\end{equation}
where $2\hbar\omega$ is the energy of the two-photon transition and 
\begin{equation}
    \label{eq:2PA cross-section}
    \acute{\sigma}_\text{coh}^{\text{2PA}}\equiv\frac{1}{2}\frac{|d_{fe}d_{eg}|^2}{(\epsilon_0c)^2}\frac{(\hbar\bar{\omega})^2}{\delta^2}\frac{1}{\gamma_{fg}},
\end{equation}
in agreement with past calculations \cite{boyd2020nonlinear,raymer2021entangled}. 
 
\subsection{Choice of parameters \label{Coherent Light: setting parameters}}
To estimate these cross-sections, and more generally the behavior of the absorption terms (\ref{eq:coherent absorption terms in full}) for different detunings, we must choose parameter values in a large parameter space. Our goal is to choose values that can be used to compare the results for the absorption of coherent light with the absorption of squeezed light, to be considered in the next section.  Since much of the recent interest in two-photon absorption of squeezed light has been generated by the possibility of increasing the efficiency of various applications that use large fluorescent molecules, we focus on parameters relevant there. However, the appropriate parameters for atoms or small molecules may be very different. 

Consider first $\gamma_{fg}$, where formally in our model the value is set by the contributions of dephasing and non-radiative population decay. However, as discussed earlier, we also want to use  $\gamma_{fg}$ to phenomenologically model the decay from the final state to the ground state due to fluorescence. So we set $\gamma_{fg}/2\pi= 1\text{THz}$, which is on the order of previously discussed values for typical fluorescent dyes \cite{raymer2020two}. For an order of magnitude estimate we set $d_{eg} = d_{fe} = |e|a_0/\hbar$, where $e$ is the electronic charge and $a_0$ is the Bohr radius. Then using $\bar\omega/2\pi = 357 \text{THz}$ ($\bar\lambda = 840 \text{nm}$), the adopted value for $\gamma_{fg}$ and setting $\delta = \bar\omega/4$ in Eq. \eqref{eq:2PA cross-section}, we calculate a two-photon absorption cross-section of $\acute{\sigma}^\text{2PA}_\text{coh} = 846 \text{GM}$. This is larger, but on the order of the experimentally determined value for Rhodamine B given by $\acute{\sigma}^\text{2PA,exp}_\text{coh} = 210\pm55 \text{GM}$, which was excited with a single-mode CW Ti:sapphire laser with an excitation power $P_\text{coh} = 1\text{W}$ at $\bar\lambda = 840\text{nm}$ focused to a beam area  $A\simeq 2.24\times10^{-13}\text{m}^2$, corresponding to the intensity $I_\text{coh}\simeq 4.4\times 10^{12} \text{W}/\text{m}^2$ \cite{xu1996measurement}.

To set an order of magnitude estimate of $\gamma_{eg}$ we note that for a clear signal of two-photon absorption we require $\mathcal{T}_\text{coh}\gg\mathcal{O}_\text{coh}$, in the far detuned limit. There the absorption terms in Eq. \eqref{eq:coherent state absorption far detuned} are valid, and this restriction simplifies to
\begin{equation}
    \gamma_{eg}\ll I_\text{coh} \frac{|d_{fe}|^2}{\epsilon_0 c }\frac{1}{\gamma_{fg}}.
\end{equation}
For the values given above this requires $\gamma_{eg}/2\pi\ll 0.17 \text{THz}$, which we use to set the order of magnitude estimate to $\gamma_{eg}/2\pi = 0.01 \text{THz}$. As a simple example we take $\gamma_{eg}\simeq\bar\Gamma_{eg}$, that is, the leading contribution to $\gamma_{eg}$ is due to $\bar\Gamma_{eg}$ and not the dephasing decay rate $\Lambda_{eg}$. Within this limit the decay of state $e$ is dominated by population decay and $R_{ee}(\omega)\simeq 1$.

Finally, we must set an order of magnitude for $\gamma_{fe}$. We begin by taking each dephasing decay rate to be approximately equal $\Lambda_{fe} \simeq \Lambda_{fg}\simeq\Lambda_{eg}\ll \bar\Gamma_{eg}$, then each $\gamma_{ij}\simeq \bar\Gamma_{ij}$ given by Eq. \eqref{eq:Gamma bar definition}. Under this assumption $\gamma_{fg}\simeq \bar\Gamma_{fg}$ and $\gamma_{fe}\simeq \bar\Gamma_{eg} + \bar\Gamma_{fg}$. Since $\bar\Gamma_{fg}\gg\bar\Gamma_{eg}$ which were set above, it necessarily follows from the definitions in Eq. \eqref{eq:Gamma bar definition} that $\gamma_{fe}\simeq\bar\Gamma_{fg}$ and we set it to be
$\gamma_{fe}/2\pi= 1\text{THz}$.

\subsection{Absorption of coherent light \label{sec:Coherent Light: plotting the absorption}} 
With these parameters set we can explore how the absorption varies with the intensity and detunings using the full expressions for each absorption term in Eq. \eqref{eq:coherent absorption terms in full}. As above we set the incident field to be resonant with the two-photon transition, and so there is only one detuning parameter $\delta$, but we allow it to range from below $\gamma_{eg}$ to above the value of $\gamma_{fe}$. As mentioned earlier, the signs of the absorption terms $\mathcal{A}^3_\text{coh}$ and $\mathcal{A}^5_\text{coh}$ depend on the linewidth parameters and detunings. For the incident field resonant with the two-photon transition ($2\bar\omega = \omega_{fg}$), as we assume here, $\mathcal{A}^5_\text{coh}$ is strictly positive, but the sign of $\mathcal{A}^3_\text{coh}$ depends on the detuning $\delta$; when $\delta < \gamma_{eg}$ it is negative, and when $\delta>\gamma_{eg}$ it is positive. For small detuning from intermediate state resonances a simple separation of the absorption into one- and two-photon contributions breaks down, and it appears to only make sense physically to talk about the total absorption. 

\begin{figure}
    \centering
    \includegraphics[width = \linewidth]{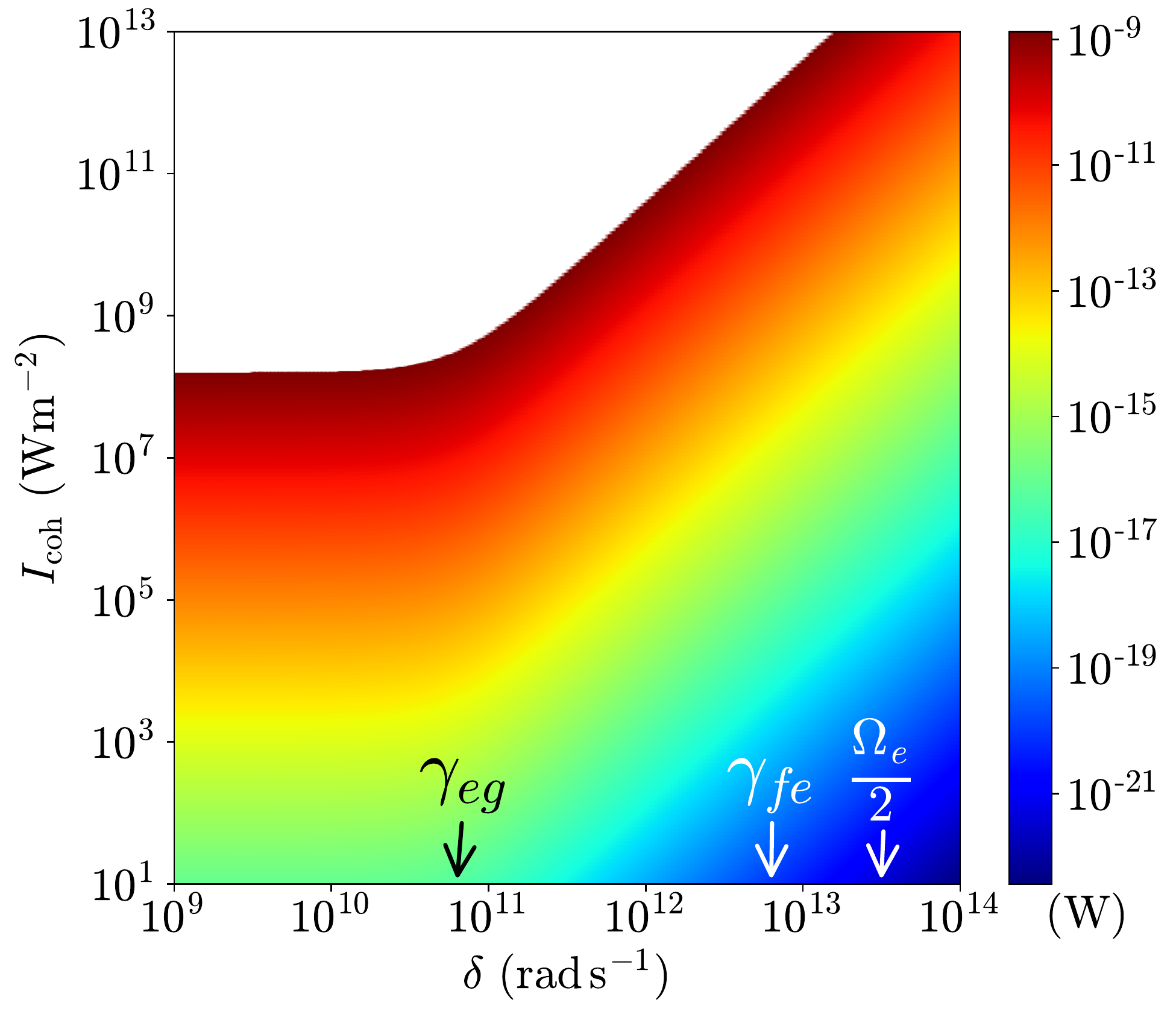}
    \caption{Plot of the rate of one-photon absorption for coherent light ($\mathcal{O}_\text{coh}/T_p$) against the detuning from resonance ($\delta$) and intensity $I_\text{coh}$. The ``white'' area represents the parameter space where the perturbation theory is not valid.}
    \label{fig:figures/Ocoh}
\end{figure}

\begin{figure}
    \centering
    \includegraphics[width = \linewidth]{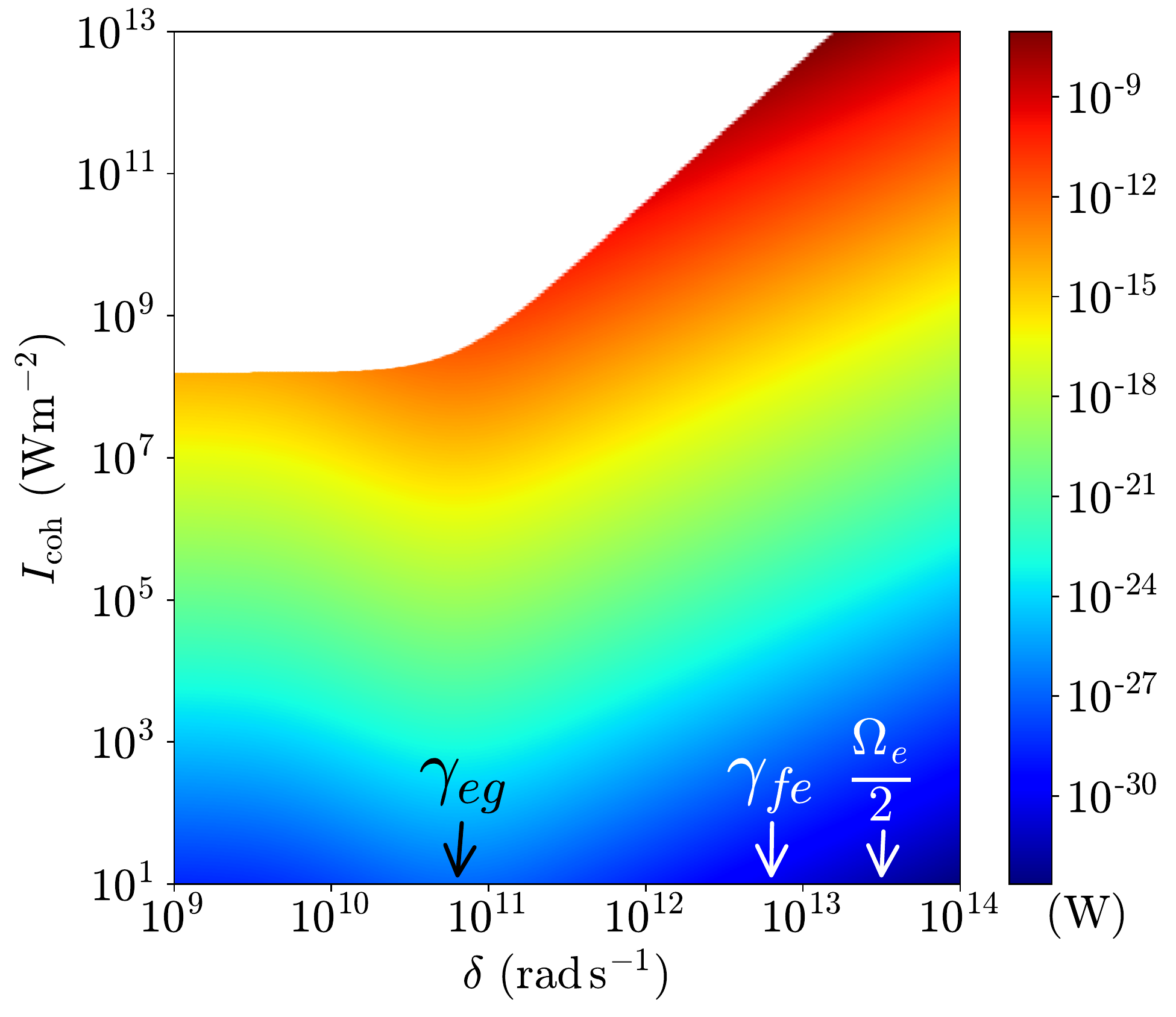}
    \caption{Plot of the rate of two-photon absorption for coherent light ($\mathcal{T}_\text{coh}/T_p$) against the detuning from resonance ($\delta$) and intensity $I_\text{coh}$. The ``white'' area represents the parameter space where the perturbation theory is not valid.}
    \label{fig:figures/Tcoh}
\end{figure}

Nonetheless, for our model parameters and intensities where the perturbation approach is valid (Eq.\ref{eq:valid PT})  we find $|\mathcal{A}^3_\text{coh}|\ll|\mathcal{A}^1_\text{coh}|$ and $|\mathcal{A}^3_\text{coh}|\ll|\mathcal{A}^2_\text{coh}|$ in the region where $\mathcal{A}^3_\text{coh}$ is negative, and thus there it is negligible compared to other contributions to the absorption. So in the region where $\mathcal{A}^3_\text{coh}$ is negative the absorption is dominated by one-photon absorption whether we group $\mathcal{A}^3_\text{coh}$ as a one- or two-photon effect. In the far detuned limit, again given by Eq. \eqref{eq:coherent state absorption far detuned}, we find $\mathcal{A}^3_\text{coh}=\mathcal{A}^5_\text{coh}$, and so in that region it clearly contributes to the two-photon absorption. Therefore for the parameters we are using, it is physically reasonable to define the general rate of one- and two-photon absorption by
\begin{subequations}
    \begin{gather}
        \label{eq:define 1PA}
        \frac{\mathcal{O}_\text{coh}}{T_p} \equiv \frac{\mathcal{A}^1_\text{coh}+\mathcal{A}^2_\text{coh}}{T_p},\\
        \label{eq:define 2PA}
        \frac{\mathcal{T}_\text{coh}}{T_p} \equiv \frac{\mathcal{A}^3_\text{coh} + \mathcal{A}^4_\text{coh}+\mathcal{A}^5_\text{coh}}{T_p},
    \end{gather}
\end{subequations}
which are plotted in Fig. \ref{fig:figures/Ocoh} and \ref{fig:figures/Tcoh} respectively, where the white areas denote the parameter region where perturbation theory is no longer valid due to Eq. \eqref{eq:valid PT} not being satisfied. In the far detuned and high intensity limit we find the absorption is dominated by $\mathcal{T}_\text{coh}$, as expected. 

\section{Squeezed Light \label{sec:Entangled States of Light}}
Next we consider degenerate squeezed light, which we model with the state
\begin{equation}
    \label{eq:squeezed state}
    \ket{\beta} = S(\beta) \ket{\text{vac}},
\end{equation}
where the unitary squeezing operator is given by
\begin{equation}
    \label{eq:squeezing operator}
    S(\beta) \equiv e^{\frac{\beta}{2}\int d\omega_1 d\omega_2\gamma(\omega_1,\omega_2)a^\dagger(\omega_1)a^\dagger(\omega_2) - \text{H.c.} },
\end{equation}
with $\beta = |\beta|e^{i\theta}$ the squeezing parameter.
The function $\gamma(\omega_1,\omega_2)$ is the joint spectral amplitude (JSA), which satisfies
\begin{equation}
    \int d\omega_1 d\omega_2 |\gamma(\omega_1,\omega_2)|^2 = 1.
\end{equation}

For a typical spontaneous parametric downconversion process, neglecting dispersion, the form of the JSA in time and frequency is 
\begin{subequations}
    \begin{gather}
        \label{eq:biphoton wf in time}
        \gamma(t_1,t_2) = \alpha\left(\frac{t_1+t_2}{2}\right)\phi(t_1-t_2),\\
        \gamma(\omega_1,\omega_2) = \alpha(\omega_1+\omega_2)\phi\left(\frac{\omega_1-\omega_2}{2}\right).
    \end{gather}
\end{subequations}
The function $\alpha(\omega)$ is related to the spectral distribution of the pump used to generate the squeezed light, and $\phi(\omega)$ is the phase-matching function describing the dispersion properties of the medium mediating the generation \cite{lerch2013tuning}.

For the simplest model of squeezed light, where the squeezing only involves a single (``super-") mode of the electromagnetic field and the JSA is unentangled, the one- and two-photon correlation functions are straightforward to evaluate using textbook results. It is well known that in this limit the two-photon correlation functions scale linearly and quadratically with photon number. For calculations where the spectral and temporal degrees of freedom are taken into account the same scaling is found \cite{schlawin2018entangled,schlawin2017entangled,dorfman2016nonlinear,dayan2007theory,schlawin2017entangled,fei1997entanglement,schlawin2013photon,oka2018two,raymer2020two,raymer2021large,landes2021quantifying,raymer2021entangled,raymer2022theory,schlawin2013photon}. The resulting linear dependence in the two-photon absorption is understood as ``photon pair absorption,'' since a squeezed state is a superposition of states involving only pairs of photons. This has been the center of experimental and theoretical discussions.

For a general squeezed state characterized by a correlated JSA, one cannot directly calculate the two-photon correlation functions because a transformation property of the form $S^\dagger(\beta)a(\omega)S(\beta)$ cannot be manipulated into a useful form. There are then three approaches that are immediately suggested. 

The first is to restrict oneself to the low flux or ``isolated pair'' limit. In this limit we can approximate the state in Eq. \eqref{eq:squeezed state} as a state that is mostly the vacuum state but includes a two-photon state with a small amplitude. This method was considered in great detail by Raymer et al. \cite{raymer2020two,raymer2021large,landes2021quantifying,raymer2021entangled} with the conclusion that, at low photon fluxes and current technologies, a linear scaling of two-photon absorption should be undetectable. 

A second approach involves performing a Schmidt decomposition of the JSA, where many Schmidt modes are present; indeed, if one approximates the JSA as a double Gaussian, the Schmidt decomposition has an analytic result \cite{schlawin2013photon}. This allows one in principle to evaluate the correlation functions, but in practice one is left with an infinite sum over Schmidt modes that are all delocalized in frequency. 

A third approach is to directly work with the field operators derived from the nonlinear light generation; this was employed by Dayan \cite{dayan2007theory}, and more recently by Raymer and Landes \cite{raymer2022theory} with similar results where they treated the narrow pump limit and included dispersion due to the nonlinear light generation, but with no resonant excitation of intermediate levels.

We employ a different approach here that is most similar to earlier work by Dayan and Raymer and Landes \cite{dayan2007theory,raymer2022theory}, where we take the squeezed light to be generated by a CW source. In this limit the spectral correlations are enhanced to such a strong degree the complexity of the JSA is \emph{reduced}, and one can carefully derive a transformation property for $S^\dagger(\beta)a(\omega)S(\beta)$, allowing for the evaluation of the correlation functions. We can then provide a calculation of the one- and two-photon absorption for a highly correlated squeezed state in the CW limit, and in both low and high intensity regimes. 

\subsection{CW squeezed light}
To model the JSA in the CW limit we take
\begin{subequations}
    \label{eq:squeezed state pump}
    \begin{gather}
    \alpha(t) = \frac{e^{-i2\bar{\omega}t}}{\sqrt{T_p}}, \hspace{5mm} -\frac{T_p}{2}\le t\le \frac{T_p}{2},\\
    \alpha(\omega) = \frac{1}{\sqrt{\Omega_p}}\sinc\left( \frac{(\omega - 2\bar{\omega})\pi}{\Omega_p}\right),
    \end{gather}
\end{subequations}
which satisfies
\begin{equation}
    \int dt |\alpha(t)|^2 =  \int d\omega |\alpha(\omega)|^2 = 1,
\end{equation}
where $T_p$ is the length of the pump pulse and $\Omega_p = 2\pi/T_p$ identifies its bandwidth. The pump frequency is centered at $2\bar\omega$ so that the photon pairs generated will be centered at $\bar\omega$. To take the CW limit we let $T_p\to\infty$, $\Omega_p\to 0$, so that
\begin{equation}
        \lim_{\Omega_p \to 0} |\alpha(\omega)|^2 = \delta (\omega - 2\bar\omega).
\end{equation}

\begin{figure}
    \centering
    \includegraphics[width = \linewidth]{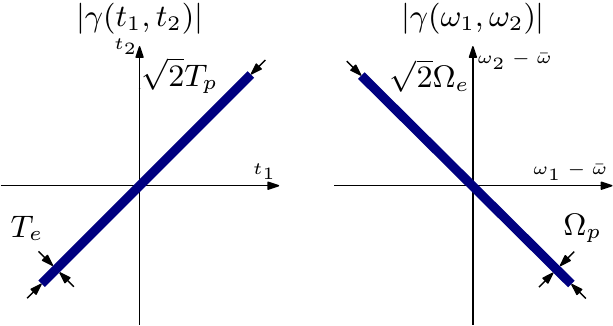}
    \caption{Schematic of JSA for the pump and phase matching function given in Eq. \eqref{eq:squeezed state pump} and \eqref{eq:model of PMF}. Our diagram is a schematic representation because we have neglected to include the sinc oscillations present, however, in the CW limit when the state is strongly correlated the schematic representation is a good approximation of the JSA being considered. }
    \label{fig:bi-photon}
\end{figure}

Although general phase-matching functions could be considered, to allow us to extract analytic results we neglect dispersion and adopt a simple model where $\phi(\omega)$ takes on a fixed value at the frequencies where it is nonzero:  
\begin{subequations}
\label{eq:model of PMF}
\begin{gather}
    \label{eq:model pmf in time}
    \phi(t) =\frac{1}{\sqrt{T_e}}\sinc\left(\frac{\pi t}{T_e}\right),\\
    \phi(\omega) = \frac{1}{\sqrt{\Omega_e}}, \hspace{5mm}-\frac{\Omega_e}{2}\le \omega\le\frac{\Omega_e}{2},
\end{gather}
\end{subequations}
which are both real functions and satisfy
\begin{equation}
    \int d\omega |\phi(\omega)|^2 = \int dt |\phi(t)|^2 =1.
\end{equation}
The phase-matching function is characterized by the bandwidth $\Omega_e$, which is the range of frequencies over which the generation is effective: The frequency components of the generated photons are centered at $\bar\omega$, but range between $\bar\omega-\Omega_e/2\le\omega\le\bar\omega+\Omega_e/2$. The related time parameter is $T_e = 2\pi/\Omega_e$ and is called the ``entanglement time''\cite{lee2006entangled,schlawin2017entangled,szoke2020entangled,villabona2018two,parzuchowski2020setting,corona2022experimental,raymer2020two,raymer2021large,schlawin2018entangled,fei1997entanglement}. The quantity $T_e$ is a measure of the ``coherence time" of pairs of photons, the time over which a pair of photons can be thought of as being correlated. Indeed, photons are generated at times $t_1$ and $t_2$ according to the JSA (Eq. \eqref{eq:biphoton wf in time}) which is proportional to $\phi(t_1-t_2)$, which from Eq. \eqref{eq:model pmf in time} is peaked for times $t_1-t_2<T_e$. The form of the JSA in time and frequency variables is schematically shown in Fig. \ref{fig:bi-photon}. For a highly correlated state the parameters satisfy $T_p\gg T_e$ or equivalently $\Omega_e\gg \Omega_p$; we consider this limit in this paper. 

In Appendix \ref{app:squeezing operator} we work out the operator $S^\dagger(\beta)a(\omega)S(\beta)$ for squeezed light from a medium pumped by a CW source with a general phase-matching function $\phi(\omega)$ (Eq.\eqref{eq:general squeezing transform 1}). Specifying to our simple model \eqref{eq:model of PMF} of the phase-matching function, that transformation simplifies to 
\begin{equation}
    \begin{split}
        \label{eq:squeezed state transform v2}
        &S^\dagger(\beta)a(\omega)S(\beta)\\
        &\to a(\omega)+ \frac{[c(\omega) - 1]}{\sqrt{\Omega_p}}\int d\omega^\prime\alpha\left(2\bar{\omega} - \omega +  \omega^\prime\right)a(\omega_2)\\
        &+ \frac{s(\omega)e^{i\theta}}{\sqrt{\Omega_p}}\int d\omega^\prime\alpha\left(\omega +  \omega_1\right)a^\dagger(\omega^\prime),
    \end{split}
\end{equation}
where we have set
\begin{subequations}
    \label{eq:s and c for simplified phi}
    \begin{gather}
        s(\omega)= s = \text{sinh}(|\beta_0|)\hspace{2mm}\text{if}\hspace{2mm} |\omega-\bar\omega|\le \frac{\Omega_e}{2},\\
        c(\omega)= c = \text{cosh}(|\beta_0|)\hspace{2mm}\text{if}\hspace{2mm} |\omega-\bar\omega|\le \frac{\Omega_e}{2},\\
        s(\omega) =0\hspace{2mm}\text{if}\hspace{2mm} |\omega-\bar\omega|> \frac{\Omega_e}{2},\\
        c(\omega)= 1\hspace{2mm}\text{if}\hspace{2mm} |\omega-\bar\omega|> \frac{\Omega_e}{2},
    \end{gather}
\end{subequations}
which satisfy $s(2\bar\omega - \omega) = s(\omega)$ and $c(2\bar\omega - \omega) = c(\omega)$ and define the squeezing parameter $\beta_0$ by setting $\beta_0 \equiv \beta\sqrt{T_e/T_p}$. As we will see, the introduction of $\beta_0$ essentially ``normalizes'' the squeezing parameter in a way analogous to the introduction of $\alpha_0$ from $\alpha$ in Sec. \ref{sec:CW Coherent state}. If we keep the intensity in the center of the pulse the same, the squeezing parameter $\beta$ will be proportional to $\sqrt{T_p/T_e}$, because $\beta$ is related to the average photon number. From the above definition of $\beta_0$ one can see that $\beta_0$ will remain fixed and independent of $T_p$ and $T_e$; although $\beta$ diverges in the CW limit, $\beta_0$ is fixed. 

With the transformation property of Eq. \eqref{eq:squeezed state transform v2} we can calculate the one- and two-photon correlation functions (Appendix \ref{app:CW Squeezed state correlation functions}). The one-photon correlation function evaluated at two different frequencies (Eq. \eqref{eq:squeezed state one-photon correlation function}) is given by
\begin{equation}
    \begin{split}
        \label{eq:squeezed correlation fn 1 v1}
        \bra{\beta}a^\dagger (\omega_2)a(\omega_1)\ket{\beta} =\frac{T_p}{2\pi}s(\omega_2)s(\omega_1)\text{sinc}\left(\frac{
        (\omega_1-\omega_2)\pi}{\Omega_p}\right)
    \end{split}
\end{equation}
which is proportional to $T_p$ as would be expected. Note that the factor $s(\omega_2)s(\omega_1)$ includes the restriction of frequencies to be within the bandwidth of the squeezed light. Using Eq. \eqref{eq:squeezed correlation fn 1 v1} we can immediately calculate the pulse energy, which is
\begin{equation}
    \label{eq:energy in a squeezed state state}
    E_{\text{sq}} = \hbar\bar{\omega}\frac{ T_p}{T_e}\text{sinh}^2\left(|\beta_0|\right).
\end{equation}
We can identify the flux of photons to be
\begin{equation}
    \label{eq:squeezed state photon flux}
    F_\text{sq} \equiv \frac{1}{AT_e}\text{sinh}^2\left(|\beta_0|\right)
\end{equation}
(and the total number of photons in the pulse to be $N_\text{sq} = T_p A F_\text{sq}$), for then we obtain the correct expression for the intensity of light in the first line below,
\begin{equation}
\begin{split}
    \label{eq:sq state intensity}
    I_\text{sq} &\equiv \frac{E_\text{sq}}{A T_p}=\hbar\bar\omega F_\text{sq}\\
    &=\frac{\hbar\bar\omega}{AT_e}\text{sinh}^2\left(|\beta_0|\right).
\end{split}
\end{equation}
while from the second line we can identify the number of photons within the entanglement time $T_e$ to be $N_{T_e}\equiv\text{sinh}^2\left(|\beta_0|\right)$. 

The squeezed state can then be understood as follows: Photons pairs are created throughout the length of the CW beam ($-T_p/2 \le t \le T_p/2$) with pairs being localized within a time $T_e$; $N_{T_e}=\text{sinh}^2\left(|\beta_0|\right)$ is the expected number of photons within a time $T_e$; and since $\beta_0$ is fixed, $N_{T_e}$ does not depend on $T_e$ or $T_p$. 

To recover the weakly squeezed (isolated pair) limit \cite{landes2021quantifying}, we take $|\beta_0|\ll1$ so that we can approximate $N_{T_e}\to|\beta_0|^2\ll 1$. Then we identify $|\beta_0|^2$ as the number of photons within a time $T_e$; and if $\beta_0$ is small enough, we expect there to be at most one pair of photons within a length of time $T_e$. Then the flux is $F_\text{sq}\to|\beta_0|^2/(AT_e)$ and the number of photons is given by $N_\text{sq} \to T_p |\beta_0|^2/T_e = |\beta|^2$, acknowledging $|\beta|^2$ to be the total number of photons in this limit.

Moving to the two-photon correlation function (Eq. \eqref{eq:squeezed state two-photon correlation function}), we have
\begin{equation}
    \begin{split}
    \label{eq:squeezed correlation fn 2 v1}
        &\bra{\beta}a^\dagger(\omega_4)a^\dagger(\omega_3)a(\omega_2)a(\omega_1)\ket{\beta} \\
        &= \frac{T_p}{2\pi} s(\omega_3)c(\omega_3)c(\omega_1)s(\omega_1)\alpha(\omega_1 + \omega_2)\alpha(\omega_3 + \omega_4)\\
        &+\frac{T_p}{2\pi} s^2(\omega_2)s^2(\omega_1)\alpha(\omega_1 - \omega_4 + 2\bar{\omega})\alpha(\omega_2 - \omega_3 + 2\bar{\omega})\\
        &+\frac{T_p}{2\pi}s^2(\omega_2)s^2(\omega_1)\alpha(\omega_1 - \omega_3 + 2\bar{\omega})\alpha(\omega_2 - \omega_4 + 2\bar{\omega}),\\
    \end{split}
\end{equation}
again proportional to $T_p$ as expected. The two-photon correlation function has three contributions, and in each the factors $s(\omega)$ ensure that the photons are within the bandwidth of the squeezed light. The first term, previously referred to as the ``coherent'' contribution  \cite{dayan2007theory,raymer2022theory} is due to photons that are anti-correlated such that $\omega_1 + \omega_2 = 2\bar\omega$. Due to the anti-correlations this contribution is dominant for $\mathcal{A}^3_\text{sq}$ and $\mathcal{A}^5_\text{sq}$, which are derived from absorption pathways which include a two-photon transition with only a virtual excitation of the intermediate level.

The second and third contributions to Eq. \eqref{eq:squeezed correlation fn 2 v1}, previously referred to as the ``incoherent contribution \cite{dayan2007theory,raymer2022theory} are correlated such that $\omega_1 = \omega_4$ and $\omega_2 = \omega_3$ for the second term and $\omega_1 = \omega_3$ and $\omega_2 = \omega_4$ for the third term. Although the third contribution is very similar to the second, the different frequency correlations can lead to large differences in how they contribute to the absorption because the broadening function $\mathcal{R}^i(\omega_1,\omega_2,\omega_3,\omega_4)$ (see Eq. \eqref{eq:resonant denominator functions}) is in general not symmetric under exchange of any of its frequency arguments. Specifically, we expect the second and third term to dominate for $\mathcal{A}^2_\text{sq}$ and $\mathcal{A}^4_\text{sq}$. But because the ``rephasing" and ``nonrephasing" pathway is doubly resonant when $\omega_2 = \omega_3$, which can be seen by examining $\mathcal{R}^2(\omega_1,\omega_2,\omega_3,\omega_4)$ and $\mathcal{R}^4(\omega_1,\omega_2,\omega_3,\omega_4)$ given in Eq. \eqref{eq:resonant denominator functions}, we expect the second contribution to be larger than the third.

Using the model we have developed and the correlation functions we derived in Eq. \eqref{eq:squeezed correlation fn 1 v1} and \eqref{eq:squeezed correlation fn 2 v1} we can immediately calculate the normalized second order correlation function in frequency, defined by
\begin{equation}
    g^{(2)}(\omega_1,\omega_2) \equiv \frac{\bra{\beta}a^\dagger(\omega_1)a^\dagger(\omega_2)a(\omega_2)a(\omega_1)\ket{\beta}}{\bra{\beta}a^\dagger(\omega_2)a(\omega_2)\ket{\beta}\bra{\beta}a^\dagger(\omega_1)a(\omega_1)\ket{\beta}}.
\end{equation}
We find
\begin{equation}
    \begin{split}
        \label{g2 freq}
        g^{(2)}(\omega_1,\omega_2) &=\frac{c^2(\omega_1)}{s^2(\omega_1)} \text{sinc}^2\left(\frac{(\omega_1+\omega_2-2\bar\omega)\pi}{\Omega_p}\right)\\
        &+1 + \text{sinc}^2\left(\frac{(\omega_1-\omega_2)\pi}{\Omega_p} \right)
    \end{split}
\end{equation}
which is in agreement with experiment when dispersion is compensated \cite{cutipa2022bright}. The anti-correlated term in which $\omega_1+\omega_2 = 2\bar\omega$ is the ``cross-correlation'' contribution and the correlation along the diagonal where $\omega_1 = \omega_2$ is the ``auto-correlation'' of each frequency mode with itself for indistinguishable photons. Here each correlation is sensitive to the pump bandwidth set by $\Omega_p$.

Alternatively we calculate the normalized second order correlation function in time, defined by
\begin{equation}
    g^{(2)}(t_1,t_2) \equiv \frac{\langle\beta| a^\dagger(t_1)a^\dagger(t_2)a(t_2)a(t_1)|\beta\rangle}{\langle\beta| a^\dagger(t_2)a(t_2)|\beta\rangle\langle\beta| a^\dagger(t_1)a(t_1)|\beta\rangle},
\end{equation}
which we calculate by taking the Fourier transforms of Eq. \eqref{eq:squeezed correlation fn 1 v1} and \eqref{eq:squeezed correlation fn 2 v1} to give
\begin{equation}
    \begin{split}
        \label{eq:g2 time}
        g^{(2)}(t_1,t_2) = 1 + \left(2+\frac{1}{N_{T_e}}\right)\text{sinc}^2\left(\frac{(t_1 - t_2)\pi}{T_e}\right)
    \end{split}
\end{equation}
which yields the usual result for $t_1=t_2$ \cite{walls1983squeezed,loudon1987squeezed,raymer2022theory,iskhakov2012superbunched,spasibko2017multiphoton} and satisfies the inequality that $g^{(2)}(0,\tau)\le g^{(2)}(0,0)$, the condition for ``bunched light''. However, we note that while in frequency space the auto-correlation term (third contribution of Eq. \eqref{g2 freq}) is peaked when $\omega_1=\omega_2$ and is characterized by the pump width $\Omega_p$, in time this contribution lead to a correlation that $t_1= t_2$ on time scales \emph{not} given by the inverse bandwidth of the pump but by the entanglement time $T_e$.

Using the correlation functions \eqref{eq:squeezed correlation fn 1 v1}, the lowest order rate of absorption is given by
\begin{equation}
    \label{eq:sq state A^1 absorption}
        \frac{\mathcal{A}^1_\text{sq}}{T_p} = I_\text{sq}\sigma^1_\text{sq},\\
\end{equation}
which scales linearly with intensity as expected, but with a modified cross-section due to the large bandwidth of the squeezed light given by
\begin{equation}
    \label{eq:squeezed state one-photon cross-section}
    \sigma^\text{1}_\text{sq} =\int\limits_{\bar\omega-\frac{\Omega_e}{2}}^{\bar\omega+\frac{\Omega_e}{2}} \frac{d\omega}{\Omega_e}\text{Im}\left[\mathcal{R}^1(\omega)\right].
\end{equation}

For each higher order absorption term we use the correlation function \eqref{eq:squeezed correlation fn 2 v1}, together with the property that in the CW limit $\alpha(\omega)$ is strongly peaked at $\omega = 2\bar\omega$, then the rate of each higher order absorption term is given by
\begin{equation}
    \label{eq:sq state A^i absorption}
        \frac{\mathcal{A}^i_\text{sq}}{T_p} = I_\text{sq}(I_\text{vac} + I_\text{sq})\sigma^{i}_\text{sq,\RN{1}}+ 2I_\text{sq}^2\sigma^{i}_\text{sq,\RN{2}},
\end{equation}
where we find it useful to define a quantity called the ``vacuum intensity'' by
\begin{equation}
    \label{eq:quantum enh}
    I_\text{vac} = \frac{\hbar\bar\omega}{AT_e},
\end{equation}
following the work of Klyshko \cite{klyshko1982transverse} where a similar quantity was defined, which we understand as the intensity of one-photon passing through an area $A$ within a time $T_e$. We find each higher order absorption term is dependent on two cross-sections $\sigma^{i}_\text{sq,\RN{1}}$ and $\sigma^{i}_\text{sq,\RN{2}}$  for $i=2,3,4,5$. The first cross-section $\sigma^{i}_\text{sq,\RN{1}}$ is given by
\begin{equation}
    \label{eq:sq state cross-section same}
    \sigma^{i}_\text{sq,\RN{1}}=\int\limits_{\bar\omega -\frac{\Omega_e}{2}}^{\bar\omega + \frac{\Omega_e}{2}}\frac{d\omega_1 d\omega_2}{\Omega_e^2} \text{Im}\left[\mathcal{R}^i(\omega_1,2\bar\omega - \omega_1,\omega_2,2\bar\omega - \omega_2)\right],
\end{equation}
where we use the subscript ``$\RN{1}$'' for $\sigma^{i}_\text{sq,\RN{1}}$ to denote that this cross-section corresponds to the anti-correlated term of the correlation function (first term in Eq. \eqref{eq:squeezed correlation fn 2 v1}). The second cross-section $\sigma^{i}_\text{sq,\RN{2}}$ is given by
\begin{equation}
    \begin{split}
    \label{eq:sq state cross-section diff}
        \sigma^{i}_\text{sq,\RN{2}}&= \frac{1}{2}\int\limits_{\bar\omega-\frac{\Omega_e}{2}}^{\bar\omega+\frac{\Omega_e}{2}}\hspace{-1mm} \frac{d\omega_1 d\omega_2}{\Omega_e^2} \text{Im}\left[\mathcal{R}^i(\omega_1,\omega_2,\omega_2,\omega_1)\right.\\
        &\hspace{40mm}+\left.\mathcal{R}^i(\omega_1,\omega_2,\omega_1,\omega_2)\right],
    \end{split}
\end{equation}
where we use the subscript ``$\RN{2}$'' for $\sigma^{i}_\text{sq,\RN{2}}$ to denote that this cross-section corresponds to the second and third term from the correlation function given by Eq. \eqref{eq:squeezed correlation fn 2 v1}. Here we combined the second and third contribution into a single term and pulled out a factor of $2$, so this can be understood as an average cross-section.

Considering the form of each higher order absorption term $\mathcal{A}^i_\text{sq}$ in Eq. \eqref{eq:sq state A^i absorption} we find two contributions. The first, which is proportional to $I_\text{sq}(I_\text{vac} + I_\text{sq})$, is multiplied by the cross-section $\sigma^{i}_\text{sq,\RN{1}}$ corresponding to photons that are anti-correlated in frequency. For weak squeezing such that $\beta_0\ll1$, then $N_{T_e}\ll1$ and we can think of only a single pair of photons within a time $T_e$ leading to a linear scaling of absorption with intensity and $I_\text{vac}$ playing the role of an enhancement factor. Then as $\beta_0$ increases, $N_{T_e}\gg 1$ where there are many photons within a time $T_e$ so the absorption scales quadratically with intensity but is still multiplied by $\sigma^{i}_\text{sq,\RN{1}}$.

While one view is that squeezed light modifies the cross-section, and so an ``entangled cross-section'' is sometimes defined \cite{lee2006entangled,szoke2020entangled,tabakaev2021energy,villabona2018two,parzuchowski2020setting,corona2022experimental,mikhaylov2021hot,fei1997entanglement,raymer2020two,raymer2021large,landes2021quantifying}, and in our notation is given by $I_\text{vac}\sigma^{i}_\text{sq,\RN{1}}$, here we take a different approach because in the current form it will be more straight forward to compare to coherent light. Further, the contribution from $\sigma^{i}_\text{sq,\RN{1}}$ involves not only a linear scaling with intensity, but also a quadratic one.

The second contribution on the right-hand side of Eq. \eqref{eq:sq state A^i absorption}, which is due to $\sigma^{i}_\text{sq,\RN{2}}$, scales as intensity squared. The result of averaging the contributions to the cross-section $\sigma^{i}_\text{sq,\RN{2}}$ in Eq. \eqref{eq:sq state cross-section diff} is a factor of 2, for the two ways of choosing two indistinguishable photons. 

To evaluate the squeezed state cross-sections and absorption terms in general we need to resort to numerical integration. However, we first consider the limiting case where the squeezed state has a narrow bandwidth $\Omega_e$. 

\subsubsection{Squeezed state absorption: narrow bandwidth limit}
From Eq. \eqref{eq:squeezed state one-photon cross-section} and \eqref{eq:sq state A^i absorption} we see that to evaluate the cross-sections for absorption we must integrate over the bandwidth of squeezed light. The integrals can be simplified if we take the limit when the bandwidth $\Omega_e$ of the squeezed light is small compared to the decay widths $\gamma_{ij}$. In this limit, the coherence time of the photon pairs is long compared to the decay times of the molecule but each photon has a well defined frequency. Since each photon has a well defined frequency, the spectral correlations within $\sigma^{i}_\text{sq,\RN{1}}$ and $\sigma^{i}_\text{sq,\RN{2}}$ (\eqref{eq:sq state cross-section same} and \eqref{eq:sq state cross-section diff}) are no longer relevant. To good approximation we can evaluate each cross-section in Eq. \eqref{eq:squeezed state one-photon cross-section}, \eqref{eq:sq state cross-section same} and \eqref{eq:sq state cross-section diff} at the centre frequency $\bar\omega$ and we find that 
\begin{subequations}
    \begin{gather}
         \left.\sigma_\text{sq}^1 \right\vert_{\frac{\Omega_e}{\gamma_{ij}}\ll1}=\sigma_\text{coh}^1\\
        \left.\sigma^{i}_\text{sq,\RN{1}},\sigma^{i}_\text{sq,\RN{2}}\right|_{\frac{\Omega_e}{\gamma_{ij}}\ll1} =\sigma_\text{coh}^i.
    \end{gather}
\end{subequations}
Here each squeezed light cross-section reduces to the coherent light cross-sections we defined in equation \eqref{eq:coherent state cross-section def.}.

The one- and two-photon absorption rates then simplify to \begin{subequations}
    \begin{gather}
        \left.\frac{\mathcal{A}^1_\text{sq}}{T_p} \right|_{\frac{\Omega_e}{\gamma_{ij}}\ll1}\hspace{-6mm}= I_\text{sq}\sigma^1_\text{coh},\\
        \label{eq:two-photon absorption narrow bandwidth limit}
        \left.\frac{\mathcal{A}^i_\text{sq}}{T_p}\right|_{\frac{\Omega_e}{\gamma_{ij}}\ll1}\hspace{-6mm}=(3 I_\text{sq}^2 + I_\text{sq}I_\text{vac})\sigma^{i}_\text{coh},
    \end{gather}
\end{subequations}
where we expect the second contribution in Eq. \eqref{eq:two-photon absorption narrow bandwidth limit} given by $I_\text{sq}I_\text{vac}\sigma^{i}_\text{coh}$ to be small. In the limit we are considering, $T_e$ is much longer than decay times of the molecule so we expect the contribution from photon pair absorption to be negligible.

In this limit we have
\begin{subequations}
    \begin{gather}
        \label{eq:narrow bandwidth ratios}
        \left.\frac{\mathcal{A}^1_\text{sq}}{\mathcal{A}^1_\text{coh}}\right|_{\frac{\Omega_e}{\gamma_{ij}}\ll1} = 1,\\
        \left.\frac{\mathcal{A}^i_\text{sq}}{\mathcal{A}^i_\text{coh}}\right|_{\frac{\Omega_e}{\gamma_{ij}}\ll1} = 3 + \frac{I_\text{vac}}{I_\text{sq}} = g^{(2)}(0,0),
    \end{gather}
\end{subequations}
where to make the comparison we took $I_\text{coh} = I_\text{sq}$; these results are in agreement with Raymer and Landes \cite{raymer2022theory}. We find the lowest order contribution is identical in this limit, which is expected since each photon from either coherent or squeezed light is at $\bar\omega$ and pair correlations are irrelevant. For each higher order term we find that for squeezed light there is a strong enhancement when $I_\text{vac}\gg I_\text{sq}$. Taking the model parameters from section \ref{Coherent Light: setting parameters} and setting $\Omega_e/2\pi = 10^{-4} \text{THz}\ll \gamma_{ij}$ ($T_e = 10^7 \text{fs}$), we calculate $I_\text{vac} = 105\text{W}/\text{m}^2$. Thus \emph{very} low intensities would be required to achieve a significant enhancement by using squeezed light rather than coherent light. 

Therefore, in this limit for reasonable intensity regimes where two-photon absorption experiments are done, $I_\text{vac}/I_\text{sq}\ll 1$ so squeezed states provide at most a factor of $3$ enhancement for each absorption term due to photon bunching. 

For a complete characterization of the absorption beyond the narrow bandwidth limit, which is valid in both the low and high intensity regimes, we resort to a numerical analysis. 

\subsubsection{Squeezed state absorption: numerical analysis}
We now calculate the $\sigma_\text{sq}^i$ cross-sections in Eq. \eqref{eq:squeezed state one-photon cross-section}, \eqref{eq:sq state cross-section same} and \eqref{eq:sq state cross-section diff} numerically using the decay parameters $\gamma_{ij}$ discussed in section \ref{Coherent Light: setting parameters} and set $\Omega_e/2\pi =10 \text{THz}$, which corresponds to the limit
\begin{equation}
    \gamma_{eg}\ll\gamma_{fe} = \gamma_{fg} < \Omega_e.
\end{equation}
We set $2\bar\omega=\omega_{fg}$, and look at the cross-sections as a function of detuning $\delta$ from resonance with the intermediate state (See Fig. \ref{fig:figures/Fig4_ipe.pdf}).

\begin{figure}[ht]
    \centering
    \includegraphics[width = \linewidth]{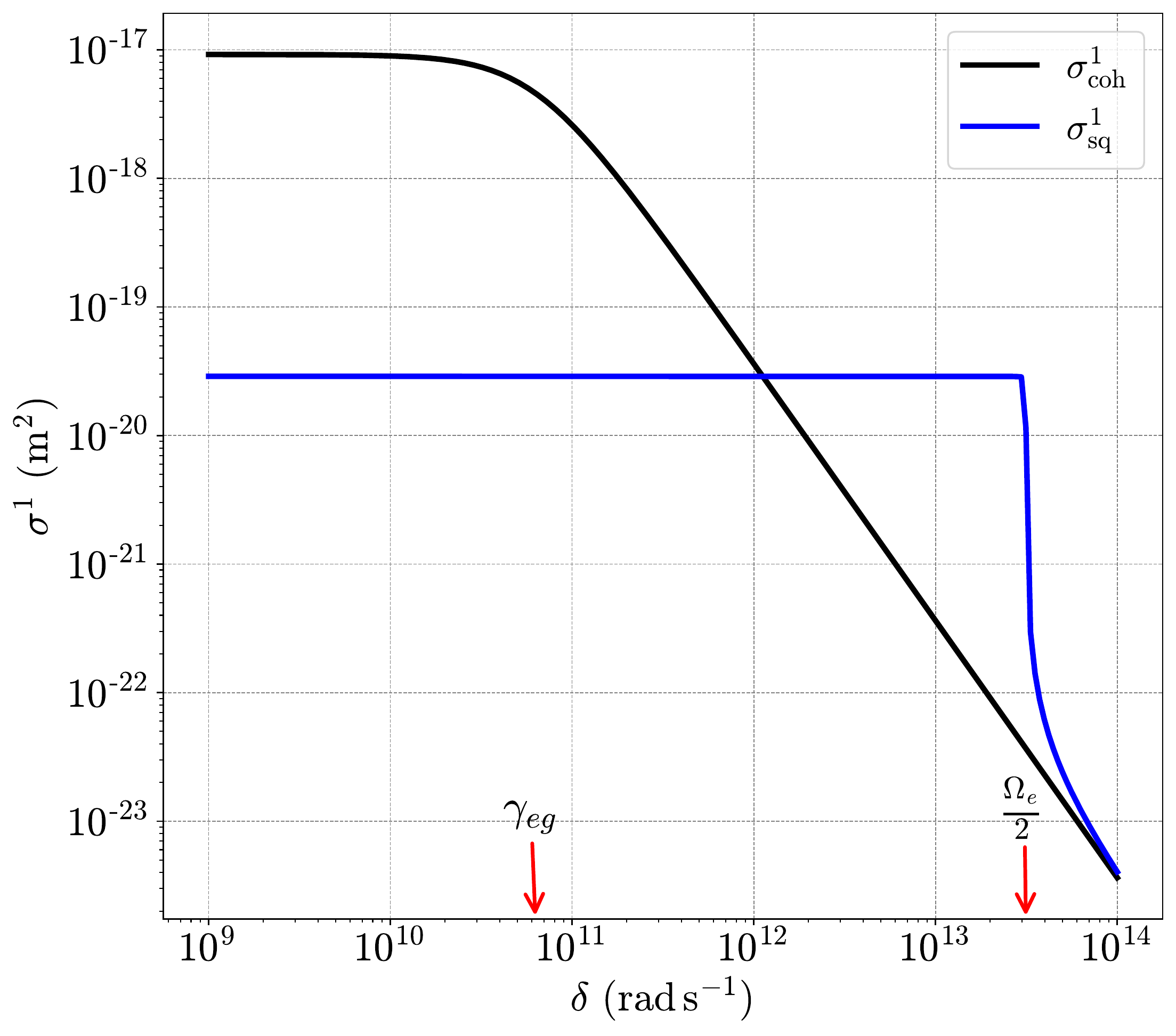}
    \caption{Plotting the $\sigma^1$ cross-sections for coherent and squeezed light vs detuning from resonance $\delta$.}
    \label{fig:CLEO-sigma1}
\end{figure}

We begin with the $\sigma^1$ cross-sections for coherent and squeezed light, plotted in Fig. \ref{fig:CLEO-sigma1}. The coherent light cross-section is fixed near resonance ($\delta< \gamma_{eg}$) and decreases for large detuning ($\delta >\gamma_{eg}$). Due to the large squeezed light bandwidth, the squeezed light cross-section is smaller but remains fixed for larger values of the detuning until the detuning is larger than half the squeezed state bandwidth ($\delta>\Omega_e/2$), at which point there is no resonant overlap. Thus for detunings  $\gamma_{eg}\ll\delta < \Omega_e/2$, there is approximately three orders of magnitude difference between the cross-section for squeezed light and that for coherent light, due simply to the large bandwidth of squeezed light. In the very far detuned limit ($\delta>\Omega_e/2\gg\gamma_{eg}$), $\delta$ is the largest quantity in the broadening function, so integrating over the squeezed light bandwidth is negligible compared to $
\delta$; therefore the squeezed light cross-section approaches the coherent limit.

\begin{figure}[ht]
    \centering
    \includegraphics[width = \linewidth]{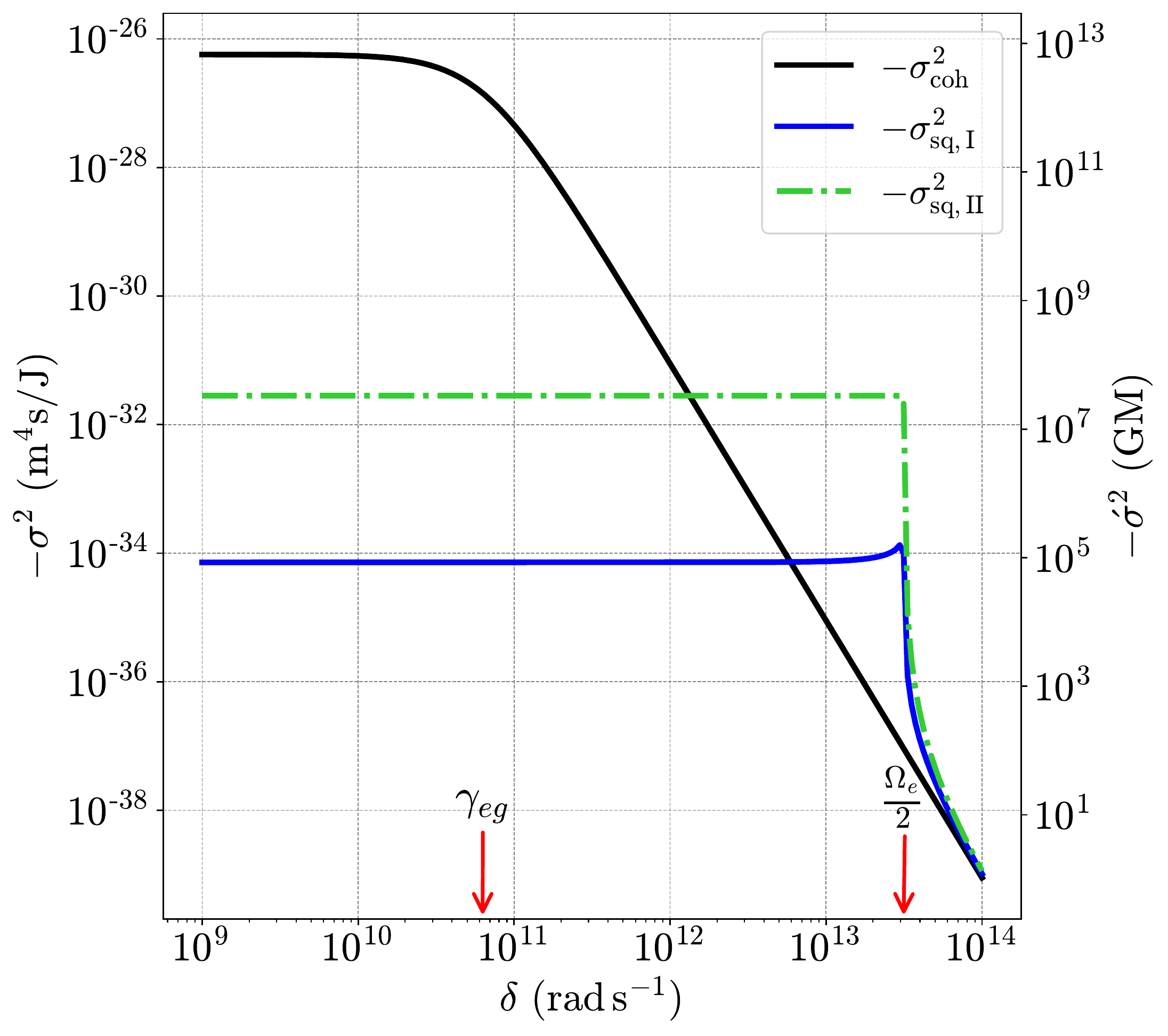}
    \caption{Plotting the $\sigma^2$ cross-sections for coherent and squeezed light vs detuning from resonance $\delta$.}
    \label{fig:CLEO-sigma2}
\end{figure}

In Fig. \ref{fig:CLEO-sigma2} we plot the negative of the coherent and squeezed light $\sigma^2$ cross-sections. The coherent light cross-section has the same behaviour as that of $\sigma^1_\text{coh}$, fixed near resonance and decreasing far from resonance. For the squeezed light cross-section we plot both $\sigma_\text{sq,\RN{1}}$ and $\sigma_\text{sq,\RN{2}}$; note that $\sigma^2_\text{sq,\RN{2}}\ge\sigma^2_\text{sq,\RN{1}}$. To see why, consider the broadening function involved in this cross-section,
\begin{equation}
    \mathcal{R}^2(\omega_1,\omega_2,\omega_3,\omega_4)\propto\frac{R_{ee}(\omega_2 - \omega_3)}{Q_{eg}(\omega_4)Q_{eg}^*(\omega_3)Q_{eg}(\omega_2)}.
\end{equation}
For anti-correlated photons such that $\omega_3+\omega_4 = 2\bar\omega$, this term is small because the two factors of the product $Q_{eg}(\omega_4)Q_{eg}^*(\omega_3)$ cannot be resonant at the same time for most frequencies within the bandwidth; the peak at $\delta \simeq \Omega_e/2$ is nontrivial, and will be discussed in greater detail below. However, if we consider photons such that $\omega_2 = \omega_3$ and $\omega_1 = \omega_4$ that product is doubly resonant, the broadening function simplifies to a squared Lorentzian, and so has the same behaviour as $\sigma^1_\text{sq}$. In both cases, when very far detuned ($\delta>\Omega_e/2$), $\delta$ is the largest quantity and each cross-section approaches the coherent light value. 

\begin{figure}[ht]
    \centering
    \includegraphics[width = \linewidth]{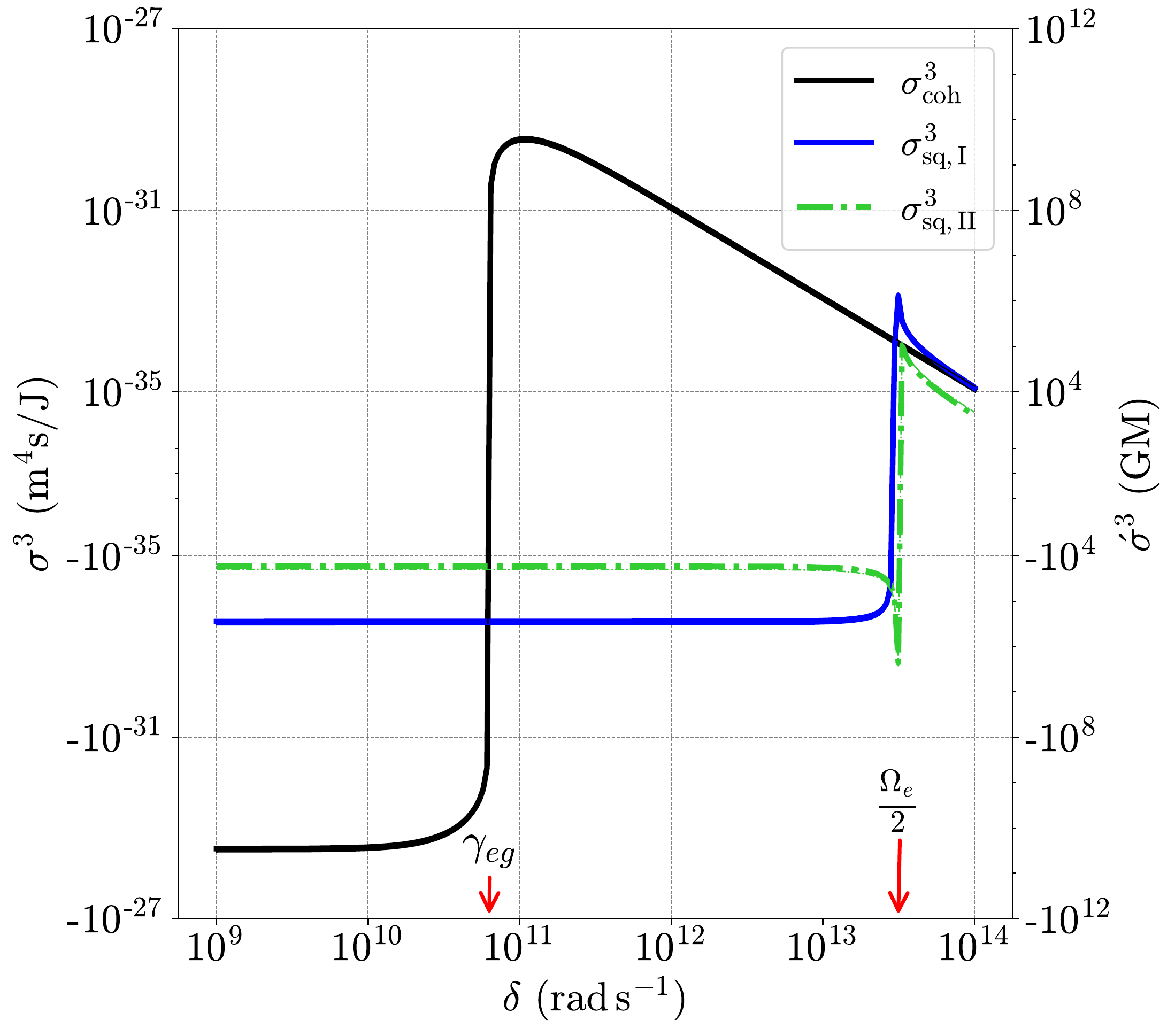}
    \caption{Plotting the $\sigma^3$ cross-section for coherent and squeezed light vs detuning from resonance $\delta$.}
    \label{fig:CLEO-sigma3}
\end{figure}

Next we consider the $\sigma^3$ cross-sections, which has a unique behaviour. When $\delta < \gamma_{eg}$ the coherent light cross-section is negative, identically zero when $\delta = \gamma_{eg}$, and positive but decreasing when $\delta> \gamma_{eg}$. The behaviour is not obvious from the diagram $\RN{3}$ in Fig. \ref{fig:exact diagrams split} and the perturbative expansion in Fig. \ref{fig:Fig7_ipe.pdf}.c). But by setting $\omega_{fg} = 2\bar\omega$ in Eq. \eqref{eq:coherent absorption terms in full A3} for $\mathcal{A}^3_\text{coh}$ one can analytically see this scaling with $\delta$. For the squeezed light cross-sections we see a similar behavior; however, the turning point from negative to positive is shifted to $\delta \simeq \Omega_e/2$ due to the large bandwidth. In Fig. \ref{fig:CLEO-sigma3} we show the different $\sigma^3$ cross-sections for the values of $\delta$ for which they are positive and negative. In the large detuned limit ($\delta > \Omega_e/2$) we find $\sigma^3_\text{sq,\RN{1}}>\sigma^3_\text{sq,\RN{2}}$ because the bandwidth is larger than the virtual two-photon absorption pathway linewidth, i.e. $\Omega_e>\gamma_{fg}$. Absorption from anti-correlated photons is therefore always resonant with $\omega_{fg}$, while for uncorrelated photons with frequency $\omega_1$ and $\omega_2$ there is wasted energy along the endpoints of the squeezed light bandwidth. Since $\sigma^3_\text{sq,\RN{1}}$ is always resonant with the virtual two-photon absorption it approaches the coherent light limit for large detuning ($\delta>\Omega_e/2$).

\begin{figure}[ht]
    \centering
    \includegraphics[width = \linewidth]{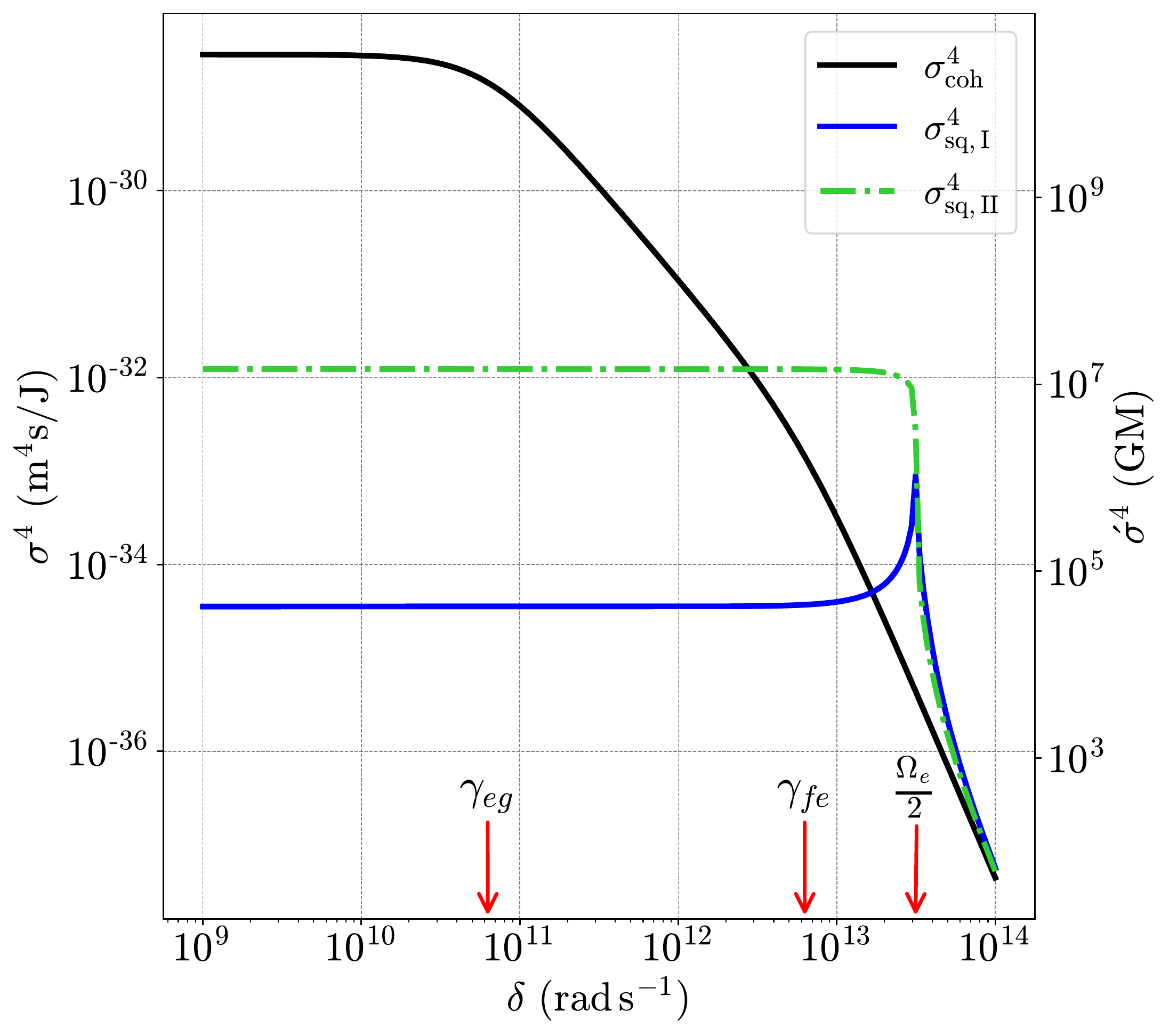}
    \caption{Plotting the $\sigma^4$ cross-sections for coherent and squeezed light vs detuning from resonance $\delta$.}
    \label{fig:CLEO-sigma4}
\end{figure}

In Fig. \ref{fig:CLEO-sigma4} we plot the $\sigma^4$ cross-sections. The coherent light cross-section is fixed near resonance and then decreases far from resonance with two different scalings. When $\delta < \gamma_{eg}$ it scales as $\delta^{-2}$, but as $\delta$ increases so that $\delta > \gamma_{fe}$ the cross-section scales as $\delta^{-4}$. The squeezed light cross-sections satisfy $\sigma^4_\text{sq,\RN{2}}\ge\sigma^4_\text{sq,\RN{1}}$, and again this can be understood from the behaviour of the relevant broadening function
\begin{equation}
    \mathcal{R}^4(\omega_1,\omega_2,\omega_3,\omega_4) \propto \frac{R_{ee}(\omega_2 - \omega_3) }{Q_{fe}(\omega_4)Q_{eg}^*(\omega_3)Q_{eg}(\omega_2)}.
\end{equation}
When the photons frequencies are given by $\omega_1 = \omega_4$ and $\omega_3 = \omega_2$, then both factors in the product $Q_{eg}^*(\omega_3)Q_{eg}(\omega_2)$ become small at the same time. This is the leading contribution to the squeezed state cross-section when $\delta < \Omega_e/2$. However, as $\delta$ approaches $\Omega_e/2$, half the squeezed light bandwidth, $\sigma^4_\text{sq,\RN{1}}$ quickly increases to its maximum value. This is a nontrivial scaling with the detuning which we discuss in detail below. In the very far detuned limit ($\delta > \Omega_e/2$), $\delta$ is the leading quantity and each cross-section approaches the coherent light cross-section. Similar to the $\sigma^1$ and $\sigma^2$ cross-sections, $\sigma^4_\text{sq,\RN{2}}$ is approximately three orders of magnitude larger than the coherent light cross-section due to the large squeezed light bandwidth. 

\begin{figure}[ht]
    \centering
    \includegraphics[width = \linewidth]{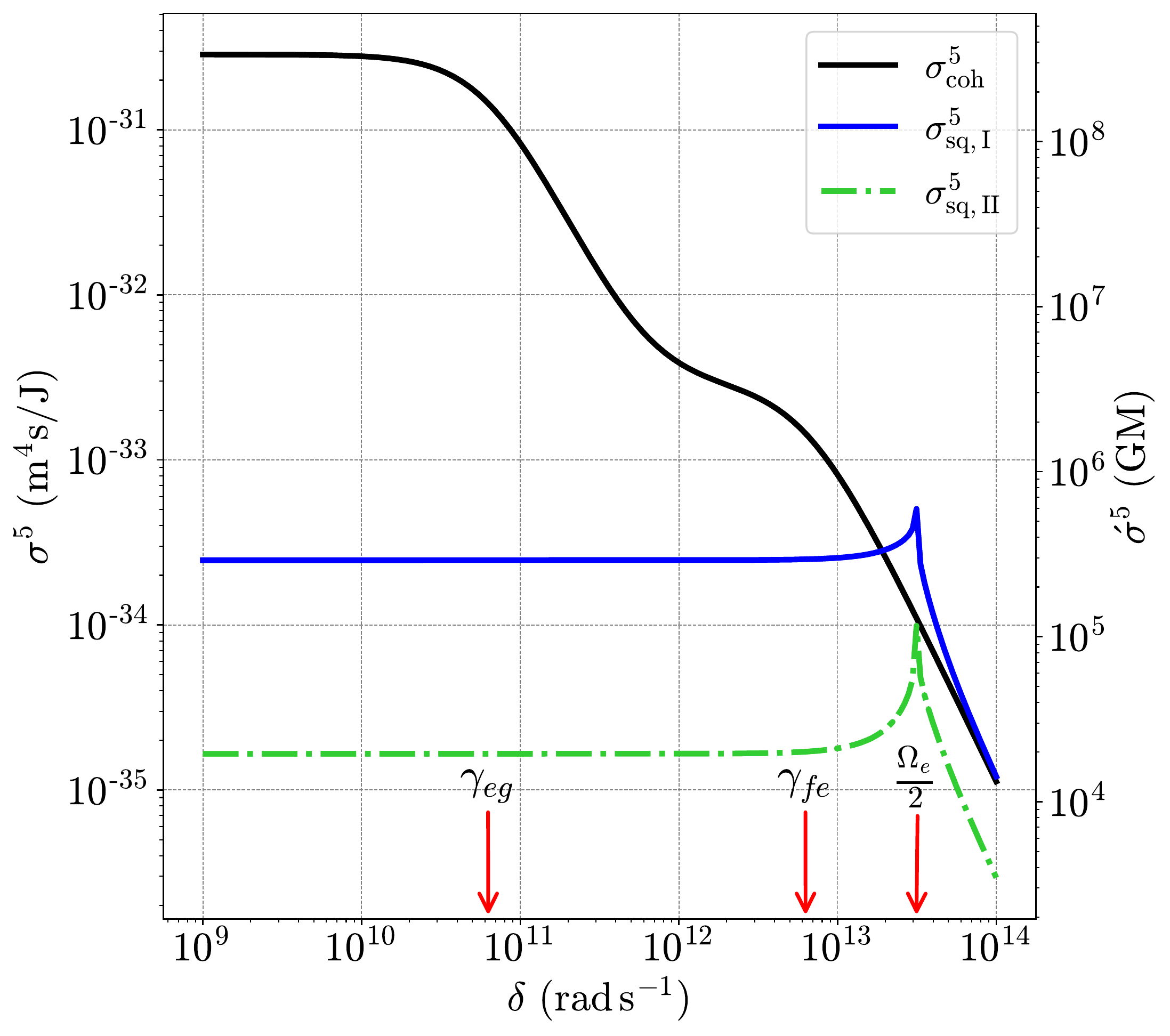}
    \caption{Plotting the $\sigma^5$ cross-sections for coherent and squeezed light vs detuning from resonance $\delta$.}
    \label{fig:CLEO-sigma5}
\end{figure}

Finally we consider the $\sigma^5$ cross-sections, which is plotted in Fig. \ref{fig:CLEO-sigma5}. The coherent light cross-section also has two behaviours between $\delta<\gamma_{eg}$ and $\delta>\gamma_{fe}$. However unlike $\sigma^4_\text{coh}$, they both scale as $\delta^{-2}$ because $\sigma^5_\text{coh}$ has a $\delta^2$ dependence in the numerator, which can be seen from Eq. \eqref{eq:coherent absorption terms in full A5}. We find $\sigma^5_\text{sq,\RN{1}}>\sigma^5_\text{sq,\RN{2}}$ since the $\sigma^5$ cross-sections corresponds to a virtual two-photon transition and $\Omega_e>\gamma_{fg}$, similar to the $\sigma^3$ cross-sections. In the large detuned limit the anti-correlated cross-section approaches the coherent limit and we find a similar cusp behaviour as in $\sigma^2_\text{sq,\RN{2}},\sigma^3_\text{sq,\RN{1}},\sigma^3_\text{sq,\RN{2}}$ and $\sigma^4_\text{sq,\RN{1}}$ which we now discuss. 

In each squeezed light cross-section there is a ``cusp'' behaviour located at $\delta \simeq \Omega_e/2$. This is due to an asymmetric scaling with frequency, and is $\delta$ dependent. Although we could choose any broadening function to illustrate this behavior we consider the broadening function $\mathcal{R}^5(\omega_1,\omega_2,\omega_3,\omega_4)$, which is the simplest mathematically. Consider $\mathcal{R}^5(\omega_1,\omega_2,\omega_3,\omega_4)$  evaluated at the anti-correlated frequencies, given by
\begin{equation}
    \begin{split}
        &\mathcal{R}^5(\omega_1,2\bar\omega-\omega_1,\omega_2,2\bar\omega-\omega_2) \propto \\
        &\hspace{30mm}\frac{-1}{Q_{fe}(2\bar\omega-\omega_2)Q_{fg}(2\bar\omega)Q_{eg}(2\bar\omega-\omega_1)},
    \end{split}
\end{equation}
which is resonant with the virtual two-photon transition as expected. Taking the imaginary part and changing variables $\sigma^5_\text{sq,\RN{1}}$ is proportional to
\begin{equation}
    \sigma^5_\text{sq,\RN{1}}\propto\frac{1}{\gamma_{fg}}\int\limits_{\delta -\frac{\Omega_e}{2}}^{\delta+ \frac{\Omega_e}{2}}\hspace{-1mm}\frac{d\omega_1 d\omega_2}{\Omega_e^2} \frac{\omega_1\omega_2+\gamma_{fe}\gamma_{eg}}{[\omega_1^2+\gamma_{fe}^2][\omega_2^2+\gamma_{eg}^2]}.
\end{equation}
Due to the behaviour of the first term there is complete cancellation when $\delta = 0$ and the contribution increases to its maximum when $\delta = \Omega_e/2$, where the range of integration is over a strictly positive function. In this case the integral is simple enough to be worked out analytically and the asymmetric dependence on frequencies in the numerator is responsible for the quick increase in cross-section values as $\delta$ approaches $\Omega_e/2$. For each resonant denominator we must evaluate it at the specified frequencies and then take the imaginary part. Taking the imaginary part leads to frequency dependent terms in the numerator which are asymmetric and lead to the resulting ``cusp'' behavior. 

\begin{figure}[ht]
    \centering
    \includegraphics[width = \linewidth]{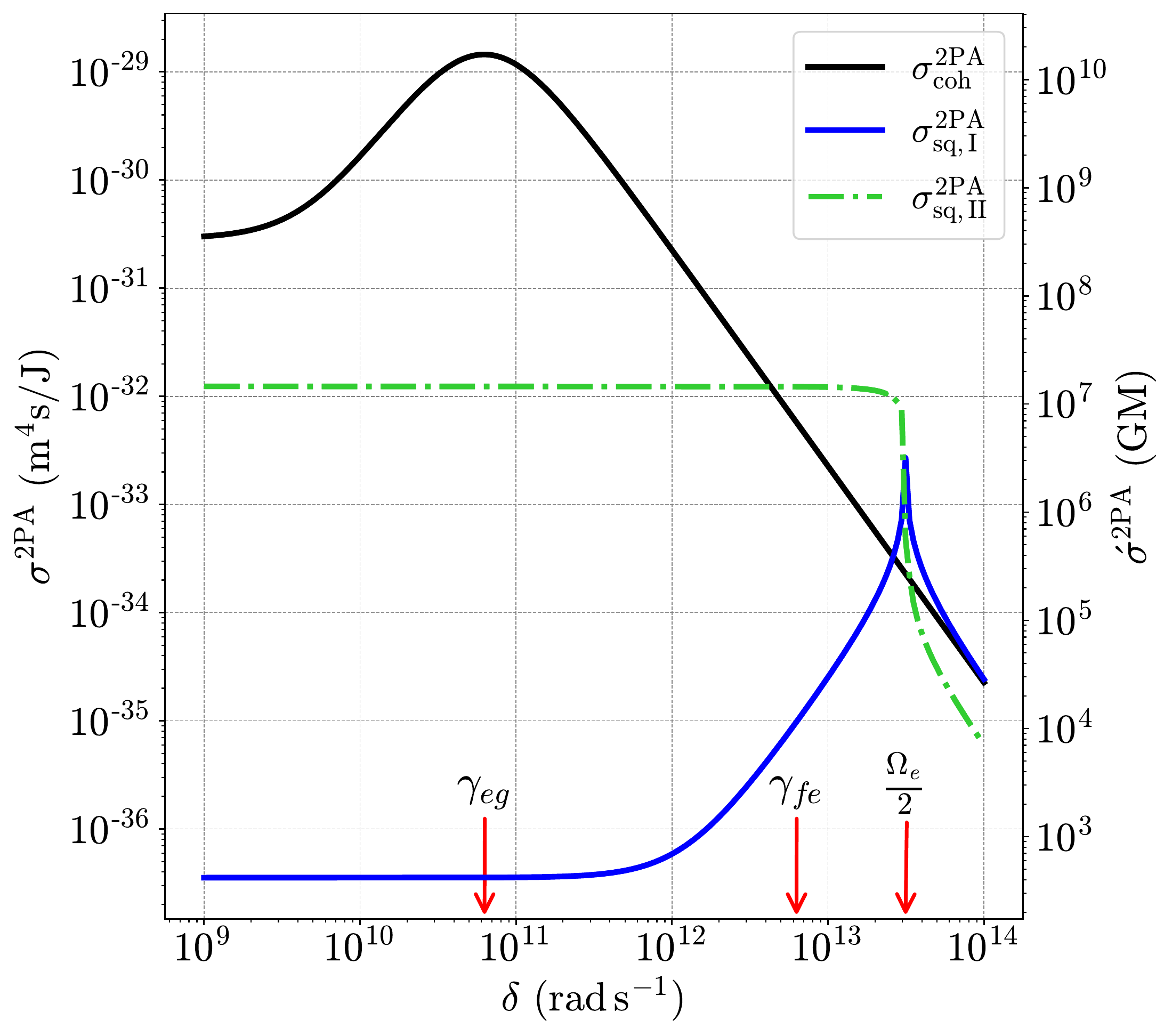}
    \caption{Plotting the $\sigma^\text{2PA}$ cross-sections for coherent and squeezed light vs detuning from resonance $\delta$.}
    \label{fig:CLEO-sigma2PA}
\end{figure}

Following the discussion in Sec. \eqref{sec:Coherent Light: plotting the absorption} we define the total squeezed state two-photon absorption cross-sections by
\begin{subequations}
    \begin{gather}
        \sigma^\text{2PA}_\text{sq,\RN{1}} = \sigma^3_\text{sq,\RN{1}}+\sigma^4_\text{sq,\RN{1}}+\sigma^5_\text{sq,\RN{1}},\\
        \sigma^\text{2PA}_\text{sq,\RN{2}} = \sigma^3_\text{sq,\RN{2}}+\sigma^4_\text{sq,\RN{2}}+\sigma^5_\text{sq,\RN{2}},
    \end{gather}
\end{subequations}
which are plotted in Fig. \ref{fig:CLEO-sigma2PA} and compared to the total coherent light two-photon absorption cross-section, given by
\begin{equation}
    \sigma^\text{2PA}_\text{coh} = \sigma^3_\text{coh}+\sigma^4_\text{coh}+\sigma^5_\text{coh},
\end{equation}
where the $\sigma^3$ cross-sections include both positive and negative contributions (see discussion at the end of section \ref{sec:Coherent Light: plotting the absorption}). 

The cross-sections in Fig. \ref{fig:CLEO-sigma2PA} are the sums of the contributions already discussed above. The coherent light two-photon absorption cross-section is fixed for $\delta\ll\gamma_{eg}$ and then increases to its maximum value at $\delta = \gamma_{eg}$; then it begins to decrease with a scaling of $\delta^{-2}$. The increase in the total coherent cross-section until its maximum at $\delta = \gamma_{eg}$ is a direct result of the coherent light cross-section $\sigma^3_\text{coh}$ being negative when $\delta<\gamma_{eg}$. When $\delta<\Omega_e/2$  $\sigma^\text{2PA}_\text{sq,\RN{2}}>\sigma^\text{2PA}_\text{sq,\RN{1}}$; which is a direct result of the large resonant enhancement due to the large bandwidth of the squeezed light. However, as soon as $\delta$ becomes larger than $\Omega_e/2$ all resonant enhancement is lost and $\sigma^\text{2PA}_\text{sq,\RN{1}}>\sigma^\text{2PA}_\text{sq,\RN{2}}$ because $\gamma_{fg}<\Omega_e$, and there is wasted energy at the endpoints of the bandwidth. In the limit where $\gamma_{fg}\ll\Omega_e$ this decrease would be even larger and we would find $\sigma^\text{2PA}_\text{sq,\RN{2}}$ to be negligible, in agreement with Raymer and Landes \cite{raymer2022theory}.

\begin{figure}
    \centering
    \includegraphics[width = \linewidth]{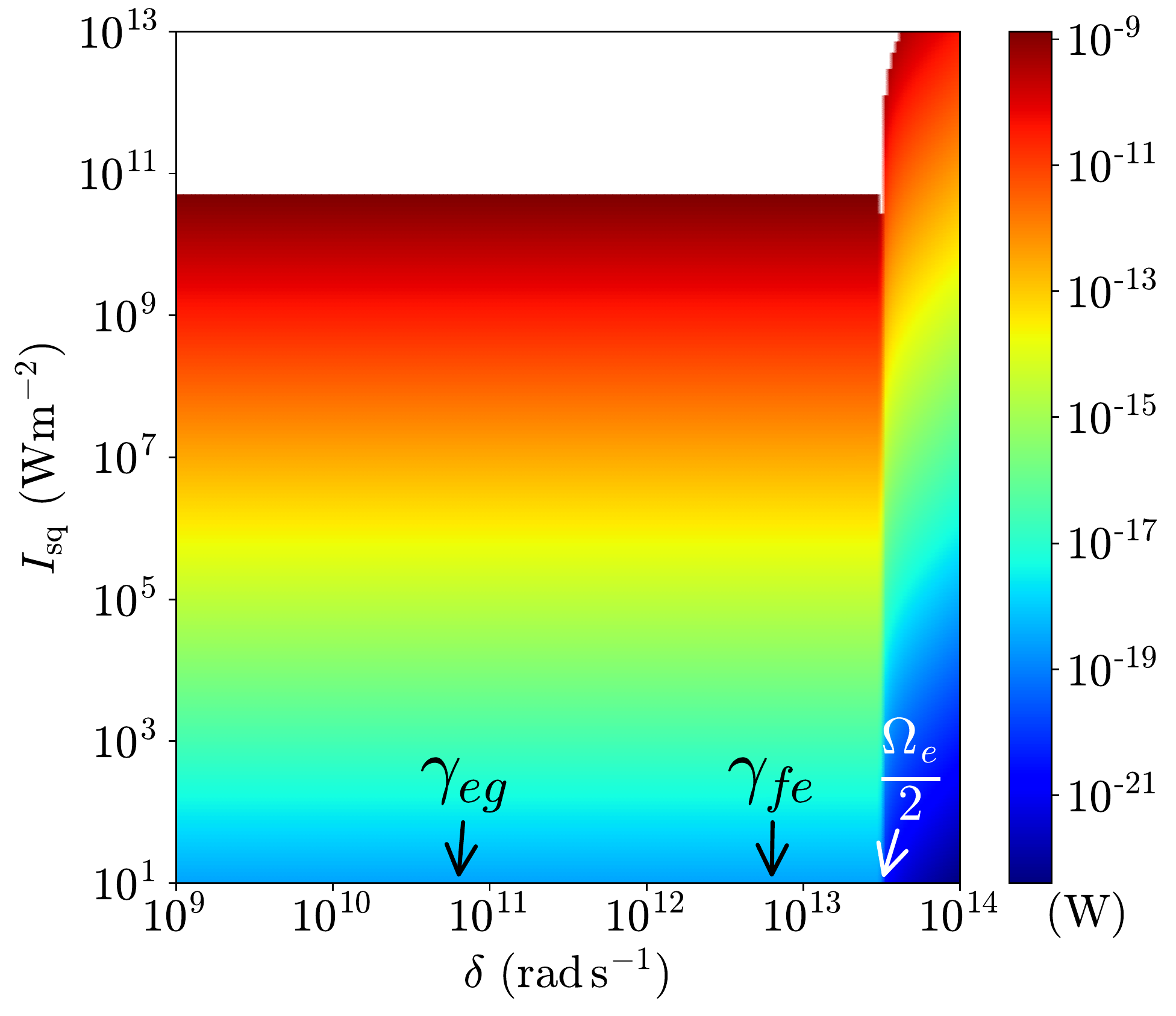}
    \caption{Plot of the rate of one-photon absorption for squeezed light ($\mathcal{O}_\text{sq}/T_p$) against the detuning from resonance ($\delta$) and intensity $I_\text{sq}$. The ``white'' area represents the parameter space where the perturbation theory is not valid.}
    \label{fig:Osq.pdf}
\end{figure}

\begin{figure}
    \centering
    \includegraphics[width = \linewidth]{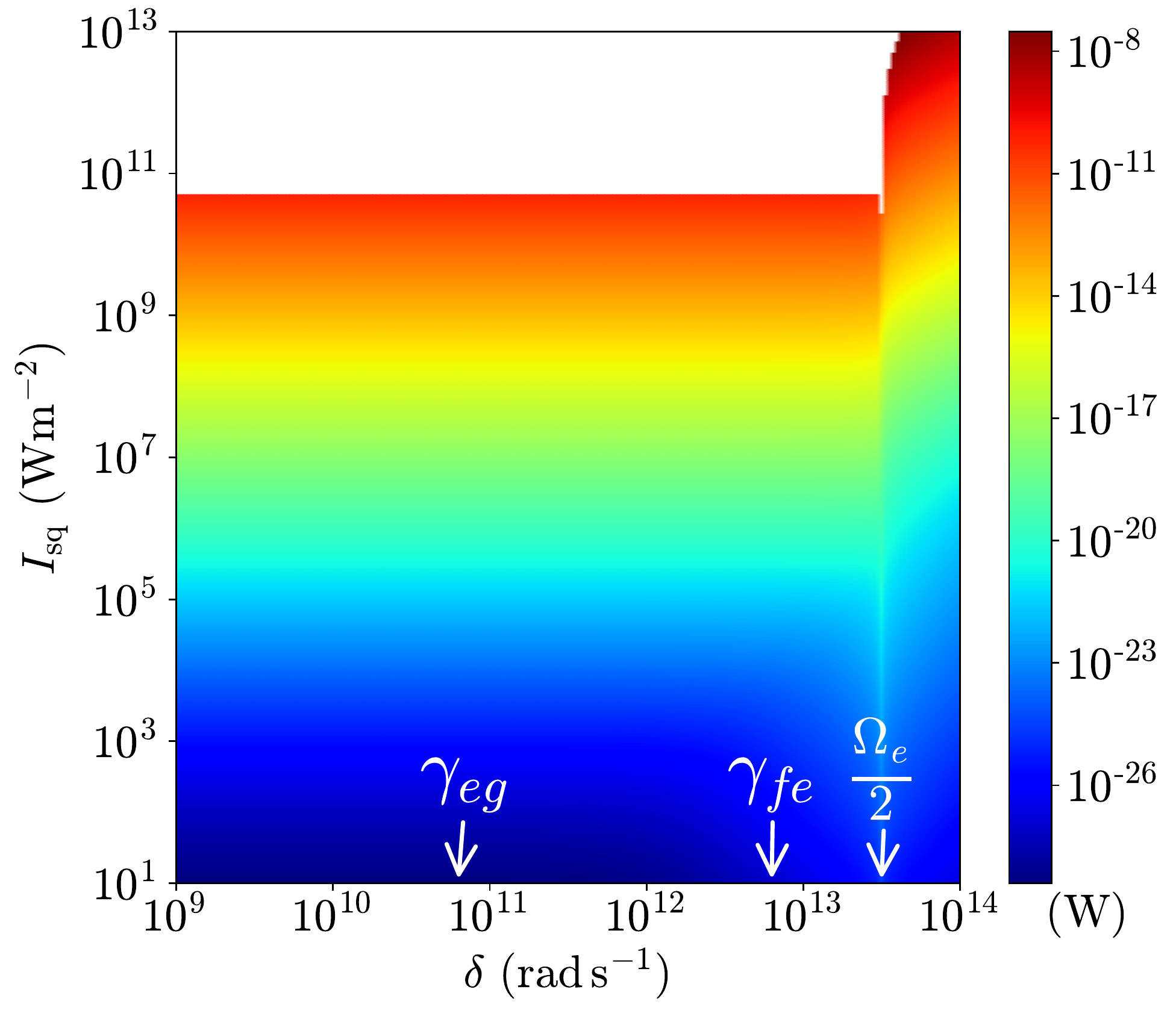}
    \caption{Plot of the rate of two-photon absorption for squeezed light ($\mathcal{T}_\text{sq}/T_p$) against the detuning from resonance ($\delta$) and intensity $I_\text{sq}$. The ``white'' area represents the parameter space where the perturbation theory is not valid.}
    \label{fig:Tsq.pdf}
\end{figure}

The largest enhancement to the squeezed state two-photon absorption cross-section is when $\delta \simeq  \Omega_e/2$; however, the one-photon absorption squeezed light cross-section (Fig. \ref{fig:CLEO-sigma1}) is also significantly enhanced when $\delta \simeq \Omega_e/2$, which - depending on the intensities - may dominate the absorption. Then we identify the squeezed light rate of one- and two-photon absorption by
\begin{subequations}
    \begin{gather}
        \frac{\mathcal{O}_\text{sq}}{T_p} = I_\text{sq}\sigma^1_\text{sq} +  I_\text{sq}(I_\text{vac} + I_\text{sq})\sigma^2_\text{sq,\RN{1}} + 2I_\text{sq}^2\sigma^2_\text{sq,\RN{2}},\\
        \frac{\mathcal{T}_\text{sq}}{T_p} =I_\text{sq}(I_\text{vac} + I_\text{sq})\sigma^\text{2PA}_\text{sq,\RN{1}} + 2I_\text{sq}^2\sigma^\text{2PA}_\text{sq,\RN{2}},
    \end{gather}
\end{subequations}
and plot the results in Fig. \ref{fig:Osq.pdf} and \ref{fig:Tsq.pdf} respectively. 
For both the one- and two-photon absorption from Fig. \ref{fig:Osq.pdf} and \ref{fig:Tsq.pdf}, we find the valid parameter space has increased due to the lowest order squeezed light contribution being smaller near resonance. For the two-photon absorption to dominate we must consider very far detunings where there is no resonant overlap with the large squeezed light bandwidth ($\delta > \Omega_e/2$) and large intensities. In this limit from each cross-section plot we find 
\begin{subequations}
    \begin{gather}
         \left.\sigma_\text{sq}^1 \right\vert_{\delta >\Omega_e/2}=\sigma_\text{coh}^1,\\
        \left.\sigma^{i}_\text{sq,\RN{1}}\right|_{\delta >\Omega_e/2} =\sigma_\text{coh}^i,\\
        \left.\sigma^{i}_\text{sq,\RN{2}}\right|_{\delta >\Omega_e/2} \lesssim\sigma_\text{coh}^i,
    \end{gather}
\end{subequations}
leading to no enhancement from the squeezed light cross-sections. And in this limit, to good approximation the one- and two-photon absorption simplify to
    \begin{subequations}
    \begin{gather}
        \left.\frac{\mathcal{O}_\text{sq}}{T_p}\right|_{\delta>\Omega_e/2} \simeq I_\text{sq}\sigma^1_\text{coh},\\
        \left.\frac{\mathcal{T}_\text{sq}}{T_p} \right|_{\delta>\Omega_e/2}\simeq  (3I_\text{sq}^2 + I_\text{sq}I_\text{vac})\sigma^\text{2PA}_\text{coh}.
    \end{gather}
\end{subequations}
Comparing to a coherent light we find
\begin{subequations}
    \begin{gather}
        \left.\frac{\mathcal{O}_\text{sq}}{\mathcal{O}_\text{coh}}\right|_{\delta>\Omega_e/2} = 1\\
        \left.\frac{\mathcal{T}_\text{sq}}{\mathcal{T}_\text{coh}}\right|_{\delta>\Omega_e/2} \simeq  3 + \frac{I_\text{vac}}{I_\text{sq}} = g^{(2)}(0,0).
    \end{gather}
\end{subequations}
In the very far detuned limit ($\delta > \Omega_e/2$), the ratio of one- and two-photon absorption is approximately the same to the ratio of each absorption term in Eq. \eqref{eq:narrow bandwidth ratios} and has the same interpretation. The factor of three we associate with photon bunching; and a large enhancement is possible when $I_\text{vac}\gg I_\text{sq}$. Had we taken a smaller $\gamma_{fg}$ such that $\gamma_{fg}\ll\Omega_e$ then $\sigma^\text{2PA}_\text{sq,\RN{1}}\gg\sigma^\text{2PA}_\text{sq,\RN{2}}$ and the ratio of absorption would be proportional to $1+I_\text{vac}/I_\text{sq}$ in agreement with Raymer and Landes \cite{raymer2022theory}.

\begin{figure}
    \centering
    \includegraphics[width = \linewidth]{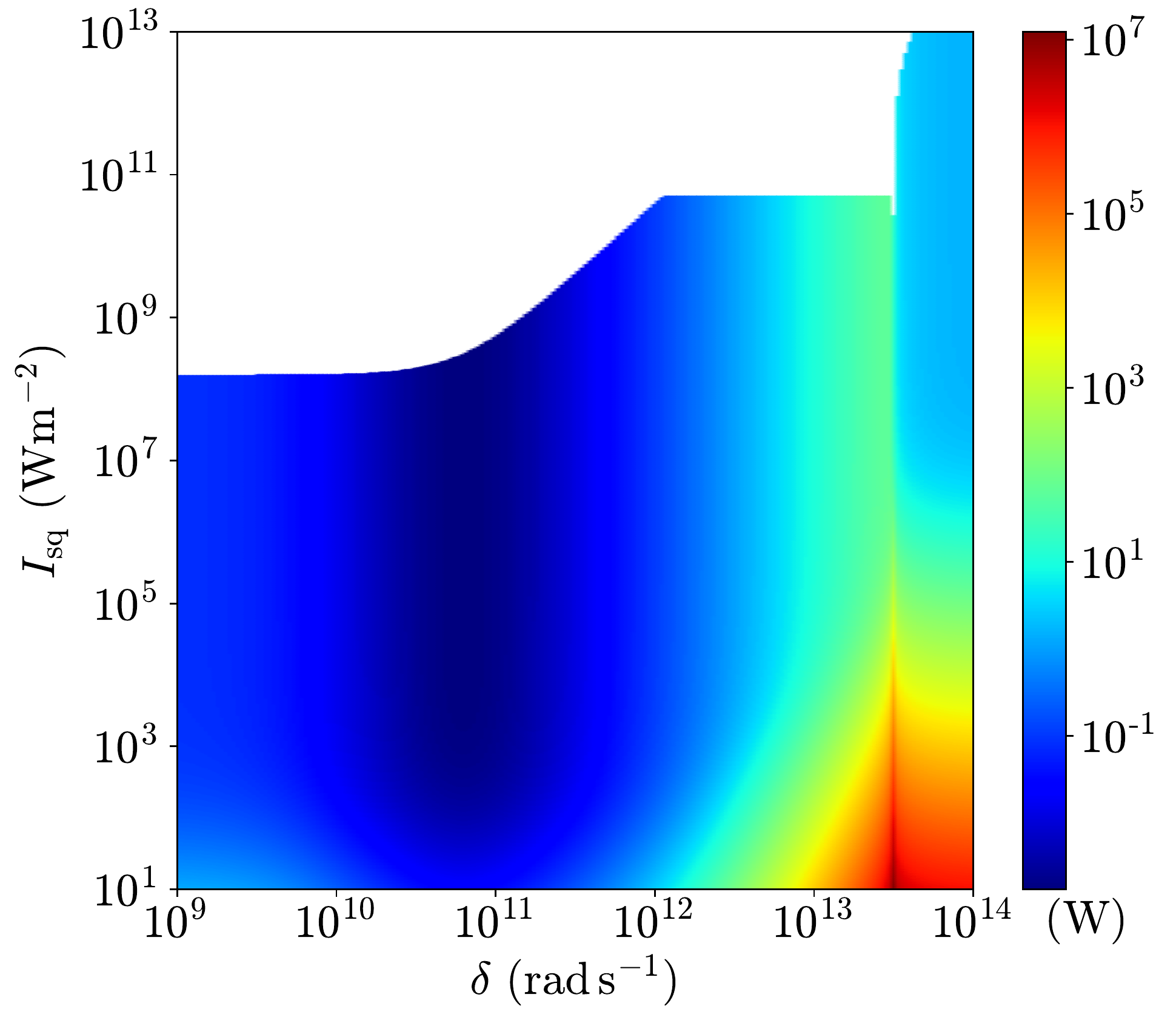}
    \caption{Plot of the ratio of two-photon absorption for squeezed and coherent light ($\mathcal{T}_\text{sq}/\mathcal{T}_\text{coh}$) detailed degree of squeezed light enhancement. The ``white'' are represents the parameter space where both coherent and squeezed light calculations are valid.}
    \label{fig:figures/R2PA.pdf}
\end{figure}

Using the parameters in Sec. \ref{Coherent Light: setting parameters} and $T_e = 100\text{fs}$, we find $I_\text{vac}\sim 10^7\text{W}/\text{m}^2$. In Fig. \ref{fig:figures/R2PA.pdf} we plot the ratio of squeezed to coherent light two-photon absorption in the region where perturbation theory is valid for both calculations. We see a large enhancement of approximately three orders of magnitude for $\delta \sim 2\times10^{13}\text{rad/s}$ due the resonant overlap with intermediate states. As the intensity decreases to $I_\text{sq}\approx I_\text{vac}$ near $\delta\sim 10^{14}\text{rad/s}$ the usual photon pair correlation enhancement begins to take effect and the enhancement near $I_\text{sq}\sim 10^3\text{W}/\text{m}^2$ corresponds to $I_\text{vac}/I_\text{sq}\sim 10^4$. Finally, when $\delta \sim \Omega_e/2$ and $I_\text{sq}\sim 10^3\text{W}/\text{m}^2$ we find the largest enhancement due to the combination of the correlations due to $g^{(2)}(0,0)$ at low intensities as well as a resonant enhancement.  

While we do witness up to seven orders of magnitude enhancement of the two-photon absorption for the set of parameters we are using, looking at Fig. \ref{fig:Osq.pdf} and \ref{fig:Tsq.pdf} in the regions of parameter space where squeezed light two-photon absorption is larger than that of coherent light, we find that the two-photon absorption is no longer dominant and it is the one-photon absorption that is larger. Therefore, for the parameters and detunings we are considering, any enhancement of two-photon absorption due to squeezed light is irrelevant because the one-photon absorption is either also enhanced or just larger at low intensities. 

Were $\gamma_{eg}$ chosen much smaller, so that the two-photon absorption always dominated even at low intensities in the far-detuned limit, then the condition for squeezed light enhancement would still be $I_\text{sq}\ll 10^7\text{W}/\text{m}^2$ or equivalently $AF_\text{sq}  \ll 1/T_e= 10^{13}\text{photons}/\text{s}$. For example, CW coherent light with $AF_\text{coh} = AF_\text{sq} = 10^{10} \text{photons}/\text{s}$ would lead to an enhancement of three orders of magnitude. However, it has been argued by Raymer et. al. that even at these intensities the rate of absorption \emph{may} be too small for current two-photon detection experiments \cite{raymer2021large,raymer2021entangled,raymer2020two} . 

Thus our results agree with past considerations \cite{raymer2021large,raymer2021entangled,raymer2020two,raymer2022theory} that for typical fluorescent experiments, either in the low or high intensity limit, CW (highly correlated) squeezed light does not provide a significant enhancement over classical light because for the parameters we chose, the one-photon absorption is the dominate process.

\section{Conclusion \label{sec:conculsion}}
In this article we built a model of a multi-level molecule with two sources of broadening and calculated the energy absorbed from an incident field that could be either pulsed or CW and near or far from resonance. Our results are closed form expressions for the cross-sections and the scaling of absorption with intensity for coherent (classical) and squeezed light.

Our analysis in this paper involved the absorption of coherent and squeezed light in the CW limit. We found nontrivial scalings with the detuning from resonance, where near resonance some terms lead to saturation of one-photon absorption, but far from resonance lead to two-photon absorption.

In the limit when the squeezed light bandwidth is very narrow compared to molecular broadening, each squeezed light cross-section approached that of coherent light, and the ratio of absorption scaled with the second order correlation function $g^{(2)}(0,0)$, in agreement with past calculations \cite{raymer2022theory}. In this limit, because there is minimal correlation \emph{very} low fluxes would be needed to enhance the absorption process above a factor of 3. 

We then considered the more general scenario where the squeezed light bandwidth could be very broad compared to the molecular broadening. This limit opens a new regime where, due to its large bandwidth, the squeezed light is able to contribute to the absorption through resonant contributions. This does lead to an approximately three orders magnitude increase in absorption for squeezed light; however, for the parameters we chose this also leads to a increase in the one-photon absorption, which overtakes the two-photon absorption. With different parameters in which the two-photon absorption is many orders of magnitude larger than the one-photon contribution, squeezed light resonant enhancements \emph{may} be useful. Similar to the coherent light cross-sections, the squeezed light cross-sections exhibit nontrivial scaling with the detuning from resonance and have a ``cusp'' behaviour when the detuning was equal to half the squeezed light bandwidth. At low intensities we find an appreciable enhancement over coherent light predicted in past treatments \textcolor{red}{\cite{dayan2007theory,raymer2022theory}}, as well as the combination of photon pair enhancement and resonant overlap; however, again for the parameters we chose, at such low intensities the one-photon absorption again dominates. In conclusion, for the parameters we chose we find a squeezed light enhancement of the two-photon absorption in the CW limit, but the one-photon absorption always dominates in the corresponding region of parameter space. Our results in the low intensity limit are in agreement with past calculations and the recent experiment by Tabakaev et al. \cite{tabakaev2022spatial} Here our main contribution is that at such low intensities linear processes such as absorption and scattering can be significant, even when expected to be small at large detunings, due to the large bandwidth of the incident squeezed light. Such linear processes may have contributed to measured signals in early photon-pair experiments, as was the case in the recent experiment by Hickam et al. \cite{hickam2022single}.

In an extension of this work, we will expand the model itself to include radiation reaction. Here we phenomenologically included scattering as decay to the reservoir, but in a more detailed treatment including radiation reaction we can more concretely categorize the absorbed as well as scattered (fluorescent) energy. 

The treatment used here models collisional dephasing with stochastic fluctuations of the molecules energy eigenstates in the ``impact limit'' leading to a Lorentzian lineshape. However, this is an approximation and over estimates the far detuned contributions and can be made more accurate by considering different models \cite{mukamel1990femtosecond,hedges1972extreme,raymer1979comparison,mukamel1999principles}. For example, collisional line broadening can be treated in general within this model by relaxing the condition that the correlation time $\tau_c$ is much shorter then the decay time of coherence, $\Lambda_{ij}\tau_c \ll 1$, which we assumed in Appendix \ref{appB:Stochastic average} \cite{mukamel1999principles,mukamel1983nonimpact}.

\begin{acknowledgments}
This work was supported by the Natural Sciences and
Engineering Research Council of Canada (NSERC).
C. D. acknowledges an Ontario Graduate Scholarship.
\end{acknowledgments}

\onecolumngrid
\newpage
\appendix

\section{Absorption in the RWA}
\label{app:sec:Absorption, Extinction and Scattering}
Consider a pulse of light incident on a molecule. To calculate the energy absorbed by the molecule from the electromagnetic field between the times $t_I$ and $t_F>t_I$, we consider the difference in energy of the electromagnetic between those times given by
\begin{equation}
\label{dE}
    \Delta E_{EM}=\bra{\Psi(t_F)}H_{EM}\ket{\Psi(t_F)}-\bra{\Psi(t_I)}H_{EM}\ket{\Psi(t_I)}
    =\int\limits_{t_I}^{t_F} \frac{d}{dt} \bra{\Psi(t)}H_{EM}\ket{\Psi(t)} dt.
\end{equation} 
The evolution of the ket $\ket{\Psi(t)}$ is given by $\ket{\Psi(t)} = \mathcal{U}(t,t_0)\ket{\Psi(t_0)}$, where the operator $\mathcal{U}(t,t_0)$ is unitary and is a solution of a Schr\"{o}dinger equation with the full Hamiltonian $H(t)$, and with $\mathcal{U}(t_0,t_0)$ equal to identity; $t_0<t_I$ is an initial time. For any Schr\"{o}dinger operator $O$ the time dependence of the corresponding Heisenberg picture operator is given by $O^H(t) = \mathcal{U}(t_0,t)O\mathcal{U}(t,t_0)$. Thus $\bra{\Psi(t)}H_{EM}\ket{\Psi(t)}=\bra{\Psi(t_0)}H^H_{EM}(t)\ket{\Psi(t_0)}$, for example, and using this in (\ref{dE}) together with the Heisenberg equation for the evolution of $H^H_{EM}(t)$ we can write 
\begin{equation}
\label{DeltaEEM}
    \Delta E_{EM} = \frac{1}{i\hbar}\int\limits_{t_I}^{t_F}dt\bra{\Psi(t_0)}[H^H_{EM}(t),H^H(t)]\ket{\Psi(t_0)}.
\end{equation}

We now evaluate the commutator $[H^H_{EM}(t),H^H(t)]$. To simplify the notation we evaluate the corresponding Schr\"{o}dinger picture commutator, $[H_{EM},H(t)]$, and then move to the Heisenberg picture. Since all field operators commute with all molecule and reservoir operators, the only contribution to that commutator is $[H_{EM},H(t)] = [H_{EM},H_{M-EM}]$. Writing the interaction Hamiltonian in completely symmetric form,
\begin{equation}
    H_{\text{M-EM}} = -\frac{1}{2}\left(\boldsymbol{\mu} \cdot \boldsymbol{E} + \boldsymbol{E} \cdot\boldsymbol{\mu}\right),
\end{equation}
and moving to the RWA, 
\begin{equation}
    H_{\text{M-EM}} = -\frac{1}{2}(\boldsymbol{\mu_+}\cdot \boldsymbol{E_-} + \boldsymbol{\mu_-}\cdot\boldsymbol{E_+} + \boldsymbol{E_-}\cdot \boldsymbol{\mu_+} + \boldsymbol{E_+}\cdot\boldsymbol{\mu_-}),
    \label{eq:H_M-EM v2}
\end{equation}
we have
\begin{equation}
\label{HEMHCommutator}
    [H_{EM},H(t)] = \frac{1}{2}(\boldsymbol{\mu_-}\cdot[\boldsymbol{E_+},H_{EM}] + \boldsymbol{\mu_+}\cdot[\boldsymbol{E_-},H_{EM}] + [\boldsymbol{E_-},H_{EM}]\cdot\boldsymbol{\mu_+} + [\boldsymbol{E_+},H_{EM}]\cdot\boldsymbol{\mu_-}).
\end{equation}
We note that at this point in the calculation each term on the right-hand side can be re-ordered since they act over different Hilbert spaces. Next we look at 
\begin{equation}
    \label{eq:commutator of E and H}
    [\boldsymbol{E_\pm},H(t)] = [\boldsymbol{E_\pm},H_{EM}] + [\boldsymbol{E_\pm},H_{M-EM}].
\end{equation}
Working out the second term on the right-hand side and rearranging the equation to solve for $[\boldsymbol{E_\pm},H_{EM}]$, we insert the results in (\ref{HEMHCommutator}) to find
\begin{equation}
    [H_{EM},H(t)] = \frac{1}{2}(\boldsymbol{\mu_-}\cdot[\boldsymbol{E_+},H(t)] + \boldsymbol{\mu_+}\cdot[\boldsymbol{E_-},H(t)] + [\boldsymbol{E_+},H(t)]\cdot\boldsymbol{\mu_-} + [\boldsymbol{E_-},H(t)]\cdot\boldsymbol{\mu_+}),
\end{equation}
where now the order of the two terms in each dot product on the right-hand side is fixed; those terms do not commute, since  $[\boldsymbol{E_\pm},H(t)]$ is a function of both molecule and field operators and in general no longer commutes with the $\boldsymbol{\mu_\pm}$ operators. Moving back into the Heisenberg picture and using the Heisenberg  equation of motion for $\boldsymbol{E_\pm}(t)$ we have
\begin{equation}
    [H^H_{EM}(t),H^H(t)] = \frac{i\hbar}{2}\left(\boldsymbol{\mu_-}^H(t)\cdot\frac{d\boldsymbol{E_+}^H(t)}{dt}+ \boldsymbol{\mu_+}^H(t)\cdot\frac{d\boldsymbol{E_-}^H(t)}{dt} + \frac{d\boldsymbol{E_+}^H(t)}{dt}\cdot\boldsymbol{\mu_-}^H(t) + \frac{d\boldsymbol{E_-}^H(t)}{dt}\cdot\boldsymbol{\mu_+}^H(t)\right).
\end{equation}
Using this in (\ref{DeltaEEM}) we find the change in total energy $\Delta E_{EM}$ in the electromagnetic field. The energy lost from the electromagnetic field is by definition the energy absorbed, $\mathcal{A}$, and so putting $\mathcal{A} = -\Delta E_{EM}$ we find
\begin{equation}
    \begin{split}
    \label{eq:absorption}
    \mathcal{A} &= \frac{1}{2}\int\limits_{t_I}^{t_F}dt\bra{\Psi(t_0)}\left(\frac{d\boldsymbol{\mu_-}^H(t)}{dt}\cdot\boldsymbol{E_+}^H(t)+ \frac{d\boldsymbol{\mu_+}^H(t)}{dt}\cdot\boldsymbol{E_-}^H(t)+ \boldsymbol{E_+}^H(t)\cdot\frac{d\boldsymbol{\mu_-}^H(t)}{dt} + \boldsymbol{E_-}^H(t)\cdot\frac{d\boldsymbol{\mu_+}^H(t)}{dt}\right) \ket{\Psi(t_0)}\\
    &+\left.\frac{1}{2}\bra{\Psi(t_0)}\left(\frac{d\boldsymbol{\mu_-}^H(t)}{dt}\cdot\boldsymbol{E_+}^H(t)+ \frac{d\boldsymbol{\mu_+}^H(t)}{dt}\cdot\boldsymbol{E_-}^H(t)+ \boldsymbol{E_+}^H(t)\cdot\frac{d\boldsymbol{\mu_-}^H(t)}{dt} + \boldsymbol{E_-}^H(t)\cdot\frac{d\boldsymbol{\mu_+}^H(t)}{dt}\right) \ket{\Psi(t_0)}\right\vert_{t_I}^{t_F}
    \end{split}
\end{equation}
where we have performed an integration by parts.

Up till now the equation we have derived is general and valid for any time $t_I$ and $t_F>t_I>t_0$. Now consider a time $t<t_\text{min}$ where $t_\text{min}$ is a very early time when the pulse is still far from the molecule, we take the initial ket of the system to be of the form $\ket{\Psi(t)}=\ket{g}_\text{M}\ket{\psi(t)}_{\text{EM}}\ket{\text{vac}}_{\text{R}}$, where $\ket{g}$ is the ground state of the molecule,  $\ket{\psi(t)}_{\text{EM}}$ is a state of the electromagnetic field, and $\ket{\text{vac}}_{\text{R}}$ is the ground state of the reservoir (the set of effective wave guides). Then for $t>t_\text{max}$ where $t_\text{max}$ is long enough after the pulse has interacted with the molecule that the molecule has returned to its ground state via interaction with the electromagnetic field and the reservoir, the full ket of the system will be of the general form $\ket{\Psi(t)}=\ket{g}_\text{M}\ket{\phi(t)}_{\text{EM-R}}$ where $\ket{\phi(t)}_{\text{EM-R}}$ is a ket in the product Hilbert space of the electromagnetic field and the reservoir. For the RWA form of the interaction Hamiltonian $H_\text{M-EM}$ and $H_\text{M-R}$ and for times $t<t_\text{min}$ and $t>t_\text{max}$ the expectation value of $H_\text{M-EM}$ and $H_\text{M-R}$ vanishes.

Since for times $t_a,t_b<t_\text{min}$ and $t_a,t_b>t_\text{max}$ any expectation value will not include a contribution from the interaction terms $H_\text{M-EM}$ and $H_\text{M-R}$, in the Heisenberg picture, this is equivalent to saying that the time evolution is given just by the free Hamiltonian $H_0$, i.e. $\mathcal{U}(t_b,t_a) = \mathcal{U}_0(t_b,t_a)$.

Moving back to the absorption in Eq. \eqref{eq:absorption}, we take the times $t_I<t_\text{min}$ and $t_F>t_\text{max}$ so that the ``boundary terms'' are zero because the time evolution is given by $H_0$ which as per the above discussion annihilates the initial ket $\ket{\Psi(t_0)}$ for $t_0<t_I<t_\text{min}$, then the absorption is given by
\begin{equation}
    \label{eq:absorption2}
    \mathcal{A} = \frac{1}{2}\int\limits_{t_I}^{t_F}dt\bra{\Psi(t_0)}\left(\frac{d\boldsymbol{\mu_-}^H(t)}{dt}\cdot\boldsymbol{E_+}^H(t)+ \frac{d\boldsymbol{\mu_+}^H(t)}{dt}\cdot\boldsymbol{E_-}^H(t)+ \boldsymbol{E_+}^H(t)\cdot\frac{d\boldsymbol{\mu_-}^H(t)}{dt} + \boldsymbol{E_-}^H(t)\cdot\frac{d\boldsymbol{\mu_+}^H(t)}{dt}\right) \ket{\Psi(t_0)}.
\end{equation}

Following the symmetric form of (\ref{eq:H_M-EM v2}), the result (\ref{eq:absorption2}) contains terms that are normally ordered and terms that are anti-normally ordered. We can write it as a sum of terms of definite order, $\mathcal{A} = \frac{1}{2}(\mathcal{A}^N + \mathcal{A}^{AN})$, where $\mathcal{A}^N$ and $\mathcal{A}^{AN}$ are respectively normally and anti-normally ordered, 
\begin{subequations}
\begin{gather}
    \mathcal{A}^N = \int\limits_{t_I}^{t_F}dt\bra{\Psi(t_0)}\left(\frac{d\boldsymbol{\mu_-}^H(t)}{dt}\cdot\boldsymbol{E_+}^H(t)+\boldsymbol{E_-}^H(t)\cdot\frac{d\boldsymbol{\mu_+}^H(t)}{dt}\right) \ket{\Psi(t_0)},\\
    \mathcal{A}^{AN} = \int\limits_{t_I}^{t_F}dt\bra{\Psi(t_0)}\left( \frac{d\boldsymbol{\mu_+}^H(t)}{dt}\cdot\boldsymbol{E_-}^H(t)+ \boldsymbol{E_+}^H(t)\cdot\frac{d\boldsymbol{\mu_-}^H(t)}{dt}\right) \ket{\Psi(t_0)}.
\end{gather}
\end{subequations}
Since the operators in each pair on the right-hand side of (\ref{eq:H_M-EM v2}) commute, that equation could in fact be written in normal or anti-normal order, and one might expect the same for (\ref{eq:absorption}). Indeed, we show below that $\mathcal{A} = \mathcal{A}^N  = \mathcal{A}^{AN}$, and thus any convenient ordering can be chosen for the absorption expression (\ref{eq:absorption}). 

We begin by considering the commutator
\begin{equation}
    \label{eq:commutator for N=AN}
    [\boldsymbol{E_+}^H(t)\cdot,\frac{d\boldsymbol{\mu_-}^H(t)}{dt}] \equiv \boldsymbol{E_+}^H(t)\cdot\frac{d\boldsymbol{\mu_-}^H(t)}{dt} - \frac{d\boldsymbol{\mu_-}^H(t)}{dt}\cdot\boldsymbol{E_+}^H(t)
\end{equation}
where we include the symbol `$\cdot$' inside the commutator to denote the dot product between vector operators. Using the Heisenberg equation of motion for the operator $\boldsymbol{\mu_+}^H(t)$ we expand the right-hand side of the commutator as 
\begin{equation}
    [\boldsymbol{E_+}^H(t)\cdot,\frac{d\boldsymbol{\mu_-}^H(t)}{dt}] = \frac{1}{i\hbar}[\boldsymbol{E_+}^H(t)\cdot,[\boldsymbol{\mu_-}^H(t),H^H(t)]].
\end{equation}
Applying the Jacobi identity for the commutator on the right-hand side we have
\begin{equation}
    [\boldsymbol{E_+}^H(t)\cdot,[\boldsymbol{\mu_-}^H(t),H^H(t)]] + [\boldsymbol{\mu_-}^H(t)\cdot,[H^H(t),\boldsymbol{E_+}^H(t)]] + [H^H(t),[\boldsymbol{E_+}^H(t)\cdot,\boldsymbol{\mu_-}^H(t)] = 0,
\end{equation}
the first term is the term we started with and the last term is zero because field and molecule operators commute. The commutator $[H^H(t),\boldsymbol{E_+}^H(t)]$ was evaluated in Eq. \eqref{eq:commutator of E and H} in the Schr\"{o}dinger picture. Moving to the Heisenberg picture, inputting the commutator, simplifying and putting it all together, the commutator in Eq. \eqref{eq:commutator for N=AN} is given by
\begin{equation}
    \label{eq:commutator of E_+ dmu_-/dt}
    [\boldsymbol{E_+}^H(t)\cdot,\frac{d\boldsymbol{\mu_-}^H(t)}{dt}]=\frac{i}{\hbar}[\boldsymbol{E_+}^H(t)\cdot,\boldsymbol{E_-}^H(t)][\boldsymbol{\mu_-}^H(t)\cdot,\boldsymbol{\mu_+}^H(t)].
\end{equation}

Now on the right-hand side we set 
\begin{equation}
    C(t) \equiv  [\boldsymbol{E_+}^H(t)\cdot,\boldsymbol{E_-}^H(t)][\boldsymbol{\mu_-}^H(t)\cdot,\boldsymbol{\mu_+}^H(t)]
\end{equation}
For each commutator appearing here we can always consider the Schr\"{o}dinger picture for its evaluation. First consider $[\boldsymbol{E_+}\cdot,\boldsymbol{E_-}]$, which at most is a complex number; but because $[\boldsymbol{E_+},\boldsymbol{E_-}]^\dagger = [\boldsymbol{E_+}\cdot,\boldsymbol{E_-}]$, the commutator $[\boldsymbol{E_+}\cdot,\boldsymbol{E_-}]$ must be a real number. The second commutator, $[\boldsymbol{\mu_-}\cdot,\boldsymbol{\mu_+}]$, is also Hermitian, and so $C(t)$ is therefore a Hermitian operator. Since these are Schr\"{o}dinger operators this is always true even after any approximations we make to the time evolution of the operators in the Heisenberg picture. Taking the Hermitian conjugate of the commutator in Eq. \eqref{eq:commutator of E_+ dmu_-/dt} and adding it to itself we expand the commutators and find
\begin{equation}
    \frac{d\boldsymbol{\mu_-}^H(t)}{dt}\cdot \boldsymbol{E_+}^H(t)+\boldsymbol{E_-}^H(t)\cdot\frac{d\boldsymbol{\mu_+}^H(t)}{dt} = \frac{d\boldsymbol{\mu_+}^H(t)}{dt}\cdot \boldsymbol{E_-}^H(t)+ \boldsymbol{E_+}^H(t)\cdot\frac{d\boldsymbol{\mu_-}^H(t)}{dt},
\end{equation}
where the non-commuting parts exactly cancel because $C(t)$ is Hermitian leading to the equality of the normally and anti-normally ordered contributions, and to  $\mathcal{A} = \mathcal{A}^N = \mathcal{A}^{AN}$. It will be computationally easier to work with the normally ordered form of the absorption which after dropping the superscript is given by
\begin{equation}
    \mathcal{A} = \int\limits_{t_I}^{t_F}dt\bra{\Psi(t_0)}\boldsymbol{E_-}^H(t)\cdot\frac{d\boldsymbol{\mu_+}^H(t)}{dt} \ket{\Psi(t_0)} + c.c.,
\end{equation}
and is manifestly real. 

\section{Perturbation Theory \label{appB:Perturbation theory}}
In this section we begin with the exact solution for each molecule operator and solve them perturbatively. We begin with the exact solution given by Eq. \eqref{eq:exact solution to sigma_ij} to each differential equation in Eq. \eqref{eq:equations of motion v2}, inputting the proper right-hand side we have
\begin{subequations}
\label{eq:app:exact soltuions}
\begin{gather}
    \bar\sigma_{ge}(t) =\hspace{-2mm} \int\limits_{-\infty}^t \hspace{-2mm}dt_1[G_{eg}(t,t_1)\bar\sigma_{gg}(t_1)\hat F_+^{eg}(t_1)+G_{ge}^*(t,t_1) \bar\sigma_{e^\prime e}(t_1)\hat F_+^{e^\prime g}(t_1) + G_{eg}(t,t_1)\hat F_-^{ef}(t_1)\bar\sigma_{gf}(t_1)],\\
    \bar{\sigma}_{ee^\prime}(t) {=} \hspace{-2mm}\int\limits_{-\infty}^t dt_1[G_{e^\prime e}(t,t_1)\bar{\sigma}_{eg}(t_1)\hat F_+^{e^\prime g}(t_1) 
    {+}G^*_{ee^\prime }(t,t_1)\hat F_-^{ge}(t_1)\bar{\sigma}_{ge^\prime}(t_1) 
    {+}G^*_{ee^\prime }(t,t_1)\bar{\sigma}_{fe^\prime}(t_1)\hat F_+^{fe}(t_1){+}G_{e^\prime e}(t,t_1)\hat F_-^{e^\prime f}(t_1)\bar{\sigma}_{e f}(t_1)]\nonumber\\
    {-}2i\Gamma_{fe}\int\limits_{-\infty}^t dt_1G_{e^\prime e}(t,t_1)\bar{\sigma}_{ff}(t_1)\delta_{ee^\prime}],\\
    \bar\sigma_{ef}(t) = \hspace{-2mm}
    \int\limits_{-\infty}^t \hspace{-2mm}dt_1[G_{f e}(t,t_1)\bar\sigma_{ee^\prime}(t_1)\hat F_+^{fe^\prime}(t_1)+G^*_{ef}(t,t_1) \hat F_-^{ge}(t_1)\bar\sigma_{gf}(t_1)+G^*_{ef}(t,t_1) \bar\sigma_{f^\prime f}(t_1)\hat F_+^{f^\prime e}(t_1)],\\
    \bar{\sigma}_{gf}(t) 
    {=} \hspace{-2mm}\int\limits_{-\infty}^t dt_1 [G_{f g}(t,t_1)\bar{\sigma}_{ge}(t_1) \hat F_+^{fe}(t_1) {+}G^*_{gf}(t,t_1) \bar{\sigma}_{ef}(t_1)\hat F_+^{eg}(t_1)],
\end{gather}
\end{subequations}
where $G_{ij}(t,t_1)$ is the Green function for each equation and is defined in Eq. \eqref{eq:Green's function}. In section \ref{sec:perturbative solution} we argued that the only nonzero term at zeroth order is $\bar{\sigma}_{gg}^{(0)}(t) = \hat{1}$, which we use to begin the perturbation theory. 

The first order solution is solved by inputting the zeroth order solution into the right-hand side of each exact solution in Eq. \eqref{eq:app:exact soltuions}. The only nonzero term at first order is given by
\begin{equation}
    \label{eq:sigma_ge first order v1}
    \bar{\sigma}_{ge}^{(1)}(t) = \int\limits_{-\infty}^{t} dt_1 G_{eg}(t,t_1)\hat F_+^{eg}(t_1),
\end{equation}
where $\hat F_+^{eg}(t_1)$ is a known operator. Note that all molecule operators on the right-hand side are absent because the zeroth order solution is the identity operator. At second order, the nonzero contributions are 
\begin{subequations}
    \begin{gather}
    \bar{\sigma}_{ee^\prime}^{(2)}(t) {=}\hspace{-1mm} \int\limits_{-\infty}^{t}dt_2G_{e^\prime e}(t,t_2) \int\limits_{-\infty}^{t_2} dt_1 G_{eg}^*(t_2,t_1)\hat F_-^{ge}(t_1)\hat F_+^{e^\prime g}(t_2)
    {+}\hspace{-1mm}\int\limits_{-\infty}^{t}dt_2G^*_{e e^\prime}(t,t_2)   \hat F_-^{ge}(t_2)\int\limits_{-\infty}^{t_2} dt_1 G_{e^\prime g}(t_2,t_1)\hat F_+^{e^\prime g}(t_1),\\
    \bar{\sigma}_{gg}^{(2)}(t)
    {=}\hspace{-1mm}\int\limits_{-\infty}^{t}dt_2G^*_{gg}(t,t_2) \int\limits_{-\infty}^{t_2} dt_1 G_{e^\prime g }^*(t_2,t_1)\hat F_-^{ge^\prime}(t_1)\hat F_+^{e^\prime g}(t_2)
    {+} \hspace{-1mm}\int\limits_{-\infty}^{t}dt_2G_{gg}(t,t_2)   \hat F_-^{ge^\prime}(t_2)\int\limits_{-\infty}^{t_2} dt_1 G_{e^\prime g}(t_2,t_1)\hat F_+^{e^\prime g}(t_1),\\
    \bar{\sigma}_{gf}^{(2)}(t) {=}\hspace{-1mm}\int\limits_{-\infty}^tdt_2G_{fg}(t,t_2)\int\limits_{-\infty}^{t_2} dt_1 G_{e^\prime g}(t_2,t_1)\hat F_+^{e^\prime g}(t_1)\hat F_+^{fe^\prime}(t_2).
    \end{gather}
\end{subequations}
Finally, we calculate the third order results for $\bar{\sigma}_{ge}(t)$ and $\bar{\sigma}_{ef}(t)$, since these are the only terms needed for the absorption; they are given by
\begin{equation}
\label{eq:sigma_ge third order v1}
\begin{split}
    \bar{\sigma}_{ge}^{(3)}(t) =& \int\limits_{-\infty}^{t} dt_3 G_{eg}(t,t_3)\int\limits_{-\infty}^{t_3}dt_2G^*_{gg}(t_3,t_2) \int\limits_{-\infty}^{t_2} dt_1 G_{e^\prime g}^*(t_2,t_1)\hat F_-^{ge^\prime}(t_1)\hat F_+^{e^\prime g}(t_2)\hat  F_+^{eg}(t_3)\\
    &+\int\limits_{-\infty}^{t} dt_3 G_{eg}(t,t_3)\int\limits_{-\infty}^{t_3}dt_2G_{gg}(t_3,t_2)  \hat  F_-^{ge^\prime}(t_2)\int\limits_{-\infty}^{t_2} dt_1 G_{e^\prime g}(t_2,t_1)\hat F_+^{e^\prime g}(t_1)\hat F_+^{eg}(t_3)\\
    %%%%
    &+\int\limits_{-\infty}^t dt_3 G^*_{ge}(t,t_3)\int\limits_{-\infty}^{t_3}dt_2G_{ee^\prime}(t_3,t_2) \int\limits_{-\infty}^{t_2} dt_1 G_{e^\prime g}^*(t_2,t_1)\hat F_-^{ge^\prime}(t_1)\hat F_+^{e g}(t_2)\hat F_+^{e^\prime g}(t_3)\\
    &+\int\limits_{-\infty}^t dt_3 G^*_{ge}(t,t_3)\int\limits_{-\infty}^{t_3}dt_2G^*_{e e^\prime}(t_3,t_2)   \hat F_-^{ge^\prime}(t_2)\int\limits_{-\infty}^{t_2} dt_1 G_{e g}(t_2,t_1)\hat F_+^{e g}(t_1)\hat F_+^{e^\prime g}(t_3)\\
    %%%%
    &+\int\limits_{-\infty}^t dt_3 G_{eg}(t,t_1)\hat F_-^{ef}(t_3)\int\limits_{-\infty}^{t_3}dt_2G_{fg}(t_3,t_2)\int\limits_{-\infty}^{t_2} dt_1 G_{e^\prime g}(t_2,t_1)\hat F_+^{e^\prime g}(t_1)\hat F_+^{fe^\prime}(t_2),
    %%%%
\end{split}
\end{equation}
and
\begin{equation}
\label{eq:sigma_ef third order v1}
    \begin{split}
    \bar{\sigma}_{ef}^{(3)}(t) &=\int\limits_{-\infty}^tdt_3G_{fe    }(t,t_3)\int\limits_{-\infty}^{t_3}dt_2G_{e^\prime e}(t_3,t_2) \int\limits_{-\infty}^{t_2} dt_1 G_{eg}^*(t_2,t_1)\hat F_-^{ge}(t_1)\hat F_+^{e^\prime g}(t_2)\hat F_+^{fe^\prime}(t_3)\\
    &+\int\limits_{-\infty}^tdt_3G_{fe    }(t,t_3)\int\limits_{-\infty}^{t_3}dt_2G^*_{e^\prime e}(t_3,t_2)   \hat F_-^{ge}(t_2)\int\limits_{-\infty}^{t_2} dt_1 G_{e^\prime g}(t_2,t_1)\hat F_+^{e^\prime g}(t_1)\hat F_+^{fe^\prime}(t_3)\\
    %%%%
    &+\int\limits_{-\infty}^t dt_3G^*_{ef}(t,t_3) \hat F_-^{ge}(t_3)\int\limits_{-\infty}^{t_3}dt_2G_{fg}(t_3,t_2)\int\limits_{-\infty}^{t_2} dt_1 G_{e^\prime g}(t_2,t_1)\hat F_+^{e^\prime g}(t_1)\hat F_+^{fe^\prime}(t_2).
\end{split}
\end{equation}
The only terms that contribute to the absorption are $\bar\sigma_{ge}^{(1)}(t),\bar\sigma_{ge}^{(3)}(t)$ and $\bar\sigma_{ef}^{(3)}(t)$.

\section{Stochastic average\label{appB:Stochastic average}}
We begin this section by taking the expectation value of the Green function $G_{ij}(t,t_1)$. To work out $[\![G_{ij}(t,t_1)]\!]$ we use the assumptions made in Section \ref{sec:model Hamiltonian} and Eq \eqref{eq:properties of stochastic variables}. We assume the random variables follow a Gaussian distribution, fluctuations of different energy levels are characterized by $c_{ij}$ with zero mean and the correlations follow a fast exponential decay with a correlation time $\tau_c$. Then using the cumulant expansion for a Gaussian distribution \cite{kardar2007statistical,mukamel1999principles} we evaluate 
\begin{equation}
\begin{split}
        [\![G_{ij}(t,t_1)]\!]& = ie^{-(\bar\Gamma_{ij} + i\omega_{ij})(t-t_1)}[\![e^{-i\int\limits_{t_1}^{t} dt^\prime\tilde{\omega}_{ij}(t^\prime)}]\!]\\
        & = ie^{-(\bar\Gamma_{ij} + i\omega_{ij})(t-t_1)}e^{-\frac{1}{2}\int\limits_{t_1}^{t}dt^\prime\int\limits_{t_1}^{t}dt^{\prime\prime}[\![\tilde{\omega}_{ij}(t^\prime)\tilde{\omega}_{ij}(t^{\prime\prime})]\!]}.
\end{split}
\end{equation}
Using Eq. \eqref{eq:stochastic correlation function} we evaluate the time integrals in the exponent to be 
\begin{equation}
    g_{ij}(t-t_1)\equiv\frac{1}{2}\int\limits_{t_1}^{t}dt^\prime\int\limits_{t_1}^{t}dt^{\prime\prime}[\![\tilde{\omega}_{ij}(t^\prime)\tilde{\omega}_{ij}(t^{\prime\prime})]\!] = \sigma_{ij}^2\tau_c^2\left(e^{-\frac{t-t_1}{\tau_c}} -1 + \frac{t-t_1}{\tau_c
    }   \right)
\end{equation}
where we defined $\sigma_{ij}^2 \equiv c_{ii}-2c_{ij}+c_{jj}\ge 0$, for which it follows that
\begin{equation}
    \label{eq:appA G_ijg_ij}
    [\![G_{ij}(t,t_1)]\!] = ie^{-(\bar\Gamma_{ij} + i\omega_{ij})(t-t_1)}e^{-g_{ij}(t-t_1)}.
\end{equation}

Taking the ``impact limit'' in which correlations decay on fast time scales compared to the coupling strength \cite{mukamel1999principles,mukamel1983nonimpact} 
\begin{equation}
    g_{ij}(t-t_1)\to \Lambda_{ij}(t-t_1),
\end{equation}
where we defined $\Lambda_{ij} = \sigma^2_{ij}\tau_c$ to be the dephasing decay rate between $i$ and $j$. Then all together the classical expectation value is given by
\begin{equation}
    [\![G_{ij}(t,t_1)]\!] = e^{-(\bar\Gamma_{ij} + \Lambda_{ij} + i\omega_{ij})(t-t_1)}
\end{equation}
and we define $G_{ij}(t-t_1)\equiv[\![G_{ij}(t,t_1)]\!]$ and the total decay rate $\gamma_{ij}\equiv\bar\Gamma_{ij} + \Lambda_{ij}$. 

\section{Absorption in time and frequency}
\label{sec:Absorption in time and frequency}
Taking the classical expectation value and inputting the coherence operators into Eq. \eqref{eq:quantum extinction approx} for the absorption we have
\begin{equation}
    \label{eq:absorption in time}
    \begin{split}
        \mathcal{A}&=\, \int_{-\infty}^{\infty}\hspace{-4mm}dt_2 \langle \boldsymbol{\mu}_{ge}\cdot \hat{\boldsymbol{E}}_{\boldsymbol{-}}(t_2)\frac{d}{dt_2} \int_{-\infty}^{t_2}\hspace{-4mm} dt_1 G_{eg}(t_2-t_1)\hat F_{+}^{eg}(t_1) \rangle + \text{c.c.},\\
        & +\int_{-\infty}^{\infty}\hspace{-4mm}dt_4 \langle  \boldsymbol{\mu}_{ge}\cdot \hat{\boldsymbol{E}}_{\boldsymbol{-}}(t_4)\frac{d}{dt_4}\int_{-\infty}^{t_4}\hspace{-4mm}dt_3G_{eg}(t_4-t_3)\int_{-\infty}^{t_3}\hspace{-4mm}dt_2G^*_{gg}(t_3-t_2)\int_{-\infty}^{t_2}\hspace{-4mm}dt_1G_{e^\prime g}^*(t_2-t_1)\hat F_-^{ge^\prime}(t_1)\hat F_+^{e^\prime g}(t_2)\hat F_+^{eg}(t_3)\rangle + \text{c.c.} \\
        &+\int_{-\infty}^{\infty}\hspace{-4mm}dt_4 \langle  \boldsymbol{\mu}_{ge}\cdot \hat{\boldsymbol{E}}_{\boldsymbol{-}}(t_4)\frac{d}{dt_4}\int_{-\infty}^{t_4}\hspace{-4mm}dt_3G_{eg}(t_4-t_3)\int_{-\infty}^{t_3}\hspace{-4mm}dt_2G_{gg}(t_3-t_2)\int_{-\infty}^{t_2}\hspace{-4mm}dt_1G_{e^\prime g}(t_2-t_1)\hat F_-^{ge^\prime}(t_2)\hat F_+^{e^\prime g}(t_1)\hat F_+^{eg}(t_3)\rangle + \text{c.c.}\\
        & +\int_{-\infty}^{\infty}\hspace{-4mm}dt_4 \langle  \boldsymbol{\mu}_{ge}\cdot \hat{\boldsymbol{E}}_{\boldsymbol{-}}(t_4)\frac{d}{dt_4}\int_{-\infty}^{t_4}\hspace{-4mm}dt_3G^*_{ge}(t_4-t_3)\int_{-\infty}^{t_3}\hspace{-4mm}dt_2G_{e e^\prime}(t_3-t_2)\int_{-\infty}^{t_2}\hspace{-4mm}dt_1G_{e^\prime g}^*(t_2-t_1)\hat F_-^{ge^\prime}(t_1)\hat F_+^{e g}(t_2)\hat F_+^{e^\prime g}(t_3)\rangle + \text{c.c.} \\
        &+\int_{-\infty}^{\infty}\hspace{-4mm}dt_4 \langle  \boldsymbol{\mu}_{ge}\cdot \hat{\boldsymbol{E}}_{\boldsymbol{-}}(t_4)\frac{d}{dt_4}\int_{-\infty}^{t_4}\hspace{-4mm}dt_3G^*_{ge}(t_4-t_3)\int_{-\infty}^{t_3}\hspace{-4mm}dt_2G^*_{e e^\prime}(t_3-t_2)\int_{-\infty}^{t_2}\hspace{-4mm}dt_1G_{e g}(t_2-t_1)\hat F_-^{ge^\prime}(t_2)\hat F_+^{e g}(t_1)\hat F_+^{e^\prime g}(t_3)\rangle + \text{c.c.}\\
        &+\int_{-\infty}^{\infty}\hspace{-4mm}dt_4 \langle  \boldsymbol{\mu}_{ge}\cdot \hat{\boldsymbol{E}}_{\boldsymbol{-}}(t_4)\frac{d}{dt_4}\int_{-\infty}^{t_4}\hspace{-4mm}dt_3G_{eg}(t_4-t_3)\int_{-\infty}^{t_3}\hspace{-4mm}dt_2G_{fg}(t_3-t_2)\int_{-\infty}^{t_2}\hspace{-4mm}dt_1G_{e^\prime g}(t_2-t_1)\hat F_-^{ef}(t_3)\hat F_+^{e^\prime g}(t_1)\hat F_+^{fe^\prime}(t_2)\rangle + \text{c.c.},\\
        &+\int_{-\infty}^{\infty}\hspace{-4mm}dt_4 \langle  \boldsymbol{\mu}_{ef}\cdot \hat{\boldsymbol{E}}_{\boldsymbol{-}}(t_4)\frac{d}{dt_4}\int_{-\infty}^{t_4}\hspace{-4mm}dt_3G_{fe}(t_4 - t_3)\int_{-\infty}^{t_3}\hspace{-4mm}dt_2G_{e^\prime e}(t_3-t_2)\int_{-\infty}^{t_2}\hspace{-4mm}dt_1G_{eg}^*(t_2-t_1)\hat F_-^{ge}(t_1)\hat F_+^{e^\prime g}(t_2)\hat F_+^{fe^\prime}(t_3)\rangle + c.c.\\
        &+\int_{-\infty}^{\infty}\hspace{-4mm}dt_4 \langle  \boldsymbol{\mu}_{ef}\cdot \hat{\boldsymbol{E}}_{\boldsymbol{-}}(t_4)\frac{d}{dt_4}\int_{-\infty}^{t_4}\hspace{-4mm}dt_3G_{fe}(t_4 - t_3)\int_{-\infty}^{t_3}\hspace{-4mm}dt_2G^*_{e^\prime e}(t_3-t_2)\int_{-\infty}^{t_2}\hspace{-4mm}dt_1G_{e^\prime g}(t_2-t_1)\hat F_-^{ge}(t_2)\hat F_+^{e^\prime g}(t_1)\hat F_+^{fe^\prime}(t_3)\rangle + c.c.\\
        &+\int_{-\infty}^{\infty}\hspace{-4mm}dt_4 \langle  \boldsymbol{\mu}_{ef}\cdot \hat{\boldsymbol{E}}_{\boldsymbol{-}}(t_4)\frac{d}{dt_4}\int_{-\infty}^{t_4}\hspace{-4mm}dt_3G^*_{ef}(t_4 - t_3)\int_{-\infty}^{t_3}\hspace{-4mm}dt_2G_{fg}(t_3-t_2)\int_{-\infty}^{t_2}\hspace{-4mm}dt_1G_{e^\prime g}(t_2-t_1)\hat F_-^{ge}(t_3)\hat F_+^{e^\prime g}(t_1)\hat F_+^{fe^\prime}(t_2) + c.c.,
    \end{split}
\end{equation}
where we have dropped the symbol for the stochastic average on the left-hand side. 

In putting the inverse Fourier transform of each $F_\pm^{ij}(t)$, computing the time integrals with the defined functions in Eq. \eqref{eq:define Q} and \eqref{eq:definition R} and combining the complex conjugate, the absorption is given by
\begin{equation}
    \label{eq:absorption in frequency}
    \begin{split}
       \mathcal{A} &=2\hspace{-1mm}\int d \omega_{1}\hbar\omega_{1}\text{Im}\left[ \frac{\langle \hat F_-^{ge}(-\omega_1)\hat F_+^{eg}(\omega_1)\rangle}{Q_{eg}(\omega_1)}\right]\\
       &{-}4\pi\hspace{-1mm}\int \dbarw{1}\dbarw{2}\dbarw{3}\dbarw{4}\hbar\omega_{4}\text{Im}\left[ \frac{\langle \hat  F_-^{ge}(-\omega_4)\hat F_-^{g e^\prime}(-\omega_3)\hat F_+^{e^\prime g}(\omega_2)\hat F_+^{eg}(\omega_1)\rangle}{Q_{eg}(\nu_{4})Q_{e^\prime g}^*(\omega_{3})Q_{e^\prime g}(\omega_{2})}R_{e^\prime e^\prime}(\omega_2 - \omega_3)\right]\delta(\omega_1 + \omega_2 - \omega_3 - \omega_4)\\
       &{-}4\pi\hspace{-1mm}\int \dbarw{1}\dbarw{2}\dbarw{3}\dbarw{4}\hbar\omega_{4}\text{Im}\left[ \frac{\langle \hat F_-^{ge}(-\omega_4) \hat F_-^{ge^\prime}(-\omega_3)\hat F_+^{eg}(\omega_2)\hat F_+^{e^\prime g}(\omega_1)\rangle}{Q_{eg}(\omega_{4})Q_{e^\prime g}^*(\omega_{3})Q_{eg}(\omega_{2})}R_{e e^\prime}(\omega_2 - \omega_3)\right]\delta(\omega_1 + \omega_2 - \omega_3 - \omega_4)\\
       &{+}4\pi\hspace{-1mm}\int \dbarw{1}\dbarw{2}\dbarw{3}\dbarw{4}\hbar\omega_{4}\text{Im}\left[\frac{\langle\hat F_-^{ge}(-\omega_4)\hat  F_-^{ef}(-\omega_3)\hat F_+^{e^\prime g}(\omega_2)\hat F_+^{f e^\prime}(\omega_1)\rangle}{Q_{eg}(\omega_4)Q_{fg}(\omega_1 + \omega_2)Q_{e^\prime g}(\omega_2)}\right]\delta(\omega_1 + \omega_2 - \omega_3 - \omega_4),\\
        &{+}4\pi\int \dbarw{1}\dbarw{2}\dbarw{3}\dbarw{4}\hbar\omega_{4}\text{Im}\left[ \frac{\langle \hat F_-^{ef}(-\omega_4)\hat F_-^{ge}(-\omega_3)F_+^{e^\prime g}(\omega_2)\hat F_+^{f e^\prime}(\omega_1)\rangle}{Q_{fe}(\omega_4)Q_{eg}^*(\omega_3)Q_{e^\prime g}(\omega_2)}  R_{e^\prime e}(\omega_2 - \omega_3)  \right]\delta(\omega_1 + \omega_2 - \omega_3 - \omega_4)\\
        &{-}4\pi\int \dbarw{1}\dbarw{2}\dbarw{3}\dbarw{4}\hbar\omega_{4}\text{Im}\left[ \frac{\langle \hat F_-^{ef}(-\omega_4)\hat F_-^{ge}(-\omega_3)\hat F_+^{e^\prime g}(\omega_2)\hat F_+^{fe^\prime}(\omega_1)\rangle}{Q_{fe}(\omega_4)Q_{fg}(\omega_1 + \omega_2)Q_{e^\prime g}(\omega_2)}\right]\delta(\omega_1 + \omega_2 - \omega_3 - \omega_4),
    \end{split}
\end{equation}
where we sum over $e,e^\prime$ and $f$ and $\text{Im}$ denotes the imaginary part of a complex number. Note that we leave the final result in terms of $\hat F_\pm^{ij}(t)$ because it is notionally simpler at this point. 

\section{Squeezing operator transformation \label{app:squeezing operator}}
We start with the squeezing operator given in Eq. \eqref{eq:squeezing operator} and define the operator
\begin{equation}
    B^\dagger = \frac{\beta}{2}\int d\omega_1d\omega_2\gamma(\omega_1,\omega_2)a^\dagger(\omega_1)a^\dagger(\omega_2)
\end{equation}
such that
\begin{equation}
    S(\beta) = e^{B^\dagger - B}.
\end{equation}
Using the Baker-Hausdorff lemma \cite{sakurai1995modern} (Equation 2.3.47), we have
\begin{equation}
    \begin{split}
        S^\dagger(\beta)a(\omega)S(\beta) =& \, a(\omega) + [B - B^\dagger,a(\omega)]+\frac{1}{2!}[B - B^\dagger,[B - B^\dagger,a(\omega)]]+\frac{1}{3!}[B - B^\dagger,[B - B^\dagger,[B - B^\dagger,a(\nu)]]]+...
    \end{split}
\end{equation}
which we can simplify by putting $C_{n+1}(\omega) = [B - B^\dagger,C_n(\omega)]$ with $C_1(\omega) = [B - B^\dagger,a(\omega)]$, then
\begin{equation}
    \label{eq:squeezed state transformation sum}
    S^\dagger(\beta)a(\omega)S(\beta) = a(\omega)  + \sum_{n=1}\frac{C_n(\omega). }{n!}
\end{equation}
We work out the commutators for $C_1(\omega)$, $C_2(\omega)$, $C_3(\omega)$ and then the general result follows. The first commutator is easily worked out to be
\begin{equation}
    \label{eq:C1 v1}
    C_1(\omega) = \beta\int d\omega_1 \alpha\left(\omega + \omega_1\right)\phi\left(\frac{\omega- \omega_1}{2}\right)a^\dagger(\omega_1).
\end{equation}
In the CW limit, $ \alpha(\omega + \omega_1)$ is strongly peaked at $\omega + \omega_1= 2\bar{\omega}$ which we can use to approximate the integrand. Then, to good approximation, $\phi\left(\frac{\omega - \omega_1}{2}\right)=\phi(\omega - \bar{\omega}
)$ which we pull out of the integrand so that
\begin{equation}
    \label{eq:C1 v2}
    C_1(\omega) = \beta\phi(\omega - \bar{\omega})\int d\omega_1 \alpha(\omega + \omega_1)a^\dagger(\omega_1).
\end{equation}
Evaluating the second commutator we have
\begin{equation}
    C_2(\omega) = |\beta|^2\phi(\omega - \bar{\omega})\int d\omega_1\alpha\left(\omega + \omega_1\right)\int d\omega_2 \gamma^*(\omega_1,\omega_2)a(\omega_2),
\end{equation}
and again in the CW limit $\alpha\left(\omega + \omega_1\right)$ is strongly peaked at $\omega + \omega_1=2\bar{\omega}$ so that to good approximation $\gamma(\omega_1,\omega_2)= \gamma(2\bar\omega - \omega,\omega_2)$. Then the second commutator simplifies to 
\begin{equation}
    C_2(\omega) = |\beta|^2\phi(\omega - \bar{\omega})\int d\omega_1\alpha\left(\omega + \omega_1\right)\int d\omega_2 \gamma^*(2\bar{\omega} - \omega,\omega_2)a(\omega_2),
\end{equation}
where all the $\omega_1$ dependence is in $\alpha\left(\omega + \omega_1\right)$. Then for frequencies $\omega$ of interest near $\bar\omega$ and within the CW limit we integrate over $\omega_1$ and find
\begin{equation}
    \int d\omega_1\alpha\left(\omega + \omega_1\right) = \sqrt{\Omega_p}.
\end{equation}
Expanding the biphoton wave function, the second commutator is
\begin{equation}
\begin{split}
    C_2(\omega) =&|\beta|^2\phi(\omega - \bar{\omega})\sqrt{\Omega_p}\int d\omega_2 \alpha^*\left(2\bar{\omega} - \omega+\omega_2\right)\phi^*\left(\frac{2\bar{\omega} - \omega-\omega_2}{2}\right)a(\omega_2).
\end{split}
\end{equation}
We repeat the process where in the CW limit $\omega= \omega_2$ and $\phi^*\left(\frac{2\bar{\omega} - \omega-\omega_2}{2}\right)= \phi^*(\omega - \bar{\omega})$. Then, 
\begin{equation}
    C_2(\omega) = \frac{1}{\sqrt{\Omega_p}}|\beta\phi(\omega - \bar{\omega})\sqrt{\Omega_p}|^2\int d\omega_2\alpha^*\left(2\bar{\omega} - \omega+\omega_2\right)a(\omega_2).
\end{equation}
Following the same steps for the third commutator we find
\begin{equation}
    C_3(\omega) = \frac{1}{\sqrt{ \Omega_p}}|\beta\phi(\omega - \bar{\omega})\sqrt{\Omega_p}|^2\beta\phi(\omega - \bar\omega)\sqrt{\Omega_p}\int d\omega_2\alpha\left(\omega  +\omega_3\right)a^\dagger(\omega_3),
\end{equation}
and in general
\begin{subequations}
    \begin{gather}
        C_n(\omega) =  \frac{1}{\sqrt{\Omega_p}}|\beta\phi(\omega - \bar{\omega})\sqrt{\Omega_p}|^{n}\int d\omega_2\alpha^*\left(2\bar{\omega} - \omega +  \omega_2\right)a(\omega_2),\hspace{5mm}\text{n even},\\
        C_n(\omega) =\frac{1}{\sqrt{ \Omega_p}}|\beta\phi(\omega - \bar{\omega})\sqrt{\Omega_p}|^{n}\frac{\beta\phi(\omega - \bar{\omega})}{|\beta\phi(\omega - \bar{\omega})|}\int d\omega_1 \alpha\left(\omega +  \omega_1\right)a^\dagger(\omega_1),\hspace{5mm}\text{n odd}.
    \end{gather}
\end{subequations}
In Eq. \eqref{eq:squeezed state transformation sum} we split the sum into odd and even contributions and then input the results for $C_n(\omega)$ so that
\begin{equation}
    \begin{split}
        S^\dagger(\beta)a(\omega)S(\beta) = a(\omega)&+ \frac{1}{\sqrt{\Omega_p}}\frac{\beta\phi(\omega - \bar{\omega})}{|\beta\phi(\omega - \bar{\omega})|}\left(\sum_{n=1}^{\text{odd}}  \frac{|\beta\phi(\omega - \bar{\omega})\sqrt{ \Omega_p}|^{n}}{n!} \right)\int d\omega_1\alpha(\omega +  \omega_1)a^\dagger(\omega_1)\\
        &+ \frac{1}{\sqrt{ \Omega_p}}\left(\sum_{n=2}^{\text{even}}  \frac{|\beta\phi(\omega - \bar{\omega})\sqrt{ \Omega_p}|^{n}}{n!} \right)\int d\omega_2\alpha^*\left(2\bar{\omega} - \omega +  \omega_2\right)a(\omega_2).
    \end{split}
\end{equation}
Using the Taylor series for $\text{sinh}x$ and $\text{cosh}x$ the sums are replaced with
\begin{equation}
    \label{eq:general squeezing transform 1}
    \begin{split}
        S^\dagger(\beta)a(\omega)S(\beta)= a(\omega)&+ \frac{[\text{cosh}(|\beta\phi(\omega - \bar{\omega})|\sqrt{\Omega_p}) - 1]}{\sqrt{\Omega_p}}\int d\omega_2\alpha^*\left(2\bar{\omega} - \omega +  \omega_2\right)a(\omega_2)\\
        &+ \frac{\beta\phi(\omega - \bar{\omega})}{|\beta\phi(\omega - \bar{\omega})|}\frac{\text{sinh}(|\beta\phi(\omega - \bar{\omega})|\sqrt{\Omega_p})}{\sqrt{\Omega_p}}\int d\omega_1\alpha\left(\omega +  \omega_1\right)a^\dagger(\omega_1),
    \end{split}
\end{equation}
which in our simple model of the phase-matching function (Eq. \eqref{eq:model of PMF}) simplifies to Eq. \eqref{eq:squeezed state transform v2}. 

Since the squeezing operator is unitary we check our approximations in moving from Eq. \eqref{eq:C1 v1} to Eq. \eqref{eq:C1 v2} by calculating its affect on the commutator $[a(\omega),a^\dagger(\omega^\prime)] = \delta(\omega-\omega^\prime)$. Then applying the squeezing transformation and evaluating the commutator we find
\begin{equation}
    \begin{split}
       &S^\dagger(\beta)[a(\omega),a^\dagger(\omega^\prime)]S(\beta)\\
       &= [S^\dagger(\beta)a(\omega)S(\beta),S^\dagger(\beta)a^\dagger(\omega^\prime)S(\beta)]\\
       &=\delta(\omega-\omega^\prime) + \frac{[\text{cosh}(|\beta\phi(\omega^\prime - \bar{\omega})|\sqrt{\Omega_p}) - 1]}{\sqrt{\Omega_p}}\alpha(2\bar\omega - \omega^\prime + \omega) + \frac{[\text{cosh}(|\beta\phi(\omega - \bar{\omega})|\sqrt{\Omega_p}) - 1]}{\sqrt{\Omega_p}}\alpha(2\bar\omega - \omega^\prime + \omega)\\
       &+\frac{[\text{cosh}(|\beta\phi(\omega - \bar{\omega})|\sqrt{\Omega_p}) - 1]}{\sqrt{\Omega_p}}\frac{[\text{cosh}(|\beta\phi(\omega^\prime - \bar{\omega})|\sqrt{\Omega_p}) - 1]}{\sqrt{\Omega_p}}\int d\omega_1 \alpha^*(2\bar\omega-\omega+\omega_1)\alpha(2\bar\omega-\omega^\prime + \omega_1)\\
       &-\frac{\beta\phi(\omega - \bar{\omega})}{|\beta\phi(\omega - \bar{\omega})|}\frac{\text{sinh}(|\beta\phi(\omega - \bar{\omega})|\sqrt{\Omega_p})}{\sqrt{\Omega_p}}\frac{\beta^*\phi^*(\omega^\prime - \bar{\omega})}{|\beta\phi(\omega^\prime - \bar{\omega})|}\frac{\text{sinh}(|\beta\phi(\omega^\prime - \bar{\omega})|\sqrt{\Omega_p})}{\sqrt{\Omega_p}}\int d\omega_1 \alpha(\omega+\omega_1)\alpha^*(\omega^\prime + \omega_1).
    \end{split}
\end{equation}
To push forward we use our definition of the pump function in Eq. \eqref{eq:squeezed state pump} and note that one representation of the delta function is given by \cite{kusse2010mathematical}
\begin{equation}
    \lim_{\Omega_p\to 0}\frac{\alpha(\omega)}{\sqrt{\Omega_p}} = \lim_{\Omega_p\to 0}\frac{1}{\Omega_p}\sinc\left( \frac{(\omega - 2\bar{\omega})\pi}{\Omega_p}\right) = \delta(\omega-2\bar\omega).
\end{equation}
Then using the above identity the commutator simplifies to
\begin{equation}
    \begin{split}
       &S^\dagger(\beta)[a(\omega),a^\dagger(\omega^\prime)]S(\beta)\\
       &=\delta(\omega-\omega^\prime) + 2[\text{cosh}(|\beta\phi(\omega - \bar{\omega})|\sqrt{\Omega_p}) - 1]\delta(\omega-\omega^\prime)+[\text{cosh}(|\beta\phi(\omega - \bar{\omega})|\sqrt{\Omega_p}) - 1]^2\delta(\omega-\omega^\prime) \\
       &- \text{sinh}^2(|\beta\phi(\omega - \bar{\omega})|\sqrt{\Omega_p})\delta(\omega-\omega^\prime),
    \end{split}
\end{equation}
then after some algebra and using the identity that $\text{cosh}^2(x)-\text{sinh}^2(x)=1$ we arrive at the final conclusion that
\begin{equation}
    S^\dagger(\beta)[a(\omega),a^\dagger(\omega^\prime)]S(\beta) = \delta(\omega-\omega^\prime),
\end{equation}
so the squeezing operator preserves the commutation relation.

\section{CW Squeezed state correlation functions \label{app:CW Squeezed state correlation functions}}
To calculate the correlation functions we begin with the simplified form of Eq. \eqref{eq:general squeezing transform 1} given in Eq. \eqref{eq:squeezed state transform v2}. Acting an annihilation operator on the squeezed state $\ket{\beta}$ and using Eq. \eqref{eq:squeezed state transform v2} and that $\alpha(\omega)$ is real we find
\begin{equation}
    \label{eq:acting 1 annihilation operator}
    \begin{split}
        a(\omega_1)\ket{\beta} &= S(\beta)S^\dagger(\beta)a(\omega_1)S(\beta)\ket{\text{vac}}\\
        &=S(\beta)\frac{s(\omega_1)e^{i\theta}}{\sqrt{\Omega_p}}\int d\omega^\prime\alpha\left(\omega_1 +  \omega^\prime\right)a^\dagger(\omega^\prime)\ket{\text{vac}},
    \end{split}
\end{equation}
which we will use to calculate the first correlation function. To calculate the second correlation function we act a second annihilation operator on the state and find
\begin{equation}
    \label{eq:calc step 1}
    \begin{split}
        a(\omega_2)a(\omega_1)\ket{\beta}&= S(\beta)\frac{s(\omega_1)e^{i\theta}}{\sqrt{\Omega_p}}\left(\alpha\left(\omega_1 + \omega_2\right) + \frac{(c(\omega_2) - 1)}{\sqrt{\Omega_p}}\int d\omega^\prime
        \alpha\left(2\bar{\omega} - \omega_2 + \omega^\prime\right)\alpha\left(\omega_1 + \omega^\prime\right)\right. \\
        &\left.\hspace{40mm}+ \frac{s(\omega_2)e^{i\theta}}{\sqrt{\Omega_p}}\int d\omega^\prime d\omega^{\prime\prime}\alpha\left(\omega_1 + \omega^\prime\right)\alpha\left(\omega_2 + \omega^{\prime\prime}\right)a^\dagger (\omega^{\prime\prime})a^\dagger (\omega^\prime) \right)\ket{\text{vac}}.
    \end{split}
\end{equation}
We can simplify Eq. \eqref{eq:calc step 1} by using the identity 
\begin{equation}
    \label{eq:sinc identity}
    \sinc(y-z) = \frac{1}{\pi}\int dx \sinc(x-y)\sinc(x-z),
\end{equation}
and find
\begin{equation}
    \int d\omega^\prime\alpha\left(2\bar{\omega} - \omega_2 + \omega^\prime\right)\alpha\left(\omega_1 + \omega^\prime\right) = \sqrt{\Omega_p}\alpha(\omega_1+\omega_2).
\end{equation}
Then Eq. \eqref{eq:calc step 1} simplifies to
\begin{equation}
\label{eq:acting 2 annihilation operators}
    \begin{split}
        &a(\omega_2)a(\omega_1)\ket{\beta} = S(\beta)\frac{s(\omega_1)e^{i\theta}}{\sqrt{\Omega_p}}\left(c(\omega_2)\alpha(\omega_1 + \omega_2) +  \frac{s(\omega_2)e^{i\theta}}{\sqrt{\Omega_p}}\int d\omega^\prime d\omega^{\prime\prime}\alpha(\omega_1 + \omega^\prime)\alpha(\omega_2 + \omega^{\prime\prime})a^\dagger (\omega^{\prime\prime})a^\dagger (\omega^\prime)      \right)\ket{\text{vac}}.
    \end{split}
\end{equation}

To calculate the first correlation function we take the adjoint of Eq. \eqref{eq:acting 1 annihilation operator} and combine it with itself. Then
\begin{equation}
\label{eq:squeezed state one-photon correlation function}
\begin{split}
    \bra{\beta}a^\dagger(\omega_2)a(\omega_1)\ket{\beta} &= \frac{s(\omega_2)s(\omega_1)}{\Omega_p}\int d\omega^{\prime\prime}\alpha(\omega_2 +  \omega^{\prime\prime})\int d\omega^\prime \alpha(\omega_1 +  \omega^\prime)\bra{\text{vac}}a(\omega^{\prime\prime})a^\dagger(\omega^\prime)\ket{\text{vac}}\\
    & = \frac{T_p}{2\pi}s(\omega_2)s(\omega_1)\text{sinc}\left(\frac{
        (\omega_1-\omega_2)\pi}{\Omega_p}\right),
\end{split}
\end{equation}
where we used $\Omega_p = 2\pi/T_p$. For the second correlation function again taking the adjoint of Eq. \eqref{eq:acting 2 annihilation operators} we have
\begin{equation}
    \begin{split}
        &\bra{\beta}a^\dagger(\omega_4)a^\dagger(\omega_3)a(\omega_2)a(\omega_1)\ket{\beta}\\
        &=\frac{s(\omega_4)c(\omega_3)c(\omega_2)s(\omega_1)}{\Omega_p} \alpha(\omega_4 + \omega_3)\alpha(\omega_2 + \omega_1)\\
        &+\frac{s(\omega_4)s(\omega_3)s(\omega_2)s(\omega_1)}{\Omega^2_p}\int d\omega^\prime d\omega^{\prime\prime}d\omega^{\prime\prime\prime}d\omega^{\prime\prime\prime\prime}\alpha(\omega_1 + \omega^\prime)\alpha(\omega_2 + \omega^{\prime\prime})\alpha(\omega_3 + \omega^{\prime\prime\prime})\alpha(\omega_4 + \omega^{\prime\prime\prime\prime})\\
        &\hspace{80mm}\times\bra{\text{vac}}a(\omega^{\prime\prime\prime\prime})a(\omega^{\prime\prime\prime})a^\dagger(\omega^{\prime\prime})a^\dagger(\omega^\prime)\ket{\text{vac}},
    \end{split}
\end{equation}
where
\begin{equation}
\label{eq:two ways to choose two photons}
    \bra{\text{vac}}a(\omega^{\prime\prime\prime\prime})a(\omega^{\prime\prime\prime})a^\dagger(\omega^{\prime\prime})a^\dagger(\omega^\prime)\ket{\text{vac}} = \delta(\omega^{\prime} - \omega^{\prime\prime\prime})\delta(\omega^{\prime\prime} - \omega^{\prime\prime\prime\prime}) + \delta(\omega^{\prime\prime\prime} - \omega^{\prime\prime})\delta(\omega^{\prime} - \omega^{\prime\prime\prime\prime}).
\end{equation}
Then using the identity in Eq. \eqref{eq:sinc identity} we simplify the second term to be
\begin{equation}
    \begin{split}
        &\int d\omega^\prime d\omega^{\prime\prime}d\omega^{\prime\prime\prime}d\omega^{\prime\prime\prime\prime}\alpha(\omega_1 + \omega^\prime)\alpha(\omega_2 + \omega^{\prime\prime})\alpha(\omega_3 + \omega^{\prime\prime\prime})\alpha(\omega_4 + \omega^{\prime\prime\prime\prime})\bra{\text{vac}}a(\omega^{\prime\prime\prime\prime})a(\omega^{\prime\prime\prime})a^\dagger(\omega^{\prime\prime})a^\dagger(\omega^\prime)\ket{\text{vac}}\\
        & = \Omega_p\alpha(\omega_1 - \omega_3 + 2\bar{\omega})\alpha(\omega_2 - \omega_4 + 2\bar{\omega}) + \Omega_p\alpha(\omega_1 - \omega_4 + 2\bar{\omega})\alpha(\omega_2 - \omega_3 + 2\bar{\omega}),
    \end{split}
\end{equation}
with the final result for the correlation function
\begin{equation}
    \begin{split}
        \label{eq:squeezed state two-photon correlation function}
        &\bra{\beta}a^\dagger(\omega_4)a^\dagger(\omega_3)a(\omega_2)a(\omega_1)\ket{\beta} \\
        &= \frac{T_p}{2\pi} s(\omega_4)c(\omega_3)c(\omega_2)s(\omega_1)\alpha(\omega_2 + \omega_1)\alpha(\omega_4 + \omega_3)\\
        &+\frac{T_p}{2\pi}s(\omega_4)s(\omega_3)s(\omega_2)s(\omega_1)[\alpha(\omega_1 - \omega_3 + 2\bar{\omega})\alpha(\omega_2 - \omega_4 + 2\bar{\omega}) + \alpha(\omega_1 - \omega_4 + 2\bar{\omega})\alpha(\omega_2 - \omega_3 + 2\bar{\omega})].
    \end{split}
\end{equation} 

\bibliography{apssamp}

\end{document}